\documentclass[11pt,a4]{book}
\usepackage{fullpage}

\usepackage[latin1]{inputenc}
\usepackage[english,french]{babel}
\usepackage[francais]{layout}
\usepackage{latexsym,amssymb,graphicx,amsmath,yhmath}
\usepackage{amsthm}
\usepackage{subfigure}
\usepackage{color}
\usepackage[cyr]{aeguill} 

\usepackage{fancyheadings}
\usepackage{paralist}
\usepackage[table]{xcolor}
\newcommand{\remove}[1]{}

\newcommand{\C}{\mbox{\tt S}}
\newcommand{\dis}{\mbox{\sf d}}
\newcommand{\id}{\mbox{\sf Id}}
\newcommand{\Label}{\mbox{$\ell$}}
\newcommand{\lab}{\mbox{$\ell$}}
\newcommand{\SC}{$\mbox{sans-cycle}$}
\newcommand{\MST}{$\mbox{\rm MST}$}

\newcommand{\nd}{\mbox{n{\oe}ud}}
\newcommand{\rob}{\mbox{\sc r}}
\newcommand{\tb}{\mbox{\sc {\small TB}}}
\newcommand{\free}{\mbox{\sc f}}
\newcommand{\GHS}{Gallager, Humblet et Spira}
\newcommand{\BFSC}{\mbox{\tt BFS}$_{SC}$}
\newtheorem{lemma}{Lemma}

\newtheorem{definition}{Définition}
\newtheorem{theorem}{Théoreme}

\newtheorem{propriete}{Propriété}
\newtheorem{problem}{Problème ouvert}

\author{L\'{e}lia Blin }
\begin{document}

\begin{titlepage}
\begin{center}
\vspace*{4cm}
\noindent \Large \textbf{\textsc{Algorithmes auto-stabilisants pour la construction d'arbres couvrants et la gestion d'entités autonomes}} \\
\medbreak
{\Large \textbf{\textsc{\textcolor{gray}{Self-stabilizing algorithms for spanning tree construction and for the management of mobile entities}}}} \\
\vspace*{1,5cm}
\noindent  Lélia Blin\\
\vspace*{1,5cm}
\noindent  Rapport scientifique présenté en vue de l'obtention\\ de l'Habilitation à Diriger les Recherches \\
\vspace*{1,5cm}
\noindent  soutenue le 1 décembre 2011\\
\vspace*{0.3cm}
\noindent {\Large à l'Université Pierre et Marie Curie - Paris 6} \\
\vspace*{0.5cm}
\end{center}
\noindent \textbf{Devant le jury composé de :} \\

\begin{tabular}{llll}
 \multicolumn{2}{l}{\textbf{Rapporteurs :}}&&\\
Paola  &\textsc{Flocchini},& Professeur, & Université d'Ottawa, Canada.\\
Toshimitsu & \textsc{Masuzawa}, & Professeur, & Université d'Osaka, Japon.\\	
Rachid &\textsc{Guerraoui}, & Professeur, & École Polytechnique Fédérale de Lausanne, Suisse.\\\\
 \multicolumn{2}{l}{\textbf{Examinateurs :}}&&\\
 Antonio &\textsc{Fernández Anta},& Professeur, & Université Rey Juan Carlos, Espagne.\\
Laurent & \textsc{Fribourg},& DR CNRS,& ENS Cachan, France.\\
Colette &\textsc{Johnen}, & Professeur, & Université de Bordeaux, France.\\
Franck & \textsc{Petit},& Professeur, & Université Pierre et Marie Curie, France.\\
Sébastien & \textsc{Tixeuil},& Professeur, & Université Pierre et Marie Curie, France.\\

\end{tabular}

\end{titlepage}
\sloppy
.
\newpage
\pagenumbering{roman}
\setcounter{page}{1}
\pagestyle{fancy}
\renewcommand{\footrulewidth}{0,02cm}

\renewcommand{\chaptermark}[1]{ %
  \markboth{\footnotesize \slshape \chaptername\ \thechapter\ : #1}{}}
\renewcommand{\sectionmark}[1]{ %
  \markright{\footnotesize \slshape \thesection\ : #1}}
\lhead[\footnotesize \thepage]{\rightmark}
\rhead[\leftmark]{\footnotesize \thepage}
\chead{}
\lfoot[\fancyplain{}{\scriptsize \slshape Habilitation à diriger les recherches, %
 2011}]{\fancyplain{}{}}

\rfoot[{\scriptsize \slshape Lélia Blin}]
\cfoot{}      


\tableofcontents

\chapter*{Summary of the document in English}
\addcontentsline{toc}{chapter}{Summary of the document in English}
\chaptermark{Summary of the document in English}
\label{chap:english}

In the context of large-scale networks, the consideration of \emph{faults} is an evident necessity. This document is focussing on the  \emph{self-stabilizing} approach which aims at conceiving algorithms ``repairing themselves'' in case of transient faults, that is of  faults implying an arbitrary modification of the states of the processes. The document focuses on two different contexts, covering the major part of my research work these last years. The first part of the document (Part~\ref{partie:un}) is dedicated to the design and analysis of self-stabilizing algorithms  for  \emph{networks of processes}. The second part of the document (Part~\ref{part2}) is dedicated  to the design and analysis of self-stabilizing algorithms  for \emph{autonomous entities} (i.e., software agents, robots, etc.) moving in a network.


\paragraph{Constrained Spanning Tree Construction.}

The first part is characterized by two specific aspects. One is the nature of the considered problems. The other is the permanent objective of optimizing the performances of the algorithms. Indeed, within the framework of spanning tree construction, self-stabilization mainly focused on the most classic constructions, namely  BFS trees,  DFS trees, or shortest path trees. We are interested in the construction of trees in a vaster framework, involving constraints of a \emph{global} nature, in both static and dynamic networks. We contributed in particular to the development of algorithms for the self-stabilizing construction of minimum-degree spanning trees, minimum-weight spanning tree (MST), and Steiner trees. Besides, our approach of self-stabilization aims not only at the feasibility but also also includes the search for  \emph{effective} algorithms. The main measure of complexity that we are considering is the memory used by every process. We however also considered  other measures, as the   convergence time and the quantity of information exchanged between the processes.

This study of effective construction of spanning trees  brings to light two facts. On one hand, self-stabilization seems to have a domain of applications as wide as distributed computing. Our work demonstrate that it is definitively case in the field of the spanning tree construction. On the other hand, and especially, our work on memory complexity seems to indicate that self-stabilization does not imply additional cost. As a typical example, distributed MST construction requires a memory of $\Omega(\log n)$ bits per process (if only to store its parent in the tree). We shall see in this document that it is possible to conceive a self-stabilizing MST construction  algorithm of using $O(\log n) $ bits of memory per process. 

\subparagraph {\rm {\em Organization of Part I.}}

Chapter~\ref{chap:SS-ST} summarizes the main lines of the theory of self-stabilization, and describes the elementary notions of graph theory used in this document. It also provides  a brief state-of-the-art of the self-stabilizing algorithms for the construction of spanning trees optimizing criteria not considered further in the following chapters.
Chapter~\ref{chap:MST} summarizes my contribution to the self-stabilizing construction of MST. My related papers are~\cite{BlinPRT09,BlinDPR10}. 
Finally, Chapter~\ref{chap:MDST} presents my works on the self-stabilizing construction of trees optimizing criteria different from minimum weight, such as minimum-degree spanning tree, and Steiner trees. My related papers are~\cite{BlinPRT09,BlinPR09b,BlinPRT10,BlinPR11}.


\paragraph{Autonomous Entities.}

The second part of the document is dedicated to the design and analysis of self-stabilizing  algorithms for \emph{autonomous entities}. This latter term refers to any computing entity susceptible to move in a space according to certain constraints. We shall consider mostly physical robots moving in a discrete or continuous space. We can make  however sometimes reference to contexts involving   software agents in a network. For the sake of simplicity, we shall use the terminology ``\emph{robot}'' in every case. Self-stabilization is a generic technique to tolerate any transient failure in a distributed system that is obviously interesting to generalize in the framework where the algorithm is executed by robots (one often rather refers to \emph{self-organization} instead  of self-stabilization). It is worth noticing strong resemblances between the self-stabilizing algorithmic for networks and the one for robots. For example, the notion of \emph{token} circulation in the former framework seems very much correlated with the circulation of \emph{robots} in the latter framework. In a similar way, we cannot miss noticing a resemblance between the traversal of graphs by messages, and  graph \emph{exploration} by a robot. The equivalence (in term of calculability) between the message passing model and the ``agent model'' was already brought to light in the literature~\cite{BarriereFFS03,ChalopinGMO06}. This document seems to indicate that this established equivalence could be extended to the framework of auto-stabilization. The current knowledge in self-stabilizing algorithms for robots is not elaborated enough to establish this generalization yet. Also, the relative youth of robots self-stabilization theory, and the lack of tools, prevent us to deal with efficiency (i.e., complexity) as it can be done in the context of network self-stabilization. In this document, we thus focused on feasibility (i.e., calculability) of elementary problems such as naming and graph exploration. To this end, we studied various models with for objective either to determine the minimal hypotheses of a model for the realization of a task, or to determine for a given model, the maximum corruption the robots can possibly tolerate.

This study of robot self-stabilization underlines the fact that it is possible to develop self-stabilizing solutions for robots within the framework of a very constrained environment, including  maximal hypotheses on the robots and system corruption, and minimal hypotheses on the strength of the model.

\subparagraph{\rm{\em Organization of Part II.}}

Chapter~\ref{chap:corruptionRobot} presents my main results obtained in a model where the faults can be generated by the robots  \emph{and} by the network. These results include impossibility results as well as determinist and probabilistic algorithms, for various problems including  \emph{naming} and \emph{election}. My related paper is~\cite{BlinPT07}. 
Chapter~\ref{chap:perpetuelle} summarizes then my work on the search for minimal hypotheses in the discrete CORDA model enabling to achieve a task (in a self-stabilizing manner). It is demonstrated that, in spite of the weakness of the model, it is possible for robots to perform  sophisticated tasks, among which is \emph{perpetual exploration}. My related paper is~\cite{BlinMPT10}.


\paragraph{Perspectives.} The document opens a certain number of long-term research directions, detailed in Chapter~\ref{chap:Conclusion}. My research perspectives get organized around the study of the tradeoff between the memory space used by the nodes of a network, the convergence time of the algorithm, and the quality of the retuned solution.

\chapter*{Introduction}
\addcontentsline{toc}{chapter}{Introduction}
\chaptermark{Introduction}
\label{chap:Intro1}
\pagenumbering{arabic}
\setcounter{page}{1}

Dans le contexte des réseaux à grande échelle,  la prise en compte des \emph{pannes} est une nécessité évidente.  Ce document s'intéresse à l'approche \emph{auto-stabilisante} qui vise à concevoir des algorithmes se \og réparant d'eux-même \fg\/ en cas de fautes transitoires, c'est-à-dire de pannes impliquant la modification arbitraire de l'état des processus. Il se focalise sur deux contextes différents, couvrant la majeure partie de mes travaux de recherche ces dernières années. La première partie du document (partie~\ref{partie:un}) est consacrée à l'algorithmique  auto-stabilisante pour les \emph{réseaux de processus}. La seconde partie du document (partie~\ref{part2}) est consacrée quant à elle à l'algorithmique  auto-stabilisante pour des \emph{entités autonomes} (agents logiciels, robots, etc.) se déplaçant dans un réseau.


\paragraph{Arbres couvrants sous contraintes.}

La première partie se caractérise par deux aspects spécifiques. Le premier est lié à la nature des problèmes considérés. Le second est lié à un soucis d'optimisation des performances des algorithmes. En effet, dans le cadre de la construction d'arbres couvrants, l'auto-stabilisation s'est historiquement principalement focalisée sur les constructions les plus classiques, à savoir arbres BFS, arbres DFS, ou arbres de plus courts chemins. Nous nous sommes intéressés à la construction d'arbres dans un cadre plus vaste, impliquant des contraintes \emph{globales}, dans des réseaux statiques ou dynamiques. Nous avons en particulier contribué au développement d'algorithmes pour la construction auto-stabilisante d'arbres couvrants de degré minimum, d'arbres couvrants de poids minimum (MST), ou d'arbres de Steiner. Par ailleurs, notre approche de l'auto-stabilisation ne vise pas seulement la faisabilité mais inclut également la recherche d'algorithmes \emph{efficaces}. La principale mesure de complexité visée est la mémoire utilisée par chaque processus. Nous avons toutefois considéré également d'autres mesures, comme le temps de convergence ou la quantité d'information échangée entre les processus. 

De cette étude de la construction efficace d'arbres couvrants, nous mettons en évidence deux enseignements. D'une part, l'auto-stabilisation semble avoir un spectre d'applications aussi large que le réparti. Nos travaux démontrent que c'est effectivement le cas dans le domaine de la construction d'arbres couvrants. D'autre part, et surtout, nos travaux sur la complexité mémoire des algorithmes semblent indiquer que l'auto-stabilisation n'implique pas de coût supplémentaire. A titre d'exemple caractéristique, construire un MST en réparti nécessite une mémoire de $\Omega(\log n)$ bits par processus (ne serait-ce que pour stocker le parent dans son arbre). Nous verrons dans ce document qu'il est possible de concevoir un algorithme auto-stabilisante de construction de MST  utilisant $O(\log n)$ bits de mémoire par processus. 

\subparagraph{\rm{\em Organisation de la partie I.}}

Le chapitre~\ref{chap:SS-ST} rappelle les grandes lignes de la théorie de l'auto-stabilisation, et décrit les notions élémentaires de théorie des graphes utilisées dans ce document. Il dresse en particulier un bref état de l'art des algorithmes auto-stabilisants pour la construction d'arbres couvrants spécifiques ou optimisant des critères non considérés dans les chapitres suivants.
Le chapitre~\ref{chap:MST} présente mes travaux sur la construction d'arbres couvrant de poids minimum (MST).
Enfin le chapitre~\ref{chap:MDST}  a pour objet de présenter mes travaux sur la construction d'arbres couvrants optimisés, différents du MST, tel que l'arbre de degré minimum, l'arbre de Steiner, etc. 


\paragraph{Entités autonomes.}

La seconde partie du document est consacrée à l'algorithmique répartie auto-stabilisante pour les \emph{entités autonomes}. Ce terme désigne toute entité de calcul susceptible de se déplacer dans un espace selon certaines contraintes. Nous sous-entendrons le plus souvent des robots physiques se déplaçant dans un espace discret ou continu. Nous pourrons toutefois parfois faire référence à des contextes s'appliquant à des agents logiciels dans un réseau. Par abus de langage, nous utiliserons la terminologie brève et imagée de \emph{robot} dans tous les cas. L'auto-stabilisation est une technique générique pour tolérer toute défaillance transitoire dans un système réparti qu'il est évidemment envisageable de généraliser au cadre où les algorithmes sont exécutés par des robots (on parle alors souvent plutôt d'\emph{auto-organisation} que d'auto-stabilisation). 
Il convient de noter de fortes similitudes entre l'algorithmique auto-stabilisante pour les réseaux de processus et celle pour les robots. Par exemple, la notion de circulation de  \emph{jetons} dans le premier cadre semble corrélée à la circulation de  \emph{robots} dans le second cadre. De manière similaire, on ne peut manquer de noter une similitude entre \emph{parcours} de graphes par des messages, et \emph{exploration} par des robots. L'équivalence (en terme de calculabilité) entre le modèle par passage de messages et celui par agents a déjà été mis en évidence dans la littérature~\cite{BarriereFFS03,ChalopinGMO06}. 
Ce document semble indiquer une généralisation de cet état de fait à l'auto-stabilisation. Les connaissances en algorithmique auto-stabilisante pour les robots ne sont toutefois pas encore suffisamment élaborées pour établir cette généralisation. De même, la relative jeunesse de l'auto-stabilisation pour les robots ne permet de traiter de questions d'efficacité (i.e., complexité) que difficilement. Dans ce document, nous nous sommes surtout focalisée sur la faisabilité (i.e., calculabilité) de problèmes élémentaires tels que le nommage ou l'exploration. A cette fin, nous avons étudié différents modèles avec pour objectif soit de déterminer les hypothèses minimales d'un modèle pour la réalisation d'une tâche, soit de déterminer, pour un modèle donné, la corruption maximum qu'il est possible de toléré. 

De cette étude de l'auto-stabilisation pour les robots, nous mettons en évidence un enseignement principal, à savoir qu'il reste possible de développer des solutions auto-stabilisantes dans le cadre d'environnements très contraignants, incluant des hypothèses maximales sur la corruption des robots et du système, et des hypothèses minimales sur la force du modèle. 

\subparagraph{\rm{\em Organisation de la partie II.}}

Le chapitre~\ref{chap:corruptionRobot} présente mes principaux résultats obtenus dans un modèle où les fautes peuvent être générées par le réseau et par les robots eux-mêmes. Ces résultats se déclinent en  résultats d'impossibilité, et en algorithmes déterministes ou probabilistes, ce pour  les problèmes du \emph{nommage} et de l'\emph{élection}. Le chapitre~\ref{chap:perpetuelle} résume ensuite mon travail sur la recherche d'hypothèses minimales dans le modèle CORDA discret permettant de réaliser une tâche. Il y est en particulier démontré que malgré la faiblesse du modèle, il reste encore possible pour des robots d'effectuer des tâches sophistiquées, dont en particulier l'\emph{exploration perpétuelle}.  


\paragraph{Perspectives.} Le document ouvre un certain nombre de perspectives de recherche à long terme, détaillées dans  le chapitre~\ref{chap:Conclusion}. Ces perspectives s'organisent  autour de l'étude du compromis entre l'espace utilisé par les n{\oe}uds d'un réseau, le temps de convergence de l'algorithme,  et la qualité de la solution retournée.  

\part{Arbres couvrants sous contraintes}
\label{partie:un}
\chapter{Algorithmes auto-stabilisants et arbres couvrants}
\label{chap:SS-ST}

Ce chapitre rappelle les grandes lignes de la théorie de l'auto-stabilisation, et décrit les notions élémentaires de théorie des graphes utilisées dans ce document. Il dresse en particulier un bref état de l'art des algorithmes auto-stabilisants pour la construction d'arbres couvrants spécifiques (BFS, DFS, etc.) ou optimisant des critères non considérés dans la suite du document (diamètre minimal, etc.). 

\section{Eléments de la théorie de l'auto-stabilisation}

Une panne (appelée aussi \emph{faute}) dans un système réparti désigne une défaillance temporaire ou définitive d'un ou plusieurs composants du système. Par composants, nous entendons essentiellement processeurs, ou liens de communications. Il existe principalement deux catégories d'algorithmes traitant des pannes~: les algorithmes \emph{robustes}~\cite{Tel94} et les algorithmes \emph{auto-stabilisants}. Les premiers utilisent typiquement des techniques de redondance de l'information et des composants (communications ou processus). Ce document s'intéresse uniquement à la seconde catégorie d'algorithmes, et donc à l'approche auto-stabilisante. Cette approche vise à concevoir des algorithmes se \og réparant eux-même \fg\/ en cas de fautes transitoires.

Dijkstra~\cite{Dijkstra74} est considéré comme le fondateur de la théorie de l'auto-stabilisation. Il définit un système auto-stabilisant comme un système qui, quelque soit son état initial, est capable de retrouver de lui même un état \emph{légitime} en un nombre fini d'étapes. Un état légitime est un état qui respecte la spécification du problème à résoudre. De  nombreux ouvrages ont été écrits dans ce domaine~\cite{Dolev00,Tixeuil06,Devisme06}. Je ne ferai donc pas une présentation exhaustive de l'auto-stabilisation, mais rappellerai uniquement dans ce chapitre  les notions qui seront utiles à la compréhension de ce document.

Un \emph{système réparti}  est un réseau composé de processeurs, ou \emph{n{\oe}ud}, (chacun exécutant un unique processus), et de mécanismes de communication entre ces n{\oe}uds. Un tel système est modélisé par un graphe non orienté. Si les n{\oe}uds sont indistingables, le réseau est dit \emph{anonyme}. Dans un système non-anonyme, les n{\oe}uds disposent d'identifiants distincts deux-à-deux. Si tous les n{\oe}uds utilisent le même algorithme, le système est dit \emph{uniforme}. Dans le cas contraire, le système est dit \emph{non-uniforme}. Lorsque  quelques n{\oe}uds exécutent un algorithme différent de l'algorithme exécuté par tous les autres, le système est dit \emph{semi-uniforme}. L'exemple le plus classique d'un algorithme semi-uniforme est un algorithme utilisant un n{\oe}ud distingué, par exemple comme racine pour la construction d'un arbre couvrant. 

L'hypothèse de base en algorithmique répartie est que chaque n{\oe}ud peut communiquer avec tous ses voisins dans le réseau. 
Dans le contexte de l'auto-stabilisation, trois grands types de mécanismes de communication sont considérés:   
\begin{inparaenum}[\itshape (i\upshape)] 
\item le modèle \emph{ à états},  dit aussi  modèle \emph{ à mémoires partagées}~\cite{Dijkstra74},
\item le modèle \emph{ à registres partagées}~\cite{DolevIM93}, et
\item le modèle \emph{par passage de messages}~\cite{Tel94,Peleg00,Santoro06}
\end{inparaenum}
Dans le modèle à état, chaque n{\oe}ud peut lire l'état de tous ses voisins et mettre à jour son propre état en une étape atomique.
Dans le modèle à registres partagés,  chaque n{\oe}ud peut  lire le registre d'un de ses voisins, ou mettre à jour son propre état, en une étape atomique, mais pas les deux à la fois. 
Dans le modèle par passage de messages un n{\oe}ud envoie un message à un de ses voisins ou reçoit un message d'un de ses voisins (pas les deux à la fois),  en une étape atomique. Les liens de communication sont généralement considérés comme FIFO, et les messages sont traités dans leur ordre d'arrivée. Dans son livre~\cite{Peleg00}, Peleg propose une classification des modèles par passage de messages. Le modèle ${\cal CONGEST}$ est le plus communément utilisé dans ce document. Il se focalise sur le volume de communications communément admis comme \og raisonnable \fg, à savoir $O(\log n)$ bits par message, où $n$ est le nombre de n{\oe}uds dans le réseau. Notons que si les n{\oe}uds possèdent des identifiants deux-à-deux distincts entre 1 et $n$, alors $\log n$ bits est la taille minimum requise pour le codage de ces identifiants.  Avec une taille de messages imposée, on peut alors comparer le temps de convergence (mesuré en nombre d'étapes de communication) et le nombre de messages échangés.

Notons qu'il existe des transformateurs pour passer d'un modèle à un autre, dans le cas des graphes non orientés~\cite{Dolev00}. L'utilisation de l'un ou l'autre des modèles ci-dessus n'est donc pas restrictive. 

Si les temps pour transférer une information d'un n{\oe}ud à un voisin (lire un registre, échanger un message, etc.) sont identiques, alors le système est dit \emph{synchrone}. Sinon, le système est dit \emph{asynchrone}. Si les temps pour transférer une information d'un n{\oe}ud à un voisin sont potentiellement différent mais qu'une borne supérieure sur ces temps  est connue, alors  le système est dit \emph{semi-synchrone}. 
Dans les systèmes asynchrones, il est important de modéliser le comportement individuel de chaque n{\oe}ud. Un n{\oe}ud est dit \emph{activable} dès qu'il peut effectuer une action dans un algorithme donné. Afin de modéliser le comportement des n{\oe}uds activables, on utilise un ordonnanceur,  appelé  parfois \emph{démon} ou \emph{adversaire},  tel que décrit dans~\cite{DattaGT00,KakugawaY02,DanturiNT09}. Dans la suite du document, le terme d'adversaire est utilisé. L'adversaire est un dispositif indépendant des n{\oe}uds, et possédant une vision globale. A chaque pas de calcul, il choisit les n{\oe}uds susceptibles d'exécuter une action parmi les n{\oe}uds activables. L'adversaire est de puissance variable selon  combien de processus activables peuvent être activés à chaque pas de calcul:
\begin{itemize}
\item l'adversaire est dit \emph{central} (ou séquentiel) s'il n'active qu'un seul n{\oe}ud activable;
\item l'adversaire est dit \emph{distribué} s'il peut activer plusieurs n{\oe}uds parmi ceux qui sont activables;
\item l'adversaire est dit \emph{synchrone} (ou parallèle) s'il doit activer tous les n{\oe}uds activables.
\end{itemize}
L'adversaire est par ailleurs contraint par des hypothèse liées à l'équité de ses choix. Les contraintes d'équité les plus courantes sont les suivantes:
\begin{itemize}
\item l'adversaire est dit \emph{faiblement équitable} s'il doit ultimement activer tout n{\oe}ud continument et infiniment activable;
\item l'adversaire \emph{fortement équitable} s'il doit ultimement activer tout n{\oe}ud infiniment activable;
\end{itemize}
L'adversaire est dit \emph{inéquitable} s'il n'est pas équitable (ni fortement, ni faiblement). 
Les modèles d'adversaires ci-dessus sont plus ou moins contraignants pour le concepteur de l'algorithme. Il peut également découler différents  résultats d'impossibilité de ces différents adversaires . 

On dit qu'un algorithme a \emph{convergé} (ou qu'il a \emph{terminé}) lorsque son état global est conforme à la spécification attendue, comme par exemple la présence d'un unique leader dans le cas du problème de l'élection. Dans le modèle à états ou celui à registres partagés, un algorithme est dit \emph{silencieux}~\cite{DolevGS96} si les valeurs des variables locales des n{\oe}uds ne changent plus après la convergence.  Dans un modèle à passage de messages,  un algorithme est dit silencieux s'il n'y a plus de circulation de messages, ou si le contenu des messages échangés ne changent pas après convergence. Dolev, Gouda et Shneider~\cite{DolevGS96} ont prouvé que, dans un modèle à registres, la mémoire minimum requise parun algorithme auto-stabilisant silencieux de construction d'arbre couvrant est $\Omega(\log n)$ bits sur chaque  n{\oe}ud.

Les performances des algorithmes se mesurent à travers de leur complexité en mémoire (spatiale) et de leur temps de convergence. On établit la performance en mémoire en mesurant l'espace mémoire occupé en chaque n{\oe}ud, et/ou en mesurant la taille des messages échangés. Le temps de convergence d'un algorithme est le  \og temps \fg\/ qu'il met à atteindre la spécification demandée après une défaillance. L'unité de mesure du temps la plus souvent utilisée en auto-stabilisation est la \emph{ronde}~\cite{DolevIM97,CournierDPV02}. Durant une ronde, tous les n{\oe}uds activables sont activés au moins une fois par l'adversaire. Notons que le définition de ronde dépend fortement de l'équité de l'adversaire.

\section{Construction d'arbres couvrants}

La construction d'une structure de communication efficace au sein de réseaux à  grande échelle (grilles de calculs, ou réseaux pair-à-pairs) ou au sein de réseaux dynamiques (réseaux ad hoc, ou réseaux de capteurs) est souvent utilisée comme brique de base permettant la réalisation de tâches élaborées. La structure de communication la plus adaptée est souvent un arbre. Cet arbre doit couvrir tout ou partie des n{\oe}uds, et posséder un certain nombre de caractéristiques dépendant de l'application. Par ailleurs, la construction d'arbres couvrants participe à la résolution de nombreux problèmes fondamentaux de l'algorithmique répartie. Le problème de la construction d'arbres couvrants a donc naturellement été très largement étudié aussi bien en réparti  qu'en auto-stabilisation. L'objectif général de la première partie du document concerne  les algorithmes auto-stabilisants permettant de maintenir un sous-graphe couvrant particulier, tel qu'un arbre couvrant, un arbre de Steiner, etc. Ce sous-graphe peut être potentiellement dynamiques: les n{\oe}uds et les arêtes peuvent apparaitre ou disparaitre, les poids des arêtes peuvent évoluer avec le temps, etc. 

\subsection{Bref rappel de la théorie des graphes}

Cette section présente quelques rappels élémentaires de  théorie des graphes. Dans un graphe $G=(V,E)$, un chemin est une suite de sommets $u_0,u_1,\dots,u_k$ où $\{u_i,u_{i+1}\}\in E$ pour tout $i=0,\dots,k-1$. Un chemin est dit élémentaire si $u_i\neq u_j$ pour tout $i\neq j$. Par défaut, les chemins considérés dans ce document sont, sauf indication contraire, élémentaires. Les sommets $u_0$ et $u_k$ sont les extrémités du chemin. Un cycle (élémentaire)  est un chemin (élémentaire) dont les deux extrémités sont identiques. En général, on notera $n$ le nombre de sommets du graphe. Un graphe connexe est un graphe tel qu'il existe un chemin entre toute paire de sommets. Un arbre est un graphe connexe et sans cycle. 
\newpage
\begin{lemma}\label{def:arbre}
Soit $T=(V,E)$ un graphe. Les propriétés suivantes sont équivalentes:
\begin{itemize}
\item $T$ est un arbre; 
\item $T$ est connexe et sans cycle;
\item Il existe un \emph{unique} chemin entre toute paire de sommets de $T$;
\item $T$ est connexe et la suppression d'une arête quelconque de $T$ suffit à déconnecter $T$; 
\item $T$ est sans cycle et l'ajout d'une arête entre deux sommets non adjacents de $T$ crée un cycle;
\item $T$ est connexe et possède $n-1$ arêtes.
\end{itemize}
\end{lemma}

On appelle \emph{arbre couvrant} de $G=(V,E)$ tout arbre $T=(V,E')$ avec $E'\subseteq E$. 
Dans un graphe $G=(V,E)$, un \emph{cocycle} est défini par un ensemble $A\in V$; il contient toutes les arêtes $\{u,v\}$ de $G$ tel que $u\in A$ et $v\notin A$. 
Les deux définitions ci-dessous sont à la base de la plupart des algorithmes de construction d'arbres couvrants. 

\begin{definition}[Cycle élémentaire associé]
\label{def:cycle_elementaire_associe}
Soit $T=(V,E_T)$ un arbre couvrant de $G=(V,E)$, et soit $e \in E \setminus E_T$. Le sous-graphe \mbox{$T'=(V,E_T\cup\{e\})$}, contient un unique cycle appelé \emph{cycle élémentaire} associé à $e$, noté $C_e$. 
\end{definition}

\begin{definition}[Echange]
Soit $T=(V,E_T)$ un arbre couvrant de $G=(V,E)$, et soit $e \notin E_T$ et $f\in C_e$, $f\neq e$. L'opération qui consiste à échanger $e$ et $f$ est appelée \emph{échange}. De cet échange résulte l'arbre couvrant $T'$ où $E_{T'}=E_T\cup\{e\}\setminus\{f\}$. 
\end{definition}

\begin{figure}[h]
\centering
\subfigure[\footnotesize{Arbre $T$ contenant l'arête $e$}]{
\includegraphics[width=.3\textwidth]{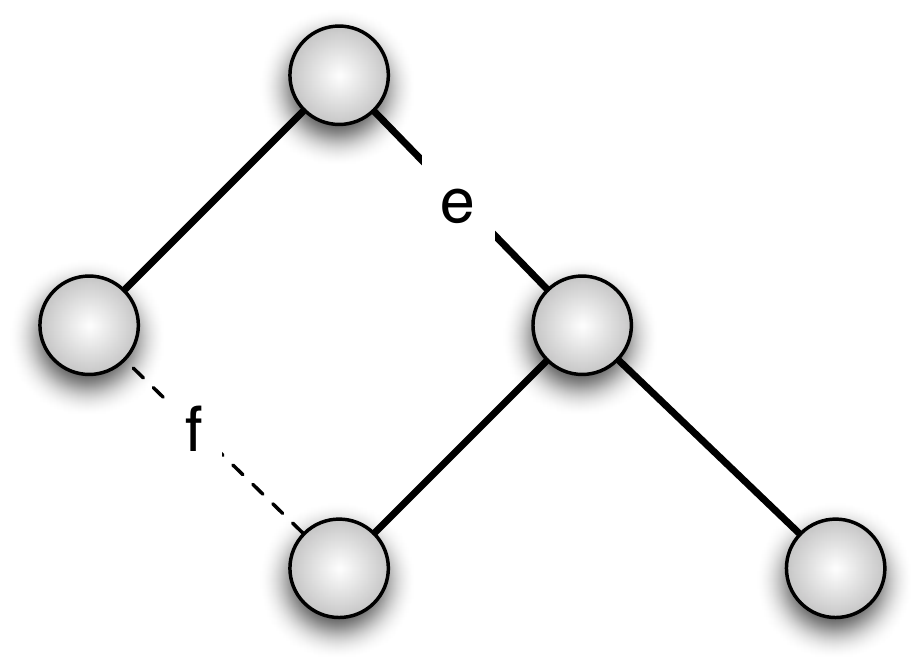}
\label{fig:beu1}
}
\subfigure[\footnotesize{Arbre $T'$ contenant l'arête $f$}]{
\includegraphics[width=.3\textwidth]{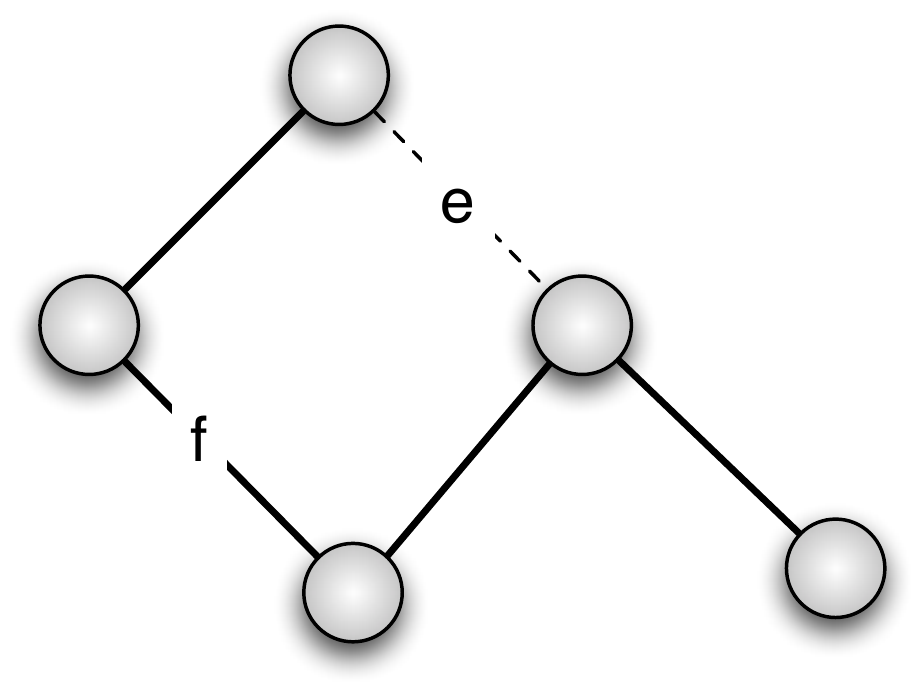}
\label{fig:beu2}
}

\caption{\small{Echange}} 
\label{fig:echange}
\end{figure}

\subsection{Bref état de l'art d'algorithmes auto-stabilisants pour la construction d'arbres couvrants}

Un grand nombre d'algorithmes auto-stabilisants pour la construction d'arbre couvrants ont été proposés à ce jour.  Gartner~\cite{Gartner03} et Rovedakis~\cite{Rovedakis09} ont proposé un état de l'art approfondi de ce domaine. Cette section se contente de décrire un état  de l'art  partiel, qui ne traite pas des problèmes abordés plus en détail dans les chapitres suivants (c'est-à-dire la construction d'arbres couvrants de poids minimum,  d'arbres couvrants de degré minimum, d'arbres de Steiner, etc.). Le tableau~\ref{tab:ST} résume les caractéristiques des algorithmes présentés dans cette section. 

\subsubsection{Arbre couvrant en largeur d'abord}

Dolev, israeli et Moran~\cite{DolevIM90, DolevIM93} sont parmi les premiers à avoir proposé un  algorithme auto-stabilisant de construction d'arbre. Leur algorithme construit un \emph{arbre couvrant en largeur d'abord} (BFS  pour \og Breadth First Search \fg\/ en anglais). Cet algorithme est semi-uniforme, et fonctionne par propagation de distance. C'est une brique de base pour la conception d'un algorithme auto-stabilisant pour le problème de l'exclusion mutuelle dans un réseau asynchrone, anonyme, et dynamique. Le modèle de communication considéré est par registres,  avec un adversaire centralisé. Les auteurs introduisent la notion de \emph{composition équitable} d'algorithmes. Ils introduisent également l'importante notion d'\emph{atomicité lecture/écriture} décrite plus haut dans ce document. 

Afek, Kutten et Yung~\cite{AfekKY91} ont proposé un algorithme construisant un BFS dans un réseau non-anonyme. La racine de l'arbre couvrant est le n{\oe}ud d'identifiant maximum. Chaque n{\oe}ud met à jour sa variable racine. Les configurations erronées vont être éliminées grâce à cette variable. Dès qu'un n{\oe}ud s'aperçoit qu'il n'a pas la bonne racine, il commence par se déclarer racine lui même, et effectue ensuite une demande de connexion en inondant le réseau. Cette connexion sera effective uniquement après accusé de réception par la racine (ou par un n{\oe}ud qui se considère de façon erronée comme une racine). Afek et Bremler-Barr~\cite{AfekB98} ont amélioré l'approche proposée dans~\cite{AfekKY91}. Dans~\cite{AfekKY91}, la racine élue pouvait ne pas se trouver dans le réseau car la variable racine peut contenir un identifiant maximum erroné après une faute. Dans~\cite{AfekB98},  la racine est nécessairement présente dans  le réseau. Datta, Larmore et Vemula~\cite{DattaLV08} ont proposé un algorithme auto-stabilisant reprenant l'approche de Afek et Bremler-Barr~\cite{AfekB98}. Ils construisent de manière auto-stabilisante un BFS afin d'effectuer une élection. Pour ce faire, ils utilisent des vagues de couleurs différentes afin de contrôler la distance à la racine, ainsi qu'un mécanisme d'accusé de réception afin d'arrêter les modifications de l'arbre. C'est donc en particulier un algorithme silencieux~\cite{DolevGS96}.

Arora et Gouda~\cite{AroraG90, AroraG94} ont présenté un système de \og réinitialisation \fg\/ après faute, dans un réseau non anonyme. Ce système en couches utilise trois algorithmes: un algorithme d'élection, un algorithme de construction d'arbre couvrant, et un algorithme de diffusion. Les auteurs présentent une solution silencieuse et auto-stabilisante pour chacun des trois problèmes. Comme un certain nombre d'autres auteurs par la suite (\cite{HighamL01,BurmanK07,BlinPR11}) ils utilisent la connaissance a priori d'une borne supérieure sur le temps de communication entre deux n{\oe}uds quelconques dans le réseau afin de pouvoir éliminer les cycles résultant d'un configuration erronées après une faute.

Huang et Chen~\cite{HuangC92}  ont proposé un algorithme semi-uniforme de construction auto-stabilisante de BFS. Leur contribution la plus importante reste toutefois les nouvelles techniques de preuves d'algorithmes auto-stabilisants qu'ils proposent dans leur article.

Enfin, dans un cadre dynamique, Dolev~\cite{Dolev93}  a proposé un algorithme auto-stabilisant de routage, et un algorithme auto-stabilisant d'élection. Pour l'élection, chaque n{\oe}ud devient racine d'un BFS. La contribution principale est le temps de convergence de chaque construction de BFS,  qui est optimal en  $O(D)$ rondes où $D$ est le diamètre du graphe. De manière indépendante, Aggarwal et Kutten~\cite{AggarwalK93} ont proposé un algorithme de construction d'un arbre couvrant enraciné au n{\oe}ud de plus grand identifiant, optimal en temps de convergence, $O(D)$ rondes. 

\subsubsection{Arbre couvrant en profondeur d'abord}

La même approche que Dolev, israeli et Moran~\cite{DolevIM90, DolevIM93} a été reprise par Collin et Dolev~\cite{CollinD94} afin de  concevoir un algorithme de construction d'arbres couvrants en profondeur d'abord (DFS, pour \og Depth Fisrt Search\fg\/ en anglais). Pour cela, ils ont utilisé un modèle faisant référence à des numéros de port pour les arêtes. Chaque n{\oe}ud $u$ connaît le numéro de port de chaque arête $e=\{u,v\}$ incidente à $u$, ainsi que le numéro de port de $e$ en son autre extrémité $v$.  Avec cette connaissance, un ordre lexicographique est créé pour construire un parcours DFS.

La construction auto-stabilisante d'arbres couvrants en profondeur d'abord va souvent de paire dans la littérature avec le parcours de jeton. Huang et Chen~\cite{HuangC93} ont proposé un algorithme auto-stabilisant pour la circulation d'un jeton dans un réseau anonyme semi-uniforme. Le jeton suit un parcours en profondeur aléatoire\footnote{Les auteurs précisent que l'algorithme peut être modifié pour obtenir un parcour déterministe}. L'algorithme nécessite toutefois la connaissance a priori de la taille du réseau. Huang et Wuu~\cite{HuangW97} ont proposé un autre algorithme auto-stabilisant pour la circulation d'un jeton, cette fois dans un réseau anonyme uniforme. Ce second algorithme nécessite également la connaissance a priori de la taille du réseau. L'algorithme de Datta, Johnen, Petit et Villain~\cite{DattaJPV00}, contrairement aux deux algorithmes précédents, ne fait aucune supposition a priori sur le réseau. De plus, cet algorithme améliore la taille mémoire de chaque n{\oe}ud en passant de $O(\log n)$ bits à $O(\log \Delta)$ bits, où $\Delta$ est le degré maximum du réseau. Notons que tous ces algorithmes de circulation de jeton  sont non-silencieux car le jeton transporte une information qui évolue le long du parcours.

\subsubsection{Arbre couvrant de plus court chemin}

Le problème de l'arbre couvrant de plus court chemin (SPT pour \og Shortest Path Tree\fg\/ en anglais) est la version pondéré du BFS: la distance à la racine dans l'arbre doit être égale à la distance à la racine dans le graphe. Dans ce cadre, Huang et Lin~\cite{HuangL02} ont proposé un algorithme semi-uniforme  auto-stabilisant pour ce problème. Chaque n{\oe}ud calcule sa distance par rapport à tous ses voisins à la Dijkstra. Un n{\oe}ud $u$ choisit pour parent son voisin $v$ qui minimise $d_v+w(u,v)$, où $d_v$ est la  distance supposée de $v$ à la racine (initialisée à zéro), et $w(u,v)$ le poids de l'arête $u,v$. 
Dans un cadre dynamique, le poids des arêtes peut changer au cours du temps. Johnen et Tixeuil~\cite{JohnenT03} ont proposé deux algorithmes auto-stabilisants de construction d'arbres couvrants. Le principal apport de leur approche est de s'intéresser à la propriété \emph{\SC}\/ introduite par~\cite{GafniB81}. Cette propriété stipule que l'arbre couvrant doit s'adapter aux changements de poids des arêtes sans se déconnecter ni créer de cycle. Cette approche est développée plus en détail dans la section~\ref{section:SC} en rapport avec mes propres contributions. Gupta et Srimani~\cite{GuptaS03} supposent le même dynamisme que Johnen et Tixeuil~\cite{JohnenT03}. Ils ont proposé plusieurs algorithmes auto-stabilisants, dont un algorithme semi-uniforme construisant un arbre SPT. Le principal apport de cette dernière contribution est de fournir un algorithme auto-stabilisant silencieux, optimal en espace et en temps de convergence.

Burman et Kutten~\cite{BurmanK07} se sont intéressés à un autre type de dynamisme: l'arrivée et/ou le départ des n{\oe}uds, et/ou du arêtes du réseau. De plus, ces auteurs ont proposé d'adapter l'atomicité lecture/écriture du modèle par registre au modèle par passage de message. Ce nouveau concept est appelé \emph{atomicité envoi/réception} (\og send/receive atomicity\fg). 
Leur algorithme de construction de SPT s'inspire de l'algorithme auto-stabilisant proposé par Awerbuch et al.~\cite{AwerbuchKMPV93} conçu pour \og réinitialiser \fg\/ le réseau après un changement de topologie (ce dernier algorithme utilise un algorithme de construction de SPT comme sous-procédure).

\subsubsection{Arbre couvrant de diamètre minimum}

Bui, Butelle et Lavault~\cite{ButelleLB95} se sont intéressés au problème de l'arbre couvrant de diamètre minimum. De manière surprenante, ce problème a été peu traité dans la littérature auto-stabilisante. L'algorithme dans~\cite{ButelleLB95} est conçu dans un cadre pondéré, où le poids des arêtes sont positifs. Les auteurs prouvent que ce problème est équivalent à trouver un \emph{centre} du réseau (un n{\oe}ud dont la distance maximum à tous les autres n{\oe}uds est minimum). Une fois un centre  identifié, l'algorithme calcule l'arbre couvrant de plus court chemin enraciné à ce centre.

 \section{Récapitulatif et problèmes ouverts}
 
Le tableau~\ref{tab:ST} résume les caractéristiques des algorithmes évoqués dans ce chapitre. Améliorer la complexité en espace ou en temps de certains algorithmes de ce tableau sont autant de problèmes ouverts. Rappelons que la seule borne inférieure non triviale en auto-stabilisation pour la construction d'arbres couvrants n'est valide que dans le cadre silencieux (voir~\cite{DolevGS99}).  

Les deux chapitres suivants sont consacrés à la construction d'arbres couvrants optimisant des métriques particulières, incluant en particulier
\begin{itemize}
\item les arbres couvrants de poids minimum,
\item les arbres couvrants de degré minimum, 
\item les arbres de Steiner, 
\item les constructions bi-critères (poids et degré), 
\item etc.
\end{itemize}

\begin{table}[t]
\begin{center}
\scalebox{1}
{
\begin{tabular}{|c||r||c|c|c|c|c|c|c|c|c|c|}\hline
\rotatebox{80}{}&Articles &\rotatebox{90}{Semi-uniforme}&\rotatebox{90}{Anonyme}&  \rotatebox{80}{Connaissance} & \rotatebox{80}{Communications}&\rotatebox{80}{Adversaire}&\rotatebox{80}{Equité}&\rotatebox{80}{Atomicité}&\rotatebox{80}{Espace mémoire} & \rotatebox{80}{\parbox[b]{0.8in}{Temps de\\ convergence}}& \rotatebox{80}{Propriété}\\\hline  \hline
ST &\cite{AggarwalK93}&&&&R&C&$f$&$\oplus$&$O(\log n)$&$O(D)$&dyn\\
\hline\hline
&\cite{HuangC93} &     $\checkmark$  & $\checkmark$  &$n$&                   R&D &&& $O(\log \Delta n)$& &\\
\cline{2-12} 
DFS&\cite{CollinD94}&      $\checkmark$  & $\checkmark$  &&                   R&C &$f$&$\oplus$&$O(n\log \Delta)$ & $O(D n\Delta)$&\\
\cline{2-12} 
&\cite{HuangW97}      &  & $\checkmark$  &$n$&                   R&C &$f$&& $O(\log n)$& &\\
\cline{2-12} 
&\cite{DattaJPV00}      &$\checkmark$  & $\checkmark$  &&                   R&D &$f $&& $O(\log \Delta)$& $O(D n\Delta)$&\\
\hline\hline
&\cite{DolevIM90}      & $\checkmark$ &$\checkmark$   &&                   R&C &$f$&$\oplus$& $O(\Delta \log n)$&$O(D)$ &dyn\\
\cline{2-12} 
&\cite{AroraG90}      &  &   &$n$&                   R&C &&& $O(\log n)$& $O(n^2)$&dyn\\
\cline{2-12} 
BFS&\cite{AfekKY91}      &  &   &&                   R&D &$f$&$\oplus$& $O(\log n)$&$O(n^2)$ &dyn\\
\cline{2-12} 
&\cite{HuangC92}      &$\checkmark$  & $\checkmark$  &$n$&                   R&D &$I$&& & &\\
\cline{2-12} 
&\cite{Dolev93}         &&&&                   R&C &$f$&$\oplus$& $O(\Delta n \log n)$& $\Theta(D)$&dyn\\
\cline{2-12} 
&\cite{AfekB98}      &&  &B\footnote{L'algorithme est semi-synchrone}&                   M&D &\cellcolor{gray}&& $O(\log n)$& $O(n)$&\\
\cline{2-12} 
&\cite{DattaLV08}      & &   &&                   R&D &$f$&$\oplus$& $O(\log n)$& $O(n)$&\\
\hline
\hline
&\cite{HuangL02}      &$\checkmark$  & $\checkmark$  &&                   R&C &$I$&& $O(\log n)$&&\\
\cline{2-12} 
SPT&\cite{JohnenT03}      & $\checkmark$ & $\checkmark$  &&                   R&D &$f$&& $O(\log n)$& &dyn\\
\cline{2-12} 
&\cite{GuptaS03}      & $\checkmark$ & $\checkmark$  &&                   M/R&D &\cellcolor{gray}&& $O(\log n)$&$O(D)$&\\
\cline{2-12} 
&\cite{BurmanK07}      &  &   &$D$&                   M&D &\cellcolor{gray}&$\oplus$& $O(\log^2n)$& $O(D)$&dyn\\
\hline
\hline
MDiam&\cite{ButelleLB95}      && $\checkmark$&   &                   M&D &\cellcolor{gray}&&$ O(n^2 \log n$& $O(n\Delta + d^2 $&\\
&&&&&&&\cellcolor{gray}&&$+ n\log W)$&$+ n \log^2 n)$&\\
\hline

\end{tabular}
\label{tab:ST}
}

\caption{\footnotesize \textbf{Algorithmes auto-stabilisants asynchrones pour la  construction d'arbres couvrants}. $D$ diamètre du graphe. $\Delta$ degré maximum du graphe.
ST: arbre couvrant; DFS: arbre en profondeur d'abord; BFS: arbre en largeur d'abord; SPT: arbre de plus court chemin; MDiam: arbre de diamètre minimum; R: registres partagés; M: passage de messages;  Adversaire: Distribué (D), central (C), inéquitable ($I$), faiblement équitable ($f$), fortement équitable ($F$). Atomicité: $\oplus$ lecture ou écriture. Propriété: dynamique (dyn), sans-cycle (SC).\vspace*{-0,8cm}}
\label{tableresume}
\end{center}
\end{table}

\chapter{Arbres couvrants de poids minimum}
\label{chap:MST}

Ce chapitre a pour objet de présenter mes travaux sur la construction d'arbres couvrants de poids minimum (Minimum Spanning Tree: \MST). Ce problème est un de ceux les plus étudiés en algorithmique séquentielle comme répartie. De façon formelle, le problème est le suivant.

\begin{definition}[Arbre couvrant de poids minimum]
Soit $G=(V,E,w)$ un graphe non orienté pondéré\footnote{Dans la partie répartie et auto-stabilisante du document nous  supposerons que les poids des arêtes sont positifs, bornés et peuvent être codés en $O(\log n)$ bits, où $n$ est le nombre de \nd s du réseau.}. On appelle arbre couvrant de poids minimum de $G$ tout arbre couvrant dont la somme des poids des arêtes est minimum.
\end{definition}

Ce chapitre est organisé de la façon suivante. La première section consiste en un très bref état de l'art résumant les principales techniques utilisées en séquentiel. La section suivante est consacrée à un état de l'art des algorithmes répartis existant pour la construction de \MST. La section~\ref{sec:SS-MST} est le c{\oe}ur de ce chapitre. Elle présente un état de l'art exhaustif des algorithmes auto-stabilisants pour le \MST, ainsi que deux de mes contributions dans ce domaine. 
\section{Approches centralisées pour le \MST}

Cette section est consacrée à un état de l'art partiel des algorithmes centralisés  pour la construction de \MST, et des techniques les plus couramment utilisées pour ce problème.

Dans un contexte centralisé, trouver un arbre couvrant de poids minimum  est une tâche qui se résout en temps polynomial, notamment au moyen d'algorithmes gloutons. Les premiers algorithmes traitant du problèmes sont nombreux. Leur historique est même sujet à débat. Bo\r{r}uvka~\cite{Boruvka26} apparait maintenant comme le premier auteur à avoir publié sur le sujet. Les travaux de \cite{Prim57,Sollin61,Kurskal64} restent cependant plus connus et enseignés. 

La construction d'un \MST\/ se fait en général sur la base de propriétés classiques des arbres, et utilise la plupart du temps au moins une des deux propriétés suivantes, mis en évidence dans~\cite{Tarjan83}:

\begin{propriete}[Bleu]
\label{propriete:bleu}
Toute arête de poids minimum d'un cocycle de $G$ fait partie d'un \MST\/ de $G$.
\end{propriete}

\begin{propriete}[Rouge] 
\label{propriete:rouge}
Toute arête de poids maximum d'un cycle de $G$ ne fait  partie d'aucun \MST\/ de $G$.
\end{propriete}

\begin{figure}[tb]
\center{
\subfigure[Cocycle]{
\includegraphics[width=.4\textwidth]{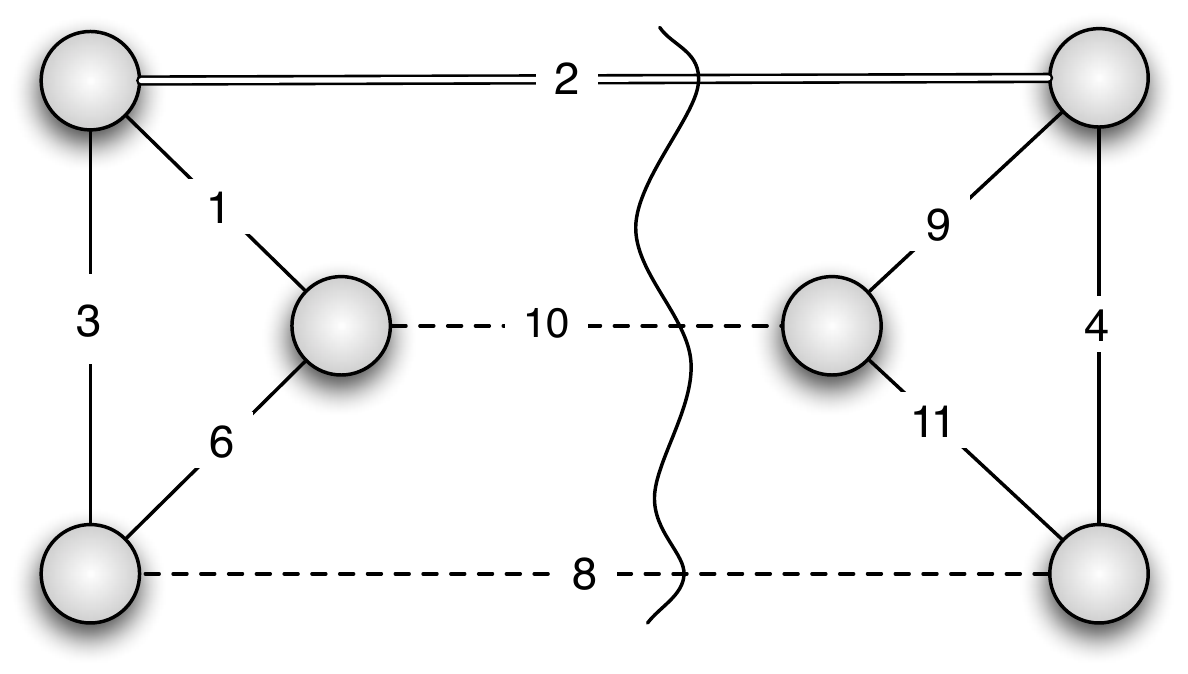}
\label{fig:cocycle}
}
\subfigure[Cycle]{
\includegraphics[width=.4\textwidth]{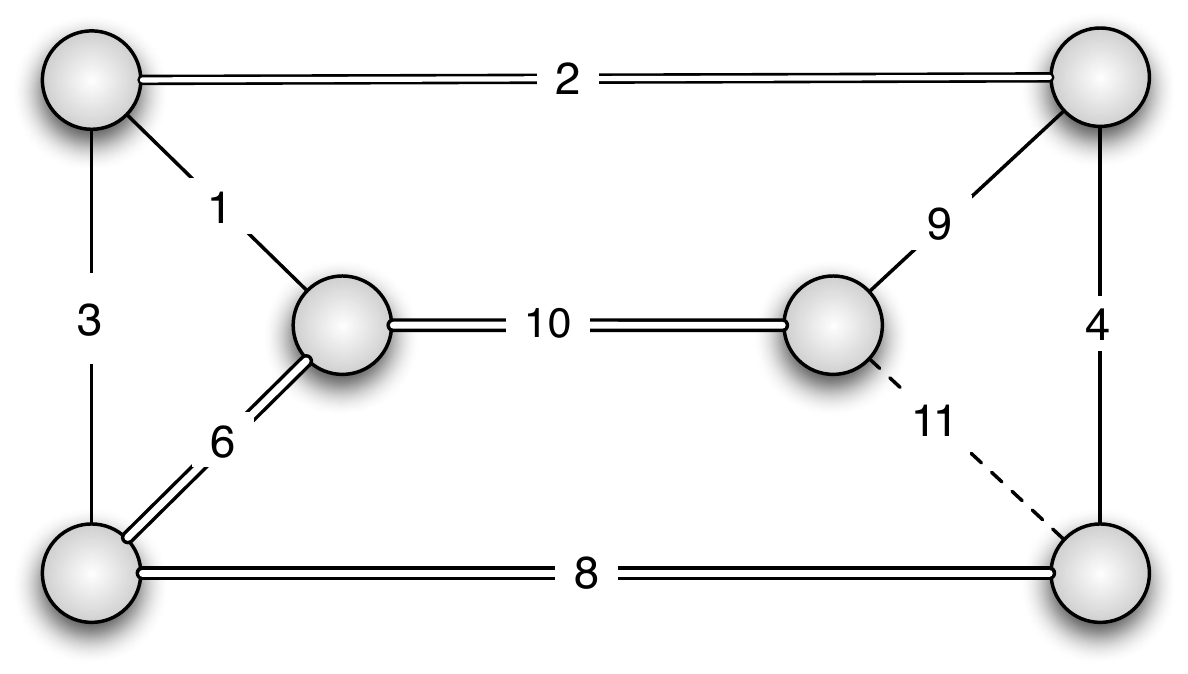}
\label{fig:cycle}
}
\caption{\small Les figures ci-dessus illustrent les propriétés bleu et rouge. Dans la figure~\ref{fig:cocycle}, le cocycle est constitué par les arêtes de poids $\{2,10,8\}$; l'arête de poids $2$ fera partie de l'unique \MST\/ de ce graphe. Dans la figure~\ref{fig:cycle}, le cycle est constitué par les arêtes de poids $\{10,11,8,6\}$; l'arête de poids $11$ ne fera pas parti de l'unique \MST. } 
}
\vspace*{-0,5cm}
\end{figure}

\begin{figure}[tb]
\center{
\subfigure[\footnotesize{Cocycle induit par $A$}]{
\includegraphics[width=.20\textwidth]{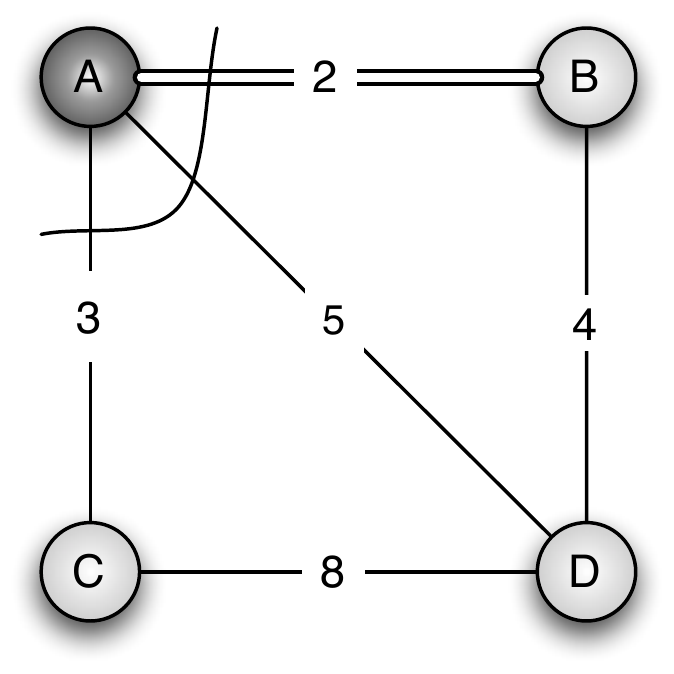}
\label{fig:beu1}
}
\subfigure[\footnotesize{Induit par $A,B$}]{
\includegraphics[width=.20\textwidth]{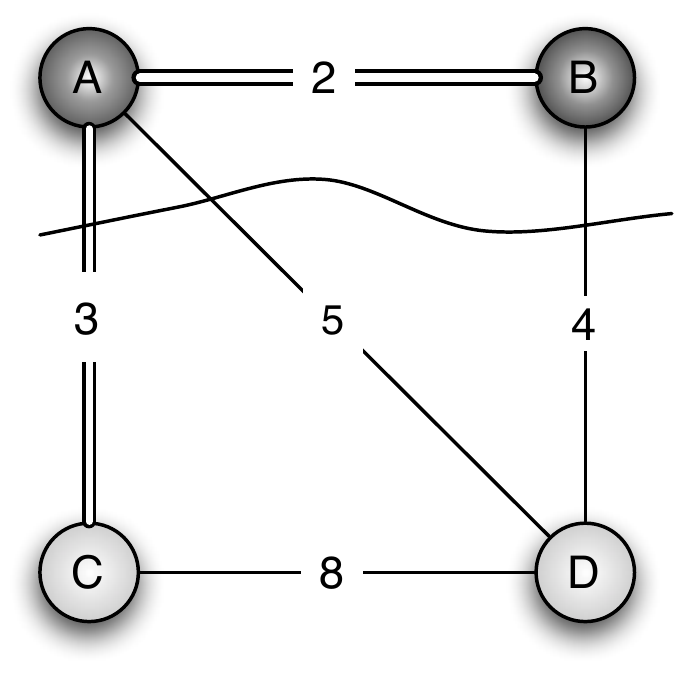}
\label{fig:beu2}
}
\subfigure[\footnotesize{\mbox{Induit par $A,B,C$}}]{
\includegraphics[width=.20\textwidth]{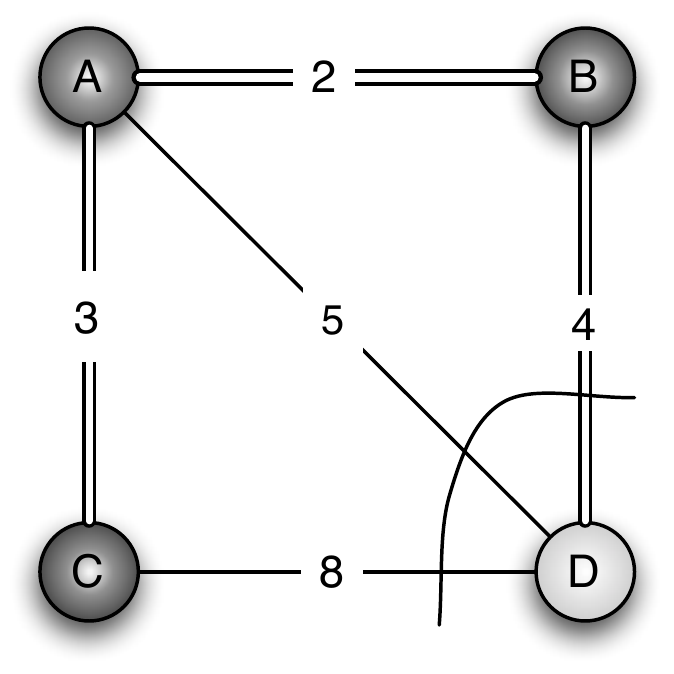}
\label{fig:beu3}
}
\caption{\small{Algorithme de Prim}} 
\label{fig:bleu}
}
\end{figure}

La plupart  des algorithmes traitant du \MST\/ peuvent être classés en deux catégories: ceux qui utilisent la propriété, ou règle, bleue, et ceux qui utilisent la règle rouge.  L'algorithme de Prim~\cite{Prim57} est l'exemple même de l'utilisation de la propriété bleue. Au départ un n{\oe}ud $u$ est choisi arbitrairement, et le cocycle séparant ce n{\oe}ud $u$ des autres n{\oe}uds est calculé; l'arête minimum de ce cocycle, notée  $\{u,v\}$, fait parti du MST final.  L'algorithme calcule ensuite  le cocycle séparant le sous-arbre induit par les n{\oe}uds $u$ et $v$ des autres n{\oe}uds du graphe. Ainsi de suite jusqu'à l'obtention d'un arbre couvrant. Cet arbre est un \MST (voir Figure~\ref{fig:bleu}). L'algorithme de Bo\r{r}uvka~\cite{Boruvka26}, redécouvert par Sollin~\cite{Sollin61}, procède de la même manière à la différence près qu'il choisit initialement de calculer les cocycles induits par chaque n{\oe}ud. Ainsi, il calcule à chaque étape des cocycles de plusieurs sous-arbres. Il apparait donc comme une solution permettant un certain degré de parallélisme. C'est par exemple cette technique qui est à la base du fameux algorithme réparti de Gallager, Humblet, et Spira~\cite{GallagerHS83}. 

\begin{figure}[tb]
\subfigure[\footnotesize{Arête 2}]{
\includegraphics[width=.17\textwidth]{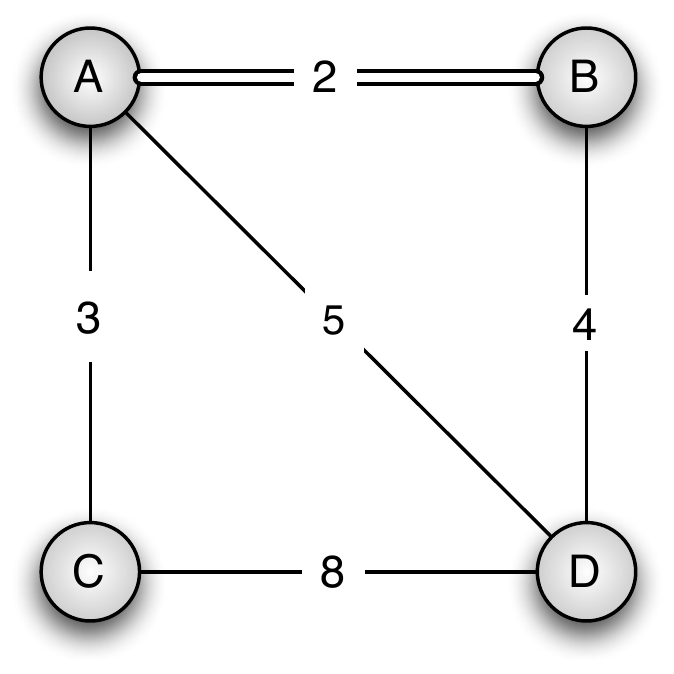}
\label{fig:rouge1}
}
\subfigure[\footnotesize{Arête 3}]{
\includegraphics[width=.17\textwidth]{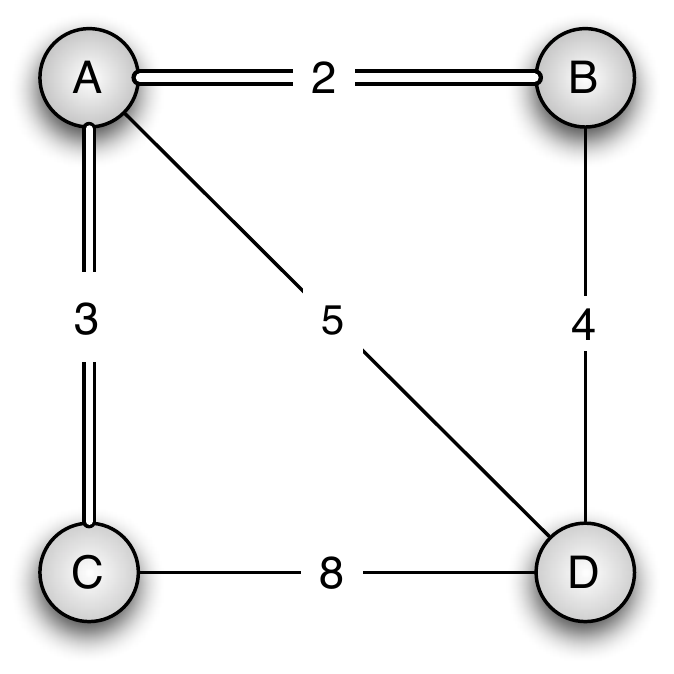}
\label{fig:rouge2}
}
\subfigure[\footnotesize{Arête 4}]{
\includegraphics[width=.17\textwidth]{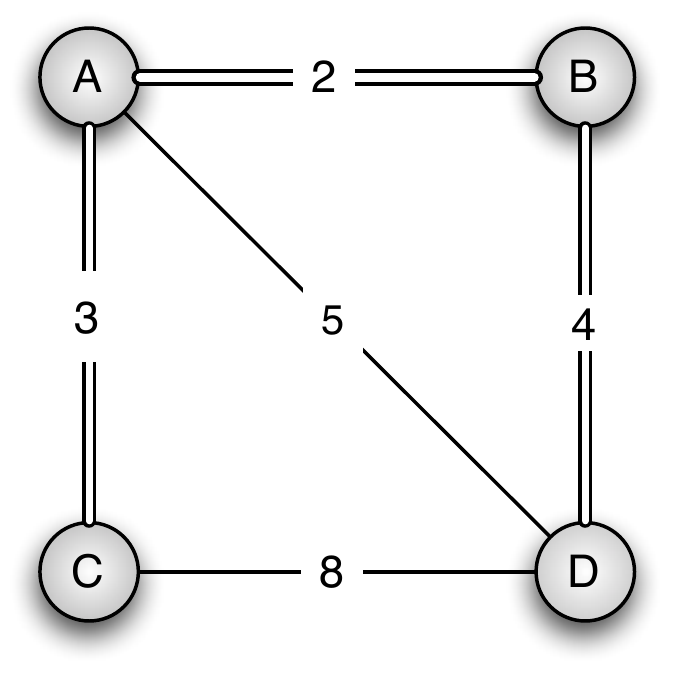}
\label{fig:rouge3}
}
\subfigure[\footnotesize{Arête 5}]{
\includegraphics[width=.17\textwidth]{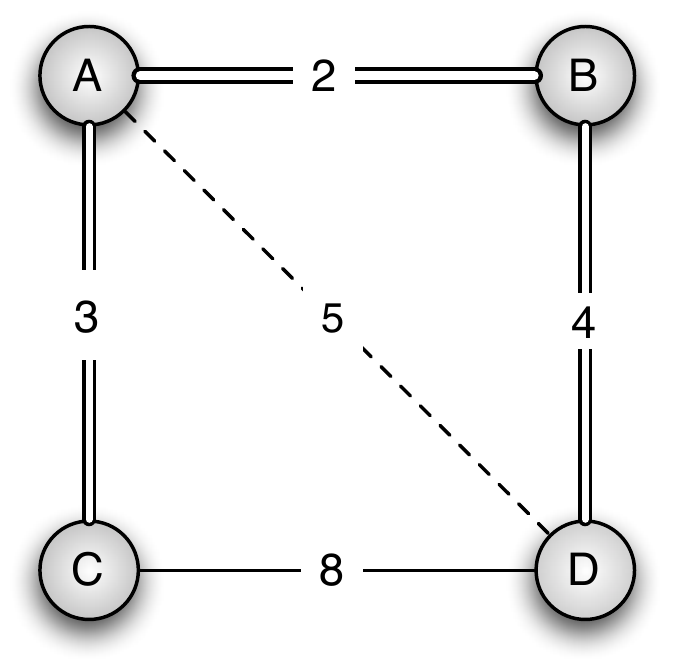}
\label{fig:rouge4}
}
\subfigure[\footnotesize{Arête 8}]{
\includegraphics[width=.17\textwidth]{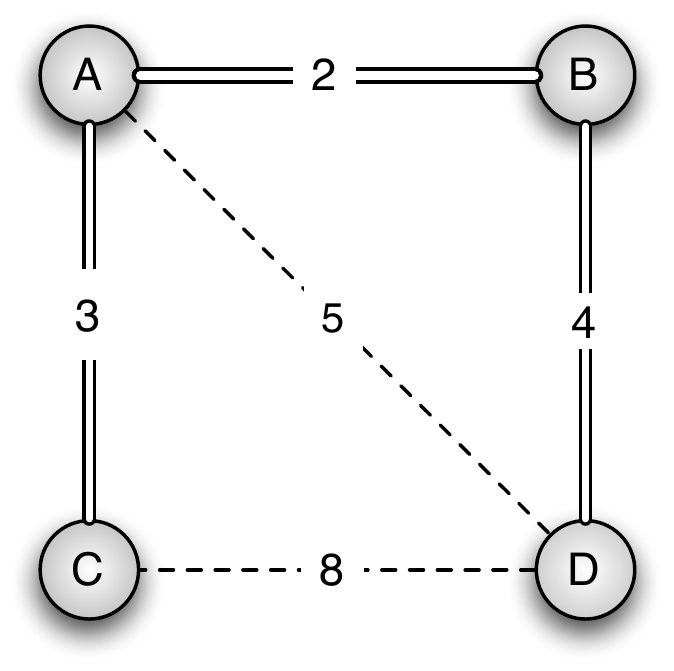}
\label{fig:rouge5}
}
\caption{\small{Algorithme de Kruskal}} 
\label{fig:rouge}
\end{figure}

L'algorithme de Kruskal est  quant à lui basé sur la propriété rouge. Il choisit des arêtes dans l'ordre croissant des poids tant que ces arêtes ne forment pas de cycle.  Notons que  lorsqu'une arête forme un cycle, celle-ci est nécessairement de poids maximum dans ce cycle puisque les poids ont été choisis dans l'orde croissant. Cette arête ne fait donc pas partie d'aucun \MST\/ (voir Figure~\ref{fig:rouge}).

A ce jour, l'algorithme de   construction centralisée de plus faible complexité est celui de Pettie et Ramachandran~\cite{PettieR02}, qui s'exécute en temps $O(|E|\alpha(|E|,n))$, où $\alpha$ est l'inverse de la fonction d'Ackermann. Des solutions linéaires en nombre d'arêtes existent toutefois, mais utilisent des approches probabilistes~\cite{FredmanW90,KargerKT95}.

\section{Approches réparties pour le \MST}

Cette section est consacrée à un état de l'art partiel des algorithmes répartis pour la construction de \MST. Dans un contexte réparti, le premier algorithme publié dans la littérature est~\cite{Dalal79,Dalal87}. Toutefois, aucune analyse de complexité n'est fourni dans cet article. La référence dans le domaine est de fait l'algorithme de Gallager, Humblet et Spira~\cite{GallagerHS83}. Cet algorithme asynchrone est optimal en nombre de messages échangés, $O(|E| + n\log n)$, et a pour complexité temporelle $O(n\log n)$ étapes en synchrone. L'optimalité de la complexité en message est une conséquence d'un résultat de~\cite{FredericksonL84,AwerbuchGPV90} qui démontre que le nombre minimum de bits échangés afin de construire un \MST\/ est de $\Omega(|E|+n\log n)$. L'algorithme de Gallager, Humblet et Spira a valu à ses auteurs le prix Dijkstra en 2004. Il a défini  les fondements des notions et du vocabulaire utilisés par la communauté du réparti pour la construction du \MST. 

L'algorithme de Gallager, Humblet et Spira est basé sur la propriété rouge, à la manière  de l'algorithme de Bo\r{r}uvka-Sollin~\cite{Boruvka26,Sollin61}. Les sous-arbres construits sont appelés \emph{fragments}. Initialement, chaque n{\oe}ud du système est un fragment. Par la suite, chaque fragment \emph{fusionne} grâce à l'arête sortante du fragment de poids minimum. Autrement dit, grâce à l'arête de poids minimum du cocycle. La difficulté en réparti est de permettre à  chaque n{\oe}ud de connaître le fragment auquel il appartient, et d'identifier les arêtes à l'extérieur du fragment et les arêtes à l'intérieur de celui-ci.  A cette fin, les n{\oe}uds ont besoin d'une vision \og globale \fg, autrement dit d'effectuer un échange d'information afin de maintenir à jour leur appartenance aux différents fragments au fur et à mesure des fusions.   

L'optimalité en message de la construction d'un \MST\/ étant obtenue par l'algorithme~\cite{GallagerHS83}, la suite des travaux dans ce domaine s'est  attachée à diminuer la complexité en temps. Les travaux dans ce domaine ont pour objet principal de contrôler la taille des fragments afin d'améliorer la rapidité de la mise à jour des n\oe uds, et donc la complexité en temps. La première amélioration  notable est celle d'Awerbuch~\cite{Awerbuch87} en temps $O(n)$ étapes, tout en restant optimal en nombre de messages échangés. Dans la fin des années 90, Garay, Kutten et Peleg relancent la recherche dans ce domaine. Dans~\cite{GarayKP98}, ils obtiennent un temps $O(D+n^{0.614})$ étapes, où $D$ est le diamètre du graphe. Cette complexité a été améliorée dans~\cite{KuttenP98} en $O(D+\sqrt{n}\log^*n)$ étapes,  au détriment du nombre de messages échangés, qui devient $O(|E|+n^{3/2})$. 
Dans~\cite{PelegR00}, il est prouvé une borne inférieure $\widetilde{\Omega}(\sqrt{n})$ étapes dans le cas particulier des graphes de diamètre $\Omega(\log n)$, où la notation  $\widetilde{\Omega}$ signifie qu'il n'est pas tenu compte des facteurs polylogarithmiques. Toujours dans le cas des graphes de petit diamètre, Lotker, Patt-Shamir et Peleg~\cite{LotkerPP01} ont obtenu une borne inférieure de  $\widetilde{\Omega}(\sqrt[3]n)$ étapes pour les graphes de diamètre $3$, et $\widetilde{\Omega}(\sqrt[4]n)$ pour graphe de diamètre $4$. Dans les graphes de diamètre $2$ il existe un algorithme s'exécutant en $O(\log n)$ étapes~\cite{LotkerPP01}. 

\section{Approches auto-stabilisantes}
\label{sec:SS-MST}

Cette section présente un état de l'art exhaustif de la construction auto-stabilisante de \MST. La présentation est faite de façon chronologique. Elle débute donc par les travaux de Gupta et Srimani~\cite{GuptaS99,GuptaS03}, et de Higham et Lyan~\cite{HighamL01}. Elle est suivie  par  deux contributions personnelles dans ce domaine~\cite{BlinPRT09,BlinDPR10}. Elle est enfin conclue par les améliorations récentes apportées par Korman, Kutten et Masuzawa~\cite{KormanKM11}. Les caractéristiques de ces différents algorithmes sont résumées dans la Table~\ref{table:SS-MST}.

\begin{table}[tb]
\begin{center}
\scalebox{1}
{
\begin{tabular}{|c||c|c|c|c|c|c|c|c|}\hline
 \rotatebox{80}{Articles} & \rotatebox{80}{Système} &   \rotatebox{80}{Connaissance} & \rotatebox{80}{Communications}&\rotatebox{80}{Taille message}&\rotatebox{80}{Espace mémoire} & \rotatebox{80}{\parbox[b]{0.8in}{Temps de\\ convergence}}& \rotatebox{80}{Non-silencieux}&\rotatebox{80}{\SC}\\\hline  \hline
 \cite{GuptaS99,GuptaS03}      &Semi&$n$               &  M                 &$O(\log n)$& $\Theta(n \log n)$ & $\Omega(n^2)$& &\\
\hline
\cite{HighamL01}     &Semi& $O(D)$         & M &$O(n\log n)$&$O(\log n)$& $O(n^3)$&$\checkmark$ &\\
\hline
\cite{BlinPRT09}     &A &                    & R &&$O(\log n)$ & $O(n^3)$&$\checkmark$ &$\checkmark$ \\
\hline
\cite{BlinDPR10}     &A&                     &R && $\Omega(\log^2 n)$ & $O(n^2)$&&\\
\hline
\cite{KormanKM11} &A&                   & R && $O(\log n)$ & $\mathbf{O(n)}$&&\\\hline
\end{tabular}
}
\caption{\small Algorithmes auto-stabilisants pour le problème du \MST. Dans cette table, \og Semi \fg\/ signifie Semi-synchrone,  et \og A \fg\/ signifie asynchrone. De même, \og M \fg\/ signifie modèle par passage de messages,  et \og R \fg\/  modèle à registres partagés.
}
\label{table:SS-MST}
\end{center}
\end{table}
\subsection{Algorithme de Gupta et Srimani}

Le premier algorithme auto-stabilisant pour la construction d'un \MST\/ a été publié par Gupta et Srimani \cite{GuptaS99,GuptaS03}.  Cet article traite essentiellement d'une approche auto-stabilisante pour le calcul des plus courts chemins entre toutes paires de n{\oe}uds. Les auteurs utilisent ensuite ce résultat pour résoudre le problème du \MST. Ils considèrent des graphes dont les poids sont uniques\footnote{Cette hypothèse n'est pas restrictive car on peut facilement transformer un graphe pondéré à poids non distincts en un graphe pondéré à poids deux-à-deux distincts. Il suffit par exemple d'ajouter au poids de chaque arête l'identifiant le plus petit d'une de ses deux extrémités.}, et se placent dans le cas de graphes dynamiques: le poids des arêtes peut évoluer avec le temps, et les \nd{s} peuvent apparaître et disparaître. Les auteurs utilisent un modèle par passage de messages similaire à un modèle par registres partagés. Dans le modèle utilisé, chaque n{\oe}ud $v$  envoie périodiquement un message à ses voisins. Si $u$ reçoit un message d'un n{\oe}ud $v$ qu'il ne connaissait pas,  il le rajoute à son ensemble de voisins. A l'inverse, si un n{\oe}ud $u$ n'a pas reçu de messages de son voisin $v$ au bout d'un certain délai, alors $u$ considère que $v$ a quitté le réseau, et il supprime donc ce n{\oe}ud de la liste de ses voisins. L'algorithme a donc besoin d'une borne sur le temps de communication entre deux n{\oe}uds. Il fonctionne donc dans un système \emph{semi-synchrone}. Si le réseau ne change pas pendant une certaine durée, alors les messages échangés contiendront toujours la même information.  L'algorithme est donc un algorithme \emph{silencieux}. Plus spécifiquement, l'algorithme de Gupta et Srimani s'exécute de la façon suivante. Fixons deux n{\oe}uds $u$ et $v$. Le n{\oe}ud $u$ sélectionne l'arête  $e$ de poids $w(e)$ minimum  parmi les arêtes de poids maximum de chaque chemin vers $v$.  Si $v$ est un n{\oe}ud adjacent à $u$ et si $w(\{u,v\})=w(e)$ alors $\{u,v\}$ est une arête du \MST\/ final. Cela revient à utiliser la propriété bleu. En d'autres termes, chaque n{\oe}ud calcule le cocycle de poids maximum, et choisit l'arête de poids minimum de ce cocycle. Pour pouvoir calculer l'arête de poids maximum de tous les chemins, les auteurs ont besoin de connaitre la taille $n$ du réseau, et ils supposent que les n{\oe}uds ont des identifiants de 1 à $n$. Comme nous l'avons vu, chaque n{\oe}ud conserve le poids de l'arête de poids maximum allant à chaque autre n{\oe}ud du réseau. Il a donc besoin d'une mémoire de taille $\Theta(n \log n)$ bits. Le temps de convergence est $\Omega(n^2)$ rondes.

\subsection{Algorithme de Higham et Lyan}

Higham et Lyan~\cite{HighamL01} ont proposé un algorithme \emph{semi}-synchrone dans le modèle  par passage de messages.  Leur algorithme suppose que chaque n{\oe}ud connaît une borne supérieure $B$ sur le délai  que met un message à traverser le réseau. Cela revient à supposer un temps maximum de traversée d'une arête, donc à considérer un réseau semi-synchrone. 

L'algorithme utilise la propriété rouge, et fonctionne de la manière suivante. Chaque arête doit déterminer si elle doit appartenir ou non au \MST. Une arête $e$ n'appartenant pas au \MST\/ inonde le graphe afin de trouver son cycle élémentaire associé, noté  $C_e$. Lorsque  $e$ reçoit le message $m_e$ ayant parcouru $C_e$, cette arête utilise les informations collectées par $m_e$, c'est-à-dire les identifiants et les poids des arêtes se trouvant sur $C_e$. Si $w_e$ n'est pas le poids le plus grand du cycle $C_e$, alors $e$ fait parti du \MST, sinon $e$ ne fait pas parti du \MST. La borne supérieure $B$ est utilisée comme un délai à ne pas dépasser. En effet si après un intervalle de temps $B$, l'arête $e$ n'a reçu aucun message en retour de son inondation, alors $e$ conclut que la structure existante n'est pas connexe. Elle décide  alors de devenir provisoirement une arête du \MST, quitte à revoir sa décision plus tard.  Si une arête $e$ faisant partie provisoirement du \MST\/ ne reçoit pas de message de recherche du cycle élémentaire $C_e$ au bout d'un temps $3B$, alors elle peut considérer que toutes les arêtes  font partie du \MST, ce qui une configuration erronée. Dans ce cas, $e$ déclenche un message de type \emph{trouver le cycle}. On remarque donc que, dans les deux cas, s'il existe au moins une arête ne faisant pas parti de l'arbre couvrant, ou si toutes les arêtes font partie de l'arbre, alors des messages sont générés. Cet algorithme est donc non-silencieux.

En terme de complexité, chaque n{\oe}uds à besoin de $O(\log n)$ bits de mémoire pour exécuter l'algorithme. La quantité d'information échangée (identifiants et poids des n{\oe}uds des chemins parcourus) est de $O(n \log n)$ bits par message.  

\subsection{Contributions à la construction auto-stabilisante de \MST}

Cette section résume mes contributions à la conception d'algorithmes auto-stabilisants de construction de \MST. Elle présente en particulier deux algorithmes. L'un est le premier algorithme auto-stabilisant pour le \MST\/ ne nécessitant aucune connaissance a priori sur le réseau. De plus, cet algorithme est entièrement asynchrone avec une taille mémoire et une taille de message logarithmique. L'autre algorithme améliore ce premier algorithme en optimisant le rapport entre le temps de convergence et la taille mémoire. Ces deux algorithmes sont décrits dans les sous-sections suivantes. 

\subsubsection{Transformation d'arbres de plus courts chemins en MST}
\label{subsubsec:tapccenMST}

Ma première contribution a été réalisée en collaboration avec Maria Potop-Butucaru, Stéphane Rovedakis et  Sébastien Tixeuil. Elle a été publié dans~\cite{BlinPRT09}. L'apport de ce travail est multiple.
D'une part, contrairement aux travaux précédents (voir~\cite{GuptaS99,HighamL01,GuptaS03}) notre algorithme n'a besoin d'aucune connaissance a priori sur le réseau. Il ne fait de plus aucune supposition sur les délais de communication, et est donc asynchrone. Par ailleurs, l'algorithme possède la propriété \emph{\SC}. Enfin il est le premier algorithme à atteindre un espace mémoire de $O(\log n)$ bits avec des tailles de message $O(\log n)$ bits.

La propriété \SC\/ est traités dans la section~\ref{section:SC}. Je ne vais donc traiter ici que des autres aspects de notre algorithme. Sans perte de généralité vis-à-vis des travaux précédents, nous considérons un réseaux anonyme, sur lequel s'exécute un algorithme \emph{semi-uniforme}. Autrement dit, nous distinguons un n{\oe}ud arbitraire parmi les n{\oe}uds du réseau. Ce dernier jouera un rôle particulier. Nous appelons ce n{\oe}ud  la \emph{racine} de l'arbre. Notons que dans un réseau avec des identifiants, on peut toujours élire une racine; réciproquement,  si le réseau dispose d'une racine alors les n{\oe}uds peuvent s'attribuer des identifiants distincts~\cite{Dolev00}.  Notons également, qu'il est impossible de calculer de manière déterministe auto-stabilisante un \MST\/ dans un réseaux anonyme (voir  \cite{GuptaS03}). L'hypothèse de semi-uniformité offre une forme de minimalité. 

Nous travaillons dans un modèle de communications par registres, avec un adversaire faiblement équitable, et une atomicité lecture/écriture. Plus précisément, quand l'adversaire active un n{\oe}ud, ce n{\oe}ud peut de façon atomique (1) lire sa mémoire et la mémoire de ses voisins, et (2) écrire dans sa mémoire. (Nous avons besoin de cette atomicité afin de garantir la propriété \SC, mais nous aurions pu nous en passer pour les autres propriétés de l'algorithme). La structure  du réseau est statique, mais les poids des arêtes peuvent changer au cours du temps, ce qui confère un certain dynamisme au réseau. Afin de préserver le déterminisme de notre algorithme, on suppose que les ports relatifs aux arêtes liant un n{\oe}ud $u$ à ses voisins sont numéroté de 1 à $\deg(u)$.  

 Notre algorithme possède deux caractéristiques conceptuelles essentielles: 
\begin{itemize}
\item d'une part, il maintient un arbre couvrant (cet arbre n'est pas nécessairement minimum, mais il se sera au final); 
\item d'autre part, les n{\oe}uds sont munis  d'\emph{étiquettes} distinctes qui, à l'inverse des identifiants, codent de l'information. 
\end{itemize}

\paragraph{Brève description de l'algorithme.}

Notre algorithme fonctionne en trois étapes:
\begin{enumerate}
\item Construction d'un arbre couvrant.
\item Circulation sur l'arbre couvrant d'un jeton qui étiquette chaque n{\oe}ud.
\item Amélioration d'un cycle élémentaire.
\end{enumerate}

Nous décrivons chacune de ces trois étapes. En utilisant l'algorithme de Johnen-Tixeuil~\cite{JohnenT03}, nous construisons un arbre de plus courts chemins enraciné à la racine $r$, tout en maintenant la propriété \SC\/ (c'est principalement le maintient de la propriété \SC\/ qui nous a motivé dans le choix de cet algorithme). Une fois l'arbre couvrant construit, la racine initie une circulation DFS d'un jeton. Pour cela, nous utilisons l'algorithme de Petit-Villain~\cite{PetitV07}. Comme l'algorithme de Higham-Liang, notre algorithme utilise la propriété rouge: il élimine du MST l'arête de poids maximum d'un cycle. Toutefois, contrairement à~\cite{HighamL01} qui nécessite d'inonder le réseau pour trouver les cycles élémentaires, notre algorithme utilise les cycles élémentaires induits par la présence d'un arbre couvrant existant. Cela est possible grâce aux deux caractéristiques essentielles de notre algorithme, à savoir: (1) maintenance d'un arbre couvrant, et (2) utilisation d'\emph{étiquettes} DFS. 
Grâce à l'arbre et aux étiquettes, les cycles élémentaires sont facilement identifiés, ce qui permet d'éviter l'inondation.

\begin{center}
\begin{figure}[h!]
\vspace*{-1cm}
\begin{center}
\subfigure[\footnotesize{Jeton au n{\oe}ud 12}]{
\includegraphics[scale=0.35]{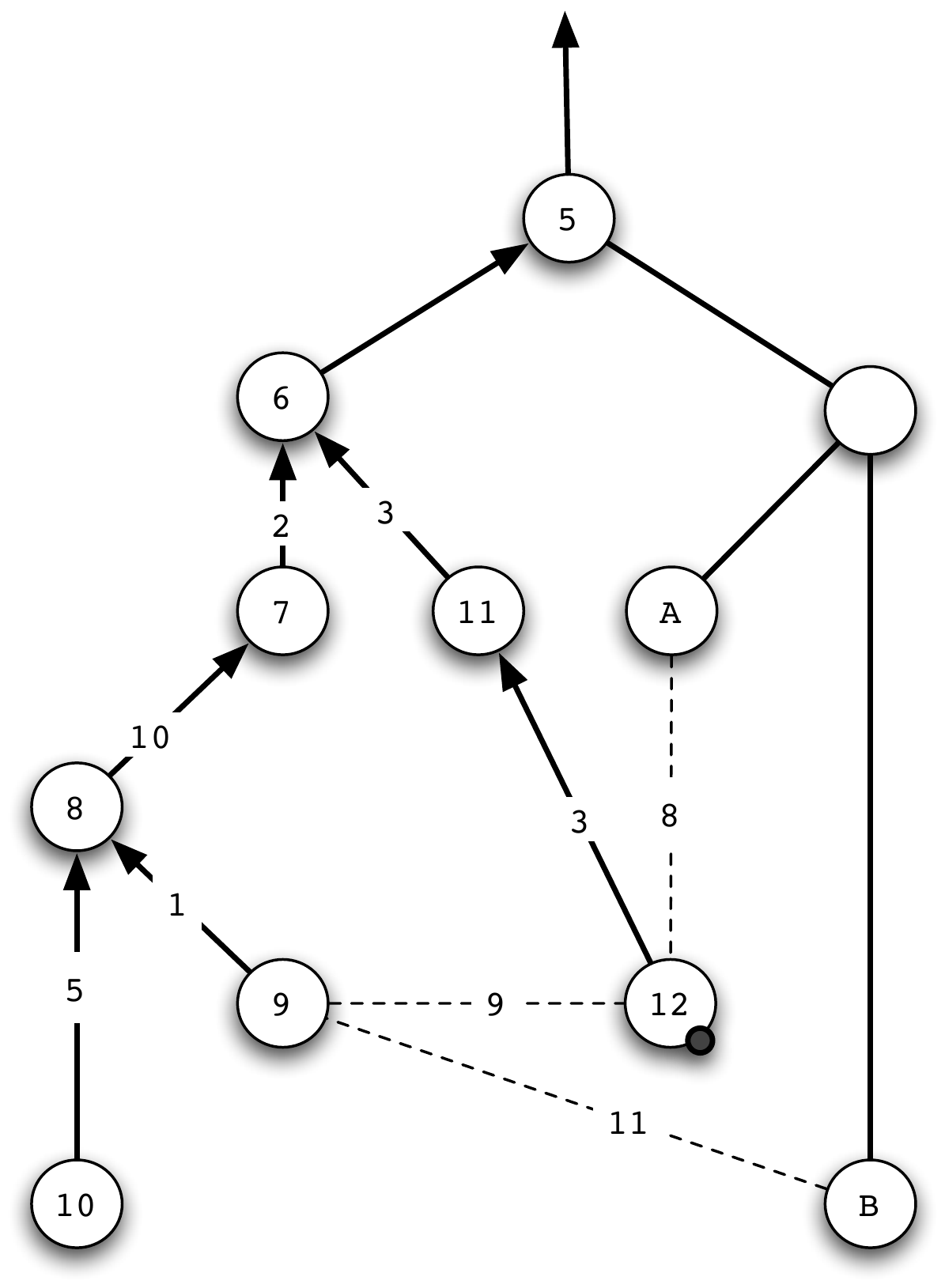}
\label{fig:ex1}
}
\subfigure[\footnotesize{Echange de 1 et 9}]{
\includegraphics[scale=0.35]{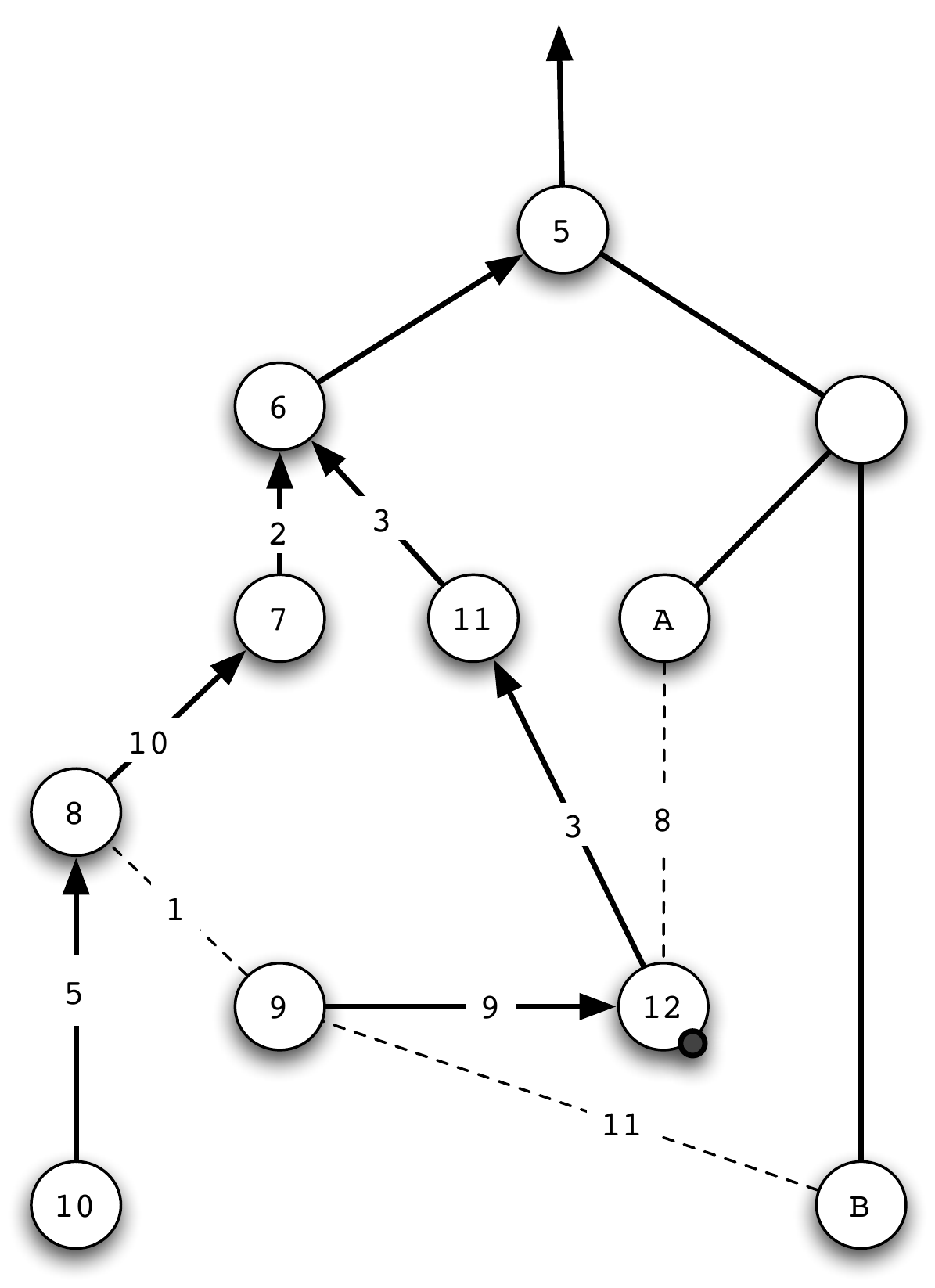}
\label{fig:ex2}
}
\subfigure[\footnotesize{Echange de 10 et 1}]{
\includegraphics[scale=0.35]{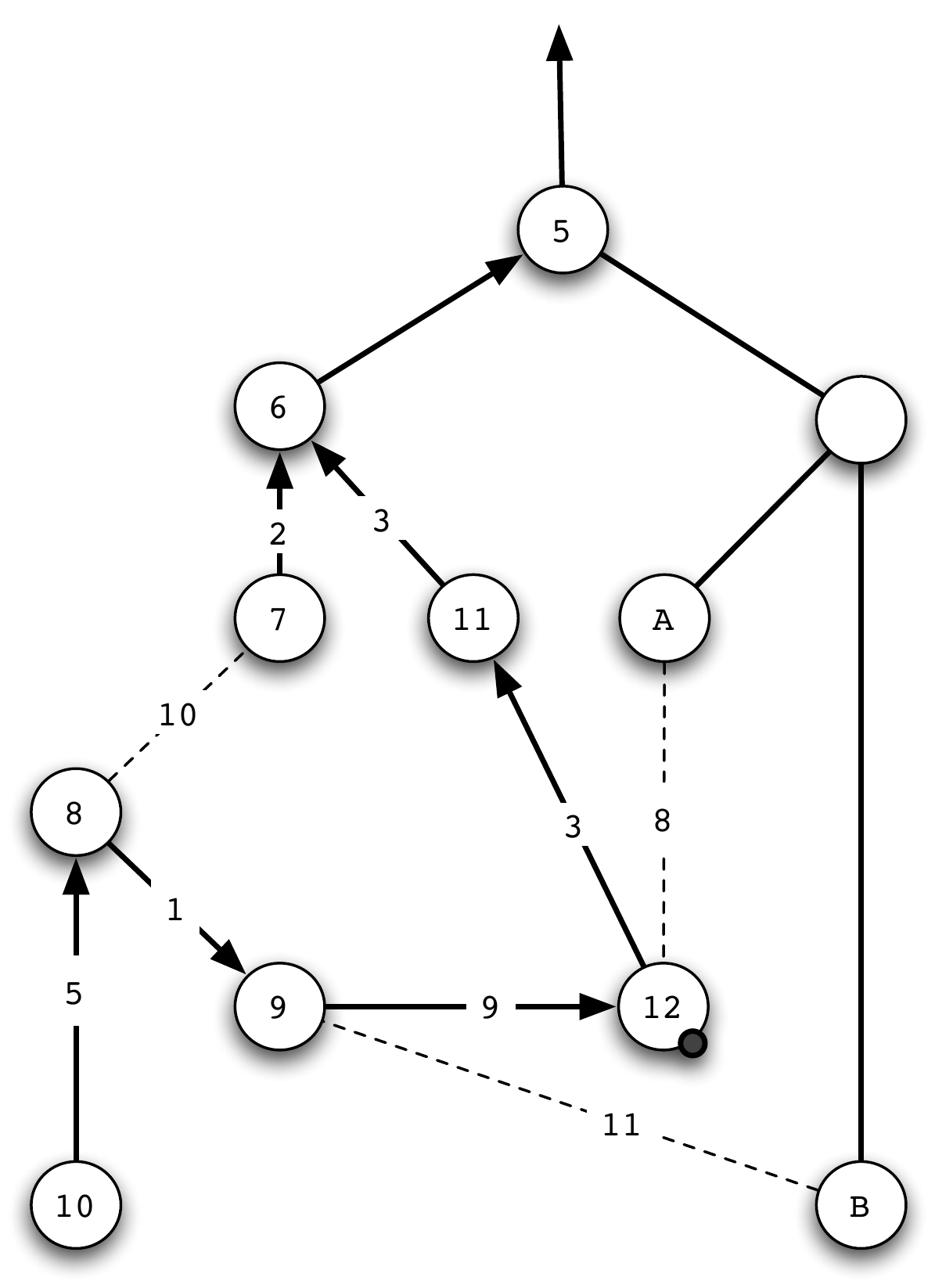}
\label{fig:ex3}
}
\end{center}
\caption{\small{Illustration de l'étiquetage lors du  déroulement de l'algorithme de la section~\ref{subsubsec:tapccenMST}}} 
\label{fig:ex}
\vspace*{-0,5cm}
\end{figure}
\end{center}

Lorsque l'arbre couvrant de plus court chemin est construit, la racine $r$ déclenche une circulation de \emph{jeton}. Ce jeton circule dans l'arbre en profondeur d'abord (DFS), en utilisant les numéros de ports. Le jeton possède un compteur. Ce compteur est initialisé à zéro lors du passage du jeton à la racine. A chaque fois que le jeton découvre un nouveau n{\oe}ud, il incrémente son compteur.  Quand le jeton arrive à un n{\oe}ud dans le sens racine-feuilles, ce n{\oe}ud prend pour étiquette le compteur du jeton (voir Figure~\ref{fig:ex}). Cette étiquette, notés $\Label_u$ (pour \og label \fg\/ en anglais), a pour objet d'identifier les cycles élémentaires associés aux arêtes ne faisant pas partie de l'arbre couvrant. Pour ce faire, lorsque le jeton arrive sur un n{\oe}ud $u$, si ce n{\oe}ud possède des arêtes qui ne font pas partie de l'arbre, alors le jeton est \emph{gelé}.  Soit $v$ le n{\oe}ud extrémité de l'arête $\{u,v\}$ ne faisant pas partie de l'arbre. Si $\Label_v<\Label_u$, alors le n{\oe}ud $u$ déclenche une procédure d'amélioration de cycle. Grâce aux étiquettes, l'algorithme construit ainsi  l'unique chemin  $P(u,v)$ entre $u$ et $v$ dans l'arbre. Si les étiquettes sont cohérentes,  chaque n{\oe}ud $w$ calcule son prédécesseur dans $P(u,v)$ de la façon suivante. Si $\Label_w>\Label_v$ alors le parent de $w$ est son prédécesseur dans $P(u,v)$, sinon son prédécesseur est le n{\oe}ud $w'$  défini comme l'enfant de $w$ d'étiquette maximum tel que $\Label_{w'}<\Label_v$. Le poids maximum d'une arête de $P(u,v)$ est collecté. Si le poids de $\{u,v\}$ n'est pas ce maximum, alors les arêtes du cycle sont échangées par échanges successifs. Le déroulement de cet échange sera spécifié dans la section~\ref{section:SC}. La figure~\ref{fig:ex} illustre l'étiquetage des n{\oe}uds, ainsi que  les échanges successifs d'arêtes. Si au cours du parcours du cycle, l'étiquette courante n'est pas cohérente par rapport à celles de ses voisins, le jeton est  libéré, et il continue sa course et l'étiquetage des n{\oe}uds. L'étiquetage sera en effet rectifié au passage suivant du jeton puisque l'algorithme maintient en permanence une structure d'arbre couvrant. 

\paragraph{Complexité.}

Notons que les algorithmes de Johnen-Tixeuil et de Petit-Villain~\cite{JohnenT03,PetitV07} utilisent un nombre constant de variables de taille $O(\log n)$ bits. Il en va de même pour la partie que nous avons développée car nous manipulons trois identifiants et un poids d'arête dans la partie amélioration de cycle. L'algorithme a donc une complexité mémoire de $O(\log n)$ bits. 
Pour fonctionner, l'algorithme a besoin d'une circulation permanente du jeton. Il n'est donc pas silencieux et ne peut donc pas être considéré comme optimal en mémoire, car, à ce jour, aucune borne inférieure n'a été donnée sur la mémoire utilisée pour la construction non silencieuse d'arbres couvrants.  Notre algorithme traite toutefois le problème dans un cadre dynamique (les poids des arêtes peuvent changer au cours du temps), et il est probablement beaucoup plus difficile de rester silencieux dans un tel cadre.

Pour la complexité temporelle, le pire des cas arrive après un changement de poids d'une arête. En effet une arête de l'arbre appartient à au plus $m-n+1$ cycles, obtenues en rajoutant à l'arbre une arête qui n'est pas dans l'arbre, où $m$ est le nombre d'arêtes. Donc, avant de déterminer si elle est effectivement dans le \MST, l'algorithme déclenche une vérification pour chacune des arêtes ne faisant pas partie de l'arbre dans chacun de ces $m-n+1$ cycles. Comme chaque amélioration nécessite $O(n)$ rondes, il découle une complexité en temps de $O(n^3)$ rondes.

\subsubsection{Utilisation d'étiquetages informatifs}

L'algorithme de la section~\ref{subsubsec:tapccenMST} utilise un étiquetage des n{\oe}uds par profondeur d'abord. Quoique trivial, nous avons vu la capacité de cet étiquetage à faciliter la conception d'algorithmes auto-stabilisants pour le \MST. Cela nous a conduit à concevoir un algorithme basé sur des schémas d'étiquetage informatifs non triviaux. Ce  travail a été réalisé en collaboration avec Shlomi Dolev, Maria  Potop-Butucaru, et Stéphane Rovedakis. Il a été publié dans \cite{BlinDPR10}. Nous sommes par ailleurs parti du constat qu'aucun des algorithmes auto-stabilisants existant n'utilisait l'approche de l'algorithme répartie le plus cité, à savoir celui de Gallager, Humblet et Spira~\cite{GallagerHS83}. La composante la plus compliquée et la plus onéreuse dans~\cite{GallagerHS83} est la gestion des fragments (i.e., permettre à un n{\oe}ud de distinguer les n{\oe}uds de son voisinage faisant partie du même fragment que lui de ceux d'un autre fragment). Dans l'approche de Gallager, Humblet et Spira, pour chaque fragment, une racine  donne l'identifiant de ce fragment. Ainsi, tout n{\oe}ud appartenant à un même fragment possède un ancêtre commun. Nous donc avons  eu  l'idée d'utiliser un étiquetage informatif donnant le plus proche ancêtre commun de deux n{\oe}uds. 

L'apport de cet algorithme est donc conceptuellement double: 

\begin{itemize}
\item d'une part, il est le premier à utiliser une approche à la Gallager, Humblet et Spira pour l'auto-stabilisation; 
\item d'autre part, il est le premier à utiliser un schéma d'étiquetage informatif non trivial pour la construction auto-stabilisante de \MST. 
\end{itemize}
Ce double apport  nous a permis d'améliorer le rapport entre le temps de convergence et l'espace mémoire, plus précisément: mémoire $O(\log^2 n)$ bits, et  temps de convergence $O(n^2)$ rondes. Par ailleurs, cet algorithme est silencieux. 
\vspace*{-0,5cm}
\paragraph{Brève description de l'algorithme.}

\begin{figure}[tb]
\begin{center}
\includegraphics[scale=0.4]{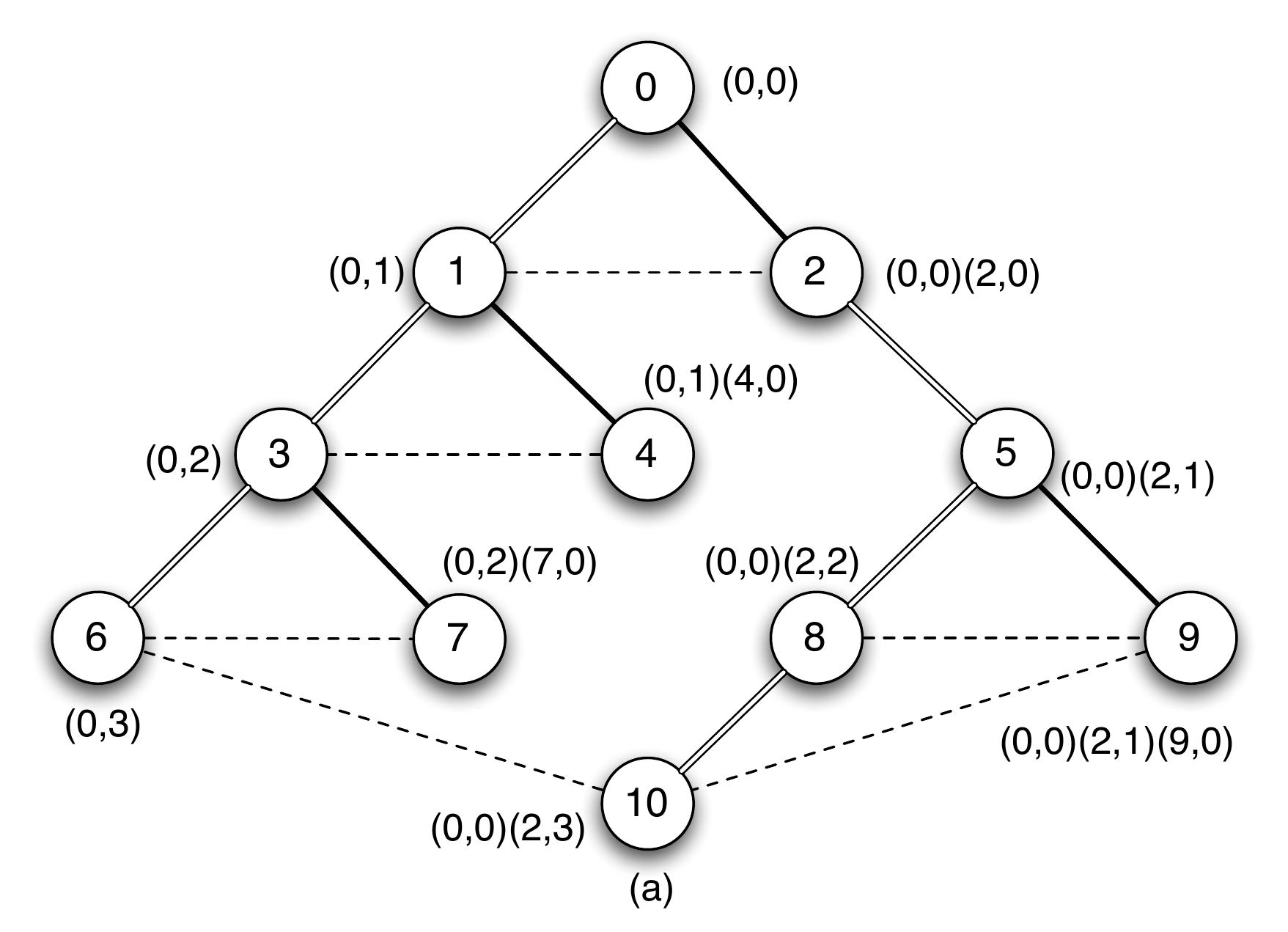}
\includegraphics[scale=0.4]{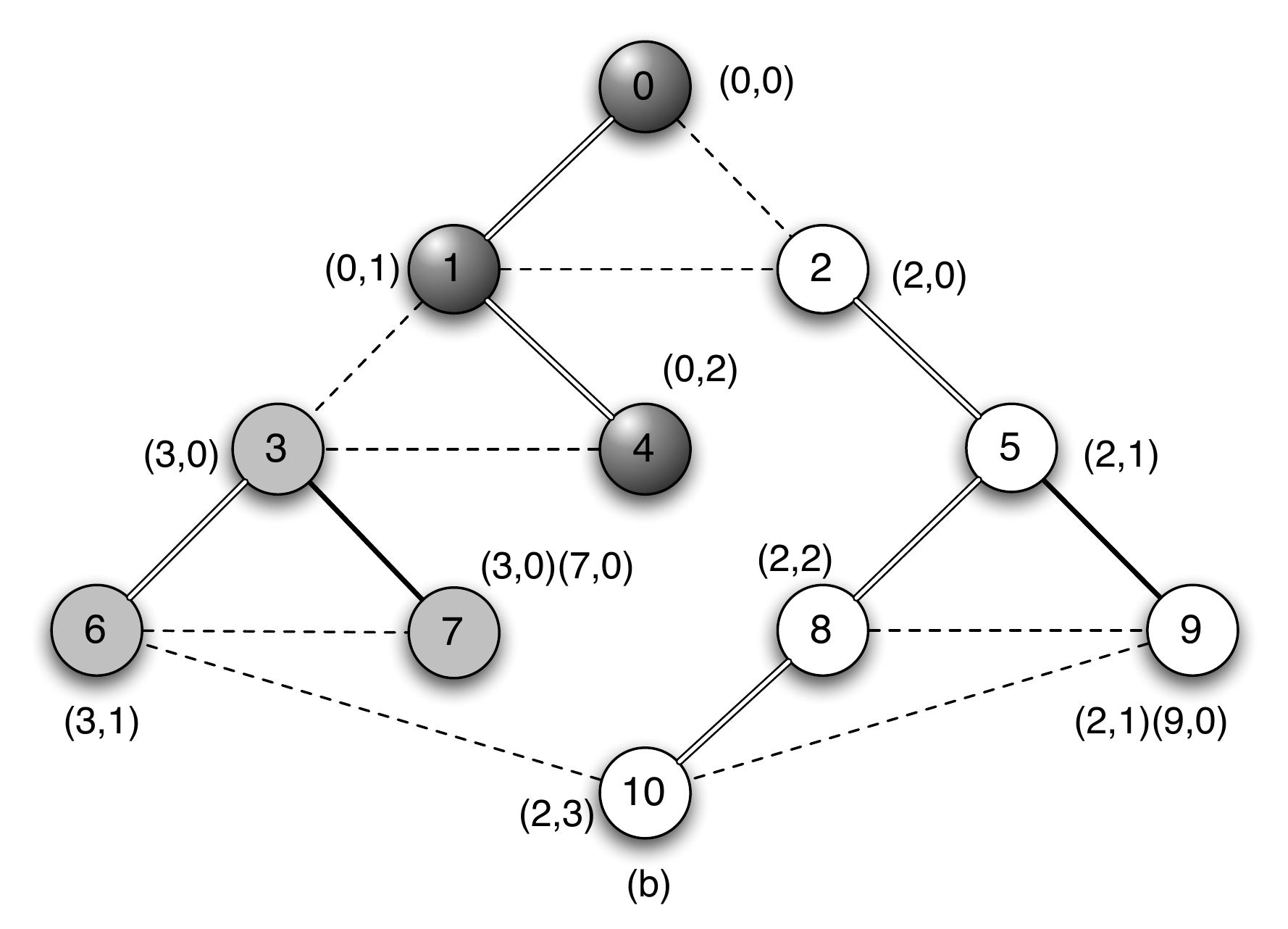}
\includegraphics[scale=0.6]{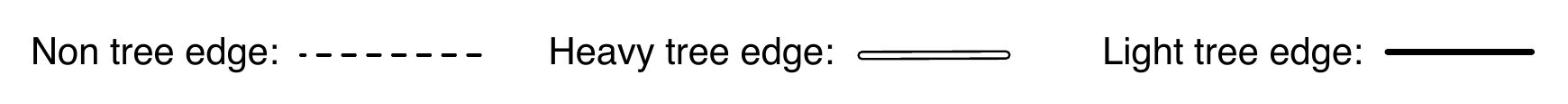}
\caption{Illustration du schéma d'étiquetage LCA} 
\label{fig:label}
\vspace*{-0,5cm}
\end{center}
\end{figure}

Chaque n{\oe}ud du réseau possède un identifiant unique. Pour l'étiquetage informatif du plus proche ancêtre commun (LCA --- pour \og least common ancestor\fg) dans un arbre enraciné, nous avons utilisé le travail de Harel et Tarjan~\cite{HarelT84} où les auteurs définissent deux sortes d'arêtes: les \emph{légères} et les \emph{lourdes}. Une arête est dite lourde si elle conduit au sous-arbre contenant le plus grand nombre de n{\oe}uds; elle est dite légère sinon. L'étiquetage est constitué d'un ou plusieurs couples. Le premier  paramètre d'un couple est l'identifiant d'un n{\oe}ud $u$. Le second est la distance au n{\oe}ud $u$. Un tel couple est noté $(\id_u,\dis_u)$. Le nombre de couples est borné par $O(\log n)$, car, comme il est remarqué dans~\cite{HarelT84}, il y a au plus $\log n$ arêtes légères sur n'importe quel chemin entre une feuille et la racine,  d'où il résulte des étiquettes de $O(\log^2 n)$ bits. Une racine $u$ sera étiquetée par $(\id_u,0)$. Un  n{\oe}ud $v$ séparé  de son parent $u$ par une arête lourde prendra l'étiquetage $(\id_r,\dis_u+1)$. Un  n{\oe}ud $v$ séparé  de son parent $u$ par une arête légère prendra l'étiquetage $(\id_r,\dis_u)(\id_v,0)$. Cet étiquetage récursif est illustré dans la Figure~\ref{fig:label}.

L'étiquette $\lab_w$ du plus petit ancêtre commun $w$ entre deux n{\oe}uds $u$ et $v$, s'il existe, est calculée de la façon suivante. Soit $\lab$ une étiquette telle que $\lab_u=\lab .(a_0,a_1).\lab_u' $ et $\lab_v=\lab.(b_0,b_1).\lab_v' $.
Alors
\[
\lab_w =  \left\{
 \begin{array}{ll}
\lab.(a_0,a_1) & \mbox{\textbf{si} } (a_0=b_0 \vee \lab \neq \emptyset)\wedge( \lab_u \prec \lab_v) \\
 \lab.(b_0,b_1) & \mbox{\textbf{si} } (a_0=b_0 \vee \lab \neq \emptyset)\wedge( \lab_v \prec \lab_u) \\ 
 \emptyset & \mbox{\textbf{sinon}}
 \end{array} \right.
\]
S'il n'existe pas d'ancêtre commun à $u$ et $v$, alors ces deux n{\oe}uds sont dans deux  fragments distincts. Par ailleurs, nous utilisons l'ordre lexicographique sur les étiquettes, noté $\prec$, dans le but de détecter la présence de cycles. Un n{\oe}ud $u$ peut détecter la présence d'un cycle en comparant son étiquette avec celle de son parent $v$. Si $\lab_u\prec \lab_v$ alors le n{\oe}ud $u$ supprime son parent et devient racine de son propre fragment.

L'auto-stabilisation impose des contraintes supplémentaires non satisfaites par l'algorithme de Gallager, Humblet et Spira: la configuration après panne peut être un arbre couvrant qui n'est pas un \MST, et il faut donc rectifier cet arbre. Nous allons donc utiliser la propriété bleue pour fusionner les fragments, et la propriété rouge pour supprimer les arêtes qui ne font pas partie du \MST\/ final\footnote{Notons que ce n'est pas la première fois que les deux propriétés bleue et rouge sont simultanément utilisées dans un même algorithme distribué pour  réseaux dynamiques (e.g., \cite{ParkMHT90,ParkMHT92}), mais jamais, à notre connaissance, sous contrainte d'auto-stabilisation.}.
Chaque fragment identifie, grâce à l'étiquetage, l'arête de poids minimum sortant de son fragment,  et l'arête de poids minimum interne au fragment ne faisant pas partie de l'arbre. La première arête sert pour la fusion de fragments, et la seconde sert pour la correction de l'arbre. Nous donnons priorité à la correction sur la fusion. Pour la correction, soit $u$ le plus petit ancêtre commun des extrémités de l'arête $e$ où $e$ est l'arête interne de plus petit poids. S'il existe une arête $f$ de poids inférieur à $e$ sur le chemin entre les deux extrémités de  $e$ dans l'arbre courant, alors $f$ est effacé de l'arbre couvrant. Pour la fusion, l'arête sortante de poids minimum est utilisée.  
\vspace*{-0,3cm}
\paragraph{Complexité.}
La complexité en mémoire découle de la taille des étiquettes, à savoir $O(\log^2n)$ bits.  Pour la complexité temporelle on considère le nombre de rondes nécessaires à casser un cycle ou à effectuer une fusion. Dans les deux cas, une partie des n{\oe}uds ont besoin d'une nouvelle étiquette. Cette opération s'effectue en $O(n)$ rondes. Comme il y a au plus $\frac{n}{2}$ cycles, il faudra $O(n^2)$ rondes pour converger. Pour ce qui concerne les fusions, le pire des cas est lorsque chaque n{\oe}ud est un unique fragment. Dans ce cas, il faut effectuer $n$ fusions, d'où l'on déduit le même nombre de rondes pour converger, à savoir $O(n^2)$.
\vspace*{-0,3cm}
\subsection{Algorithme de Korman, Kutten et Masuzawa}

Nous concluons cette section en mentionnant que nos contributions~\cite{BlinPRT09,BlinDPR10} ont été récemment améliorées par Korman, Kutten et Masuzawa~\cite{KormanKM11}.   Ces auteurs se placent dans le même modèle que nos travaux. Leur article traite à la fois  la \emph{vérification}  et la construction d'un \MST. Le problème de la vérification a été introduit par Tarjan~\cite{Tarjan79} et est défini de la façon suivante. Etant donnés un graphe pondéré et un sous-graphe de ce graphe, décider si le sous-graphe forme un \MST\/ du graphe. La vérification de \MST\/ possède sa propre littérature. Il est à noter que le problème est considéré comme plus facile que la construction de \MST, en tout cas de façon centralisée. Fort de leur expérience en répartie  aussi bien pour le \MST\/ que pour l'étiquetage informatif~(\cite{KormanKP10,KormanK07}), Korman, Kutten et Masuzawa reprennent l'idée de l'étiquetage informatif pour le \MST\/ auto-stabilisant introduit dans~\cite{BlinDPR10}. Ils l'optimisent afin d'être  optimal en  mémoire $O(\log n)$~bits, et afin d'obtenir une convergence en temps  $O(n)$~rondes.
Pour cela ils reprennent l'idée de \emph{niveau de fragments} introduite par \GHS\/ pour limiter la taille des fragments, et pour que les fusions se fassent entre des fragments contenant à peu près le même nombre de n{\oe}uds. Cette dernière démarche permet d'obtenir au plus $O(\log n)$~fusions. L'étiquetage est composé de l'identifiant du fragment, du niveau du fragment, et des deux identifiants des extrémités de l'arête de poids minimum ayant servi à la fusion. Grâce à  cette étiquette, il est possible d'effectuer la vérification, ce qui  leur permet de corriger le \MST\/ courant, et donc d'être auto-stabilisant. Une des difficultés de leur approche est de maintenir une mémoire de $O(\log n)$~bits. En effet, lorsque le \MST\/ est construit, chaque n{\oe}ud a participé à au plus $\log n$~fusions. Les étiquettes peuvent donc atteindre $O(\log^2n)$~bits. Pour maintenir une mémoire logarithmique, les auteurs  distribuent l'information et la font circuler à l'aide de ce qu'ils appellent un \emph{train}, qui pipeline le transfert d'information entre les n{\oe}uds.
\vspace*{-0,3cm}
\section{Conclusion}

Ce chapitre a présenté mes contributions à l'auto-stabilisation visant à construire des \MST\/ en visant une double optimalité, en temps et en mémoire. A ce jour, en ce qui concerne les algorithmes silencieux, l'algorithme de Korman et al.~\cite{KormanKM11} est optimal en mémoire. Le manque de bornes inférieures sur le temps d'exécution d'algorithmes auto-stabilisants ne permet pas de conclure sur l'optimalité en temps de~\cite{KormanKM11}. Nous pointons donc le problème ouvert suivant~: 

\begin{problem}
Obtenir une borne inférieure non triviale du temps d'exécution d'algorithmes auto-stabilisants silencieux (ou non) de construction de \MST. 
\end{problem}

Pour ce qui concerne la taille mémoire, il n'existe pas de borne dans le cas d'algorithmes non-silencieux pour le \MST, ni même pour la construction d'arbres couvrants en général. 

\begin{problem}
Obtenir une borne inférieure non triviale de la taille mémoire d'algorithmes auto-stabilisants non-silencieux de construction d'arbres couvrants. 
\end{problem}

Enfin,  un des défis de l'algorithmique répartie est d'obtenir des algorithmes optimaux à la fois en mémoire et en temps. Dans le cas du \MST, nous soulignons le problème suivant~: 

\begin{problem} 
Concevoir un algorithme réparti de construction de \MST, optimal en temps et en nombre de messages, dans le modèle ${\cal CONGEST}$.
\end{problem}

\chapter{Autres constructions d'arbres couvrants sous contraintes}
\label{chap:MDST}

Ce chapitre a pour objet de présenter mes travaux sur la construction d'arbres couvrants optimisés, différents du \MST\/. La première section est consacrée à l'approche \emph{\SC}\/  évoquée dans le chapitre précédent. Cette même section présente  deux de mes contributions mettant en {\oe}uvre cette propriété. La première est appliquée au \MST\/ dynamique auto-stabilisant, la seconde  est appliquée à la généralisation de la propriété \SC\/ à \emph{toute} construction d'arbres couvrants dans les réseaux dynamiques. La section~\ref{sec:steiner} est consacrée  au problème de l'arbre de Steiner, c'est-à-dire une généralisation du problème \MST\/ à la couverture d'un sous-ensemble quelconque de n\oe uds. Enfin les deux dernières sections du chapitre traitent du problème de la minimisation du degré de l'arbre couvrant,  l'une dans le cas des réseaux non-pondérés, l'autre dans le cas des réseaux pondérés. Dans le second cas, on vise la double minimisation du degré et du poids de l'arbre couvrant.

\section{Algorithmes auto-stabilisants \SC}
\label{section:SC}

Dans cette section, nous nous intéressons à la propriété \SC\/  (\og loop-free \fg\/ en anglais).  Cette propriété est particulièrement intéressante dans les réseaux qui supportent un certain degré de dynamisme. Elle garantit qu'une structure d'arbre couvrant est préservée pendant tout le temps de l'algorithme, jusqu'à convergence vers l'arbre couvrant optimisant la métrique considérée. Les algorithmes \SC\/ ont été introduits par~\cite{GafniB81,Aceves93}. 

Cette section, présente un état de l'art des algorithmes auto-stabilisant \SC, suivi d'un résumé de mes deux contributions à ce domaine~: un algorithme auto-stabilisant  \SC\/ pour le \MST, et une méthode de transformation de tout algorithme auto-stabilisant de construction d'arbres couvrants en un algorithme \SC. 


\subsection{Etat de l'art en auto-stabilisation}

\begin{figure}[tb!]
\center{

\subfigure[\footnotesize{}]{
\includegraphics[scale=0.35]{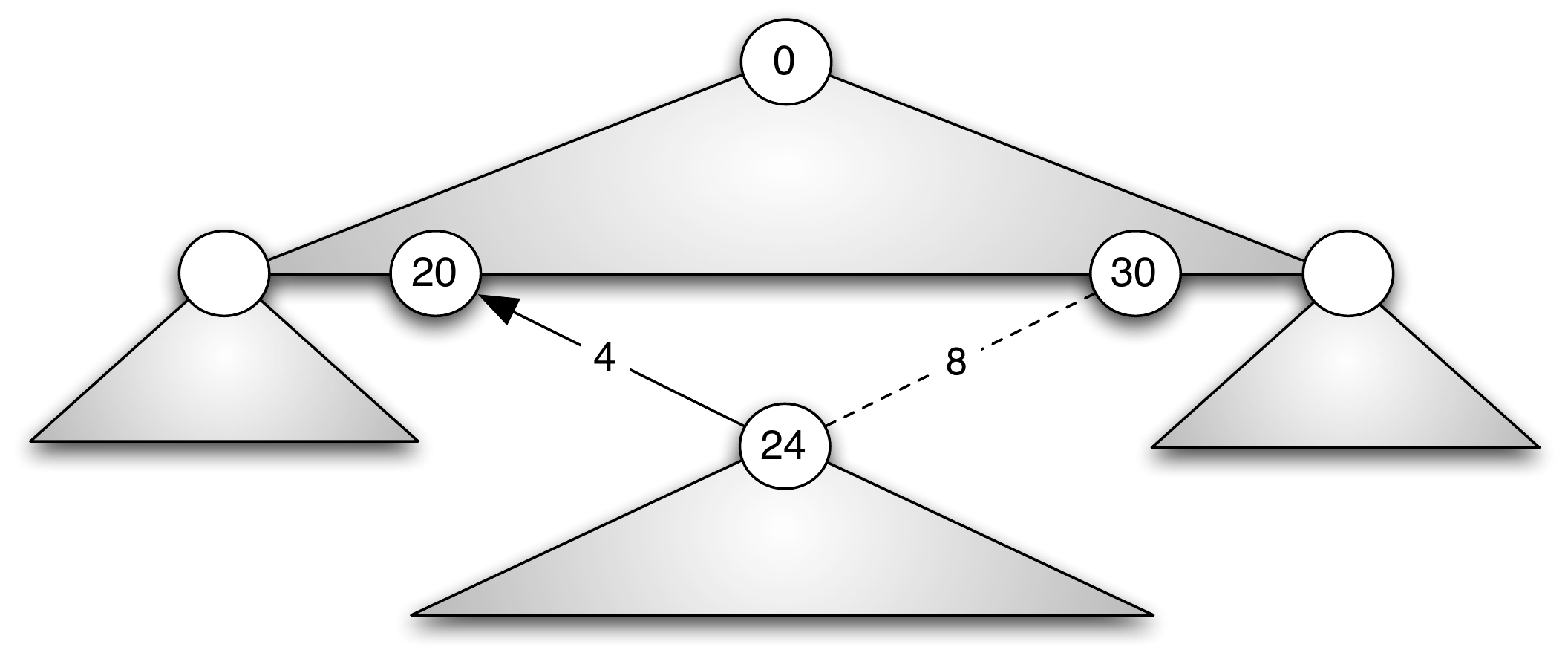}
\label{fig:MDST1}
}
\subfigure[\footnotesize{}]{
\includegraphics[scale=0.35]{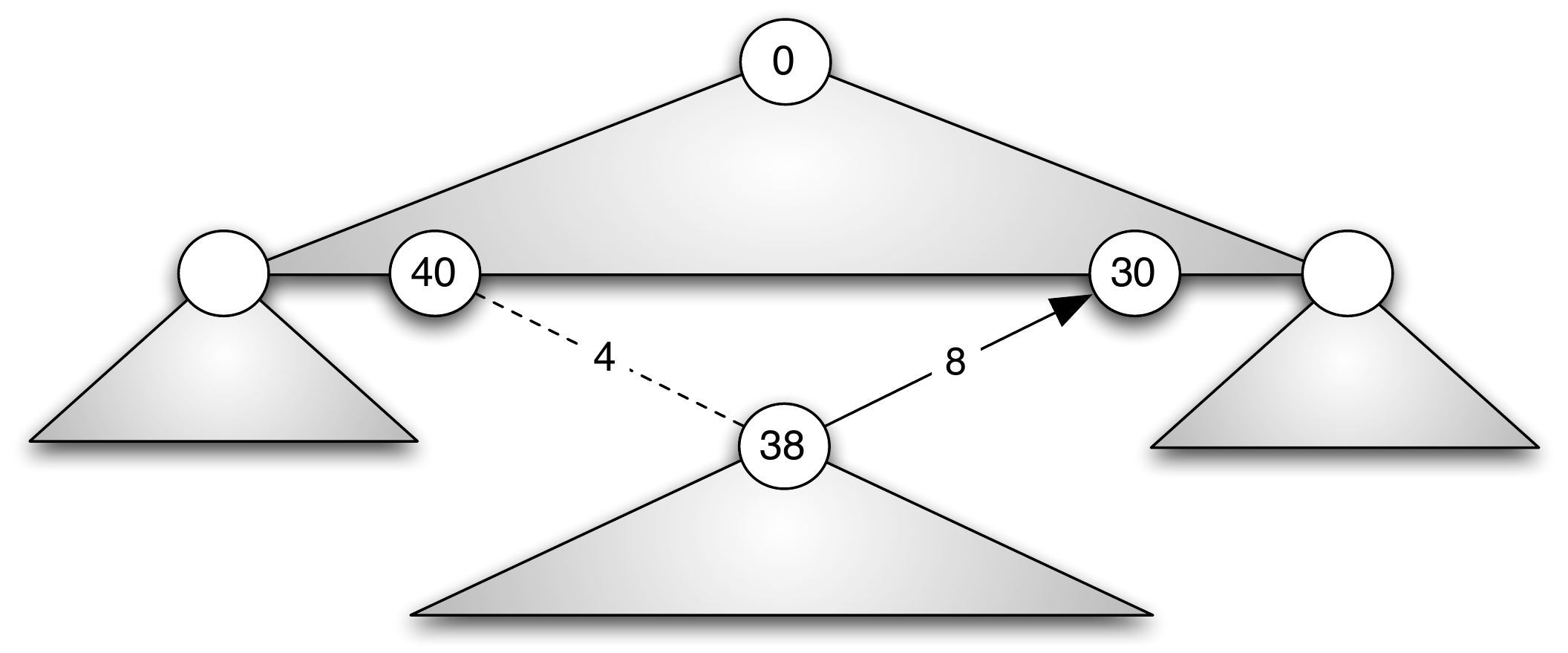}
\label{fig:ex2}
}
\caption{\small{Changement de parent de manière atomique et locale dans un arbre de plus courts chemins.}} 
\label{fig:SP}
}
\end{figure}

A notre connaissance, il n'existe que  deux contributions à l'auto-stabilisation faisant référence à la notion de \SC~: \cite{CobbG02} et \cite{JohnenT03}. Ces deux travaux s'intéressent à la construction d'arbres couvrants de plus court chemin enracinés. L'article de Johnen et Tixeuil~\cite{JohnenT03} améliore les résultats de Cobb et Gouda~\cite{CobbG02}. En effet, contrairement  à~\cite{CobbG02}, \cite{JohnenT03} ne nécessite aucune connaissance a priori du réseau. Dans ces deux articles, le dynamisme considéré est l'évolution  au cours du temps des valeurs des arêtes. L'algorithme s'exécute de la façon suivante. Chaque n{\oe}ud $u$ maintient sa distance à la racine $r$, et pointe vers un voisin (son parent) qui le conduit par un plus court à cette racine. Pour maintenir la propriété \SC, un n{\oe}ud $u$ qui s'aperçoit d'un changement de distance dans son voisinage qui implique un changement de parent, vérifie que le  n{\oe}ud voisin $v$ qui annonce la plus courte distance vers la racine n'est pas un de ses descendants. Si $v$ n'est pas un descendant de  $u$, alors $u$ change (de façon atomique et locale --- voir Figure~\ref{fig:SP}) son pointeur vers $v$. Sinon, il reste en attente de la mise à jour de son sous-arbre. 

\subsection{Algorithme auto-stabilisant \SC\/ pour le \MST}

Les contributions ci-dessus soulèvent immédiatement la question de savoir s'il est possible de traiter de façon auto-stabilisante \SC\/ des problèmes de construction d'arbres dont le critère d'optimisation est  \emph{global}. Le problème du plus court chemin peut être qualifié de \og local \fg~\cite{GoudaS99} car les changements nécessaires à la maintenance d'un arbre de plus courts chemins sont locaux (changement de pointeurs entre arêtes incidentes).   En revanche, des problèmes comme le \MST\/ ou l'arbre couvrant de degré minimum, implique des changement entre arêtes arbitrairement distantes dans le réseau. Pour aborder des problèmes globaux, nous nous sommes intéressés à la construction auto-stabilisante \SC\/ d'un \MST.

Ma première contribution dans le domaine a été effectuée en  collaboration avec Maria Potop-Butucaru, Stéphane Rovedakis et  Sébastien Tixeuil~\cite{BlinPRT10}. Nous traitons le problème du \MST\/ dans un réseau où les poids des arêtes sont dynamiques. 
Un algorithme auto-stabilisant \SC\/ pour le \MST\/ a été décrit dans la section~\ref{subsubsec:tapccenMST}. Cette description a cependant omis la vérification du respect de la propriété \SC, que nous traitons maintenant. L'idée principale de notre algorithme consiste à travailler sur les cycles fondamentaux engendrés par les arêtes ne faisant pas partie de l'arbre. Si une arête $e$ ne faisant pas partie de l'arbre possède un poids plus petit qu'une arête $f$ faisant partie de l'arbre et du cycle engendré par $e$, alors $e$ doit être échangée avec $f$. Ce changement ne peut être fait directement sans violer la propriété \SC. Nous avons donc mis en place un mécanisme de changement atomique, arête par arête, le long d'un cycle engendré par~$e$. 

Notre algorithme possède donc deux caractéristiques conceptuelles essentielles~: 

\begin{itemize}
\item Application de la propriété \SC\/ à des optimisations globales.
\item Mécanisme de changement atomique arête par arête le long d'un cycle.
\end{itemize}

\paragraph{Brève description du mécanisme de changement atomique.}

Rappelons que le modèle dans lequel l'algorithme s'exécute est un modèle à registres partagés avec une atomicité lecture/écriture. Autrement dit, lorsqu'un n{\oe}ud est activé il peut en même temps lire sur les registres de ses voisins et écrire dans son propre registre. 

Comme nous l'avons vu  dans la section~\ref{subsubsec:tapccenMST}, lorsque le jeton arrive sur un n{\oe}ud $u$ qui possède une arête $e=\{u,v\}$ ne faisant pas partie de l'arbre couvrant courant, ce jeton est gelé et un message circule dans le cycle fondamental $C_e$ à l'aide des étiquettes sur les n{\oe}uds. Ce message récolte le poids de l'arête de poids maximum $f$, ainsi que les étiquettes de ses extrémités.  Supposons qu'au terme de cette récolte, l'arête $e=\{u,v\}$ doive être échangée avec $f$ (voir Figure~\ref{fig:ex}). Soit $w$ le plus proche ancêtre commun à $u$ et $v$. Supposons sans perte de généralité que l'arête $f$ est comprise entre $u$ et $w$. On considère alors le chemin $u,x_1,x_2,...,x_k,u'$ entre $u$ est $u'$ où $u'$ est l'extrémité de $f$ la plus proche de $u$. Le n{\oe}ud $u$ va changer son pointeur parent en une étape atomique (ce qui maintient la propriété \SC) afin de pointer vers $v$, puis le n{\oe}ud $x_1$ va changer son pointeur vers $u$, puis le n{\oe}ud $x_2$ va changer son pointeur vers $x_1$, ainsi de suite jusqu'à ce que $u'$ prenne pour parent $x_k$. Tous ces changements sont effectués  de manière atomique. La propriété \SC\/ est donc préservée.   

\subsection{Généralisation}

Dans ce travail en collaboration avec Maria Potop-Butucaru, Stéphane Rovedakis et  Sébastien Tixeuil, publié dans~\cite{BlinPRT10}, nous proposons un algorithme  généralisant la propriété \SC\/ à toute construction d'arbres couvrants sous contrainte. Contrairement aux algorithmes auto-stabilisants précédents~\cite{CobbG02,JohnenT03,BlinPRT10}, dont le dynamisme est dû uniquement aux changements de poids des arêtes, le dynamisme considéré dans cette sous-section est l'arrivée et le départ arbitraire de n{\oe}uds.

Soit  $T$ un arbre couvrant un réseau $G$, optimisant une critère $\mu$ donné, et construit par un algorithme $\cal{A}$. Supposons que le réseau $G$ subisse des changements topologiques jusqu'à obtenir un réseau $G'$. L'arbre $T$ n'est pas forcément optimal pour le réseau $G'$. Il faut donc  transformer l'arbre $T$ en un arbre optimisé pour le réseau $G'$. Cette transformation doit respecter la propriété \SC\/ pour passer de $T$ à $T'$. Pour cela nous utilisons de la composition d'algorithmes, dont, plus spécifiquement, la \emph{composition équitable} introduite par Dolev,  Israeli, et Moran~\cite{DolevIM90,DolevIM93}.  La composition équitable fonctionne de la manière suivante. Soit deux algorithmes $\cal M$ et $\cal E$, le premier est dit \og maître\fg\/ et le second est dit \og esclave \fg. 
\begin{itemize}
\item L'algorithme $\cal E$ est un algorithme \og statique \fg\/ qui calcule, étant  donné un graphe $G'$, un arbre couvrant $T'$ de $G'$ optimisant le critère $\mu$. 
\item L'algorithme $\cal M$ est un algorithme \og dynamique \fg\/ qui prend en entrée $T'$ et effectue le passage de $T$ à $T'$ en respectant la propriété \SC.
\end{itemize}
Dolev,  Israeli, et Moran~\cite{DolevIM90,DolevIM93} ont prouvé que la composition équitable de $\cal M$ avec $\cal E$ dans un contexte dynamique résulte en un algorithme qui respecte la propriété \SC\/ et dont l'arbre couvrant satisfera le critère d'optimisation $\mu$. Notre apport conceptuel dans ce cadre est le suivant~: 

\begin{itemize}
\item Conception d'un mécanisme générique pour la construction d'algorithmes auto-stabilisants \SC\/ pour la construction d'arbres couvrants optimisant un critère quelconque\footnote{Plus exactement, un algorithme réparti auto-stabilisant dans un environnement statique doit exister pour ce critère.}, dans des réseaux dynamiques. 
\end{itemize}

Afin d'utiliser la composition équitable, il faut concevoir un algorithme maître $\cal M$ qui nous permettra, par sa composition avec un algorithme de construction d'arbre optimisé, de supporter le dynamisme en respectant la propriété \SC. L'algorithme $\cal M$ que nous avons proposé est un algorithme (respectant la propriété \SC) de construction d'arbres couvrants en largeur. Cet algorithme est appelé \BFSC  (BFS pour \og Breadth-first search \fg\/ en anglais et SC pour sans-cycle). Son exécution repose sur l'existence d'une racine $r$. 

Soit $\cal E$ un algorithme de construction d'arbre optimisé pour le critère $\mu$. Cet algorithme sert d'oracle pour notre algorithme \BFSC. Soit un réseau $G$, et $T$ l'arbre couvrant de $G$ obtenu par $\cal E$.  \BFSC\/ effectue un parcours en largeur d'abord de $T$ à partir de la racine $r$. Ce parcours induit une orientation des arêtes de $T$, qui pointent vers la racine.  Après un changement topologique de $G$ en $G'$, un appel à $\cal E$ permet de calculer un nouvel arbre $T'$. \BFSC\/ utilise la combinaison de l'orientation des arêtes de $T$ et de la connaissance de $T'$ pour passer de $T$ à $T'$ par une succession d'opérations locales atomiques, comme dans la section précédente. 

Le coup additionnel temporel de cette composition est de $O(n^2)$ rondes, à ajouter à l'algorithme esclave $\cal E$. L'algorithme \BFSC\/ utilise une espace mémoire de $O(\log n)$ bits.

\section{Arbre de Steiner}
\label{sec:steiner}

Comme suite logique au \MST, je me suis intéressée au problème de l'arbre de Steiner. Ce problème est une généralisation du \MST\/ consistant à connecter un ensemble $S$ quelconque de n{\oe}uds  en minimisant le poids de l'arbre de connexion.  Les éléments de cet ensemble $S$ sont appelés les \emph{membres} à connecter.  Ce problème classique de la théorie  des graphes est un des problèmes NP-difficiles fondamentaux. Il est donc normal que les dernières décennies aient été consacrées à trouver la meilleure approximation possible à ce problème. De façon formelle le problème se présente de la façon suivante~:

\begin{definition}[Arbre de Steiner]
Soit $G=(V,E,w)$ un graphe non orienté pondéré, et soit $S\subset V$. On appelle arbre de Steiner de $G$ tout arbre couvrant  les n{\oe}uds de $S$ en minimisant la somme des poids de ses arêtes. 
\end{definition}

Un algorithme est une $\rho$-approximation de l'arbre de Steiner s'il calcule un arbre dont la somme des poids des arêtes est au plus $\rho$ fois le poids d'un arbre de Steiner (i.e., optimal pour ce critère de poids). 
Cette section présente un état de l'art pour le problème de Steiner ainsi que ma contribution dans le domaine.

\subsection{Etat de l'art}

D'un point de vu combinatoire, la première approximation est une 2-approximation dû  à Takahashi et Matsuyama~\cite{TakahashiM80}. Dans leur article, les auteurs fournissent trois algorithmes pour le problème de Steiner, qui illustrent des techniques utilisées par la suite. Le premier algorithme est un algorithme qui construit un \MST. Cet \MST\/ est ensuite \og élagué \fg\/, autrement dit les branches de l'arbre qui ne conduisent pas à un membre sont supprimées pour obtenir  un arbre de Steiner approché. Cette approche permet d'obtenir une  $(n -|S| + 1)$-approximation. Le deuxième algorithme construit un arbre de plus courts chemins enraciné à un n{\oe}ud membre, vers tous les autres n{\oe}uds membres. Par cette approche, on obtient une $(|S|+1)$-approximation. Enfin le dernier algorithme donne une 2-approximation. Il se base sur la même idée que l'algorithme de Prim. Autrement dit, on choisit un n{\oe}ud membre $u$ arbitraire, que l'on connecte par un plus court chemin à un n{\oe}ud membre $v$ le plus proche de $u$. Ensuite, la composante connexe créé par $u,v$ est connectée par un plus court chemin à un troisième n{\oe}ud membre $w$ le plus proche de cette composante connexe, et ainsi de suite jusqu'à obtenir une  composante connexe incluant tous les membres. 
L'approche combinatoire a une longue histoire~\cite{WuWW86,Zelikovsky93,BermanR94,PromelS97}. La meilleure approximation connue par des techniques combinatoires est de $1,55$. Elle est dû à Robins et Zelikovsky~\cite{RobinsZ05}.

La programmation linéaire est une autre approche pour trouver la meilleure approximation à l'arbre de Steiner. En 1998, Jain~\cite{Jain98} utilise une méthode d'arrondis successifs pour obtenir une 2-approximation. En 1999, Rajagopalan et Vazirani~\cite{RajagopalanV99} posent la question suivante: est-il possible d'obtenir une approximation significativement plus petite que~2  par une approche de programmation linéaire.
L'an dernier, \cite{ByrkaGRS10} ont répondu positivement  en obtenant une approximation bornée supérieurement par~$1,33$. 

Dans le cadre de l'algorithmique répartie, Chen Houle et Kuo~\cite{ChenHK93} ont produit une version répartie de l'algorithme de~\cite{WuWW86}, dont le rapport d'approximation est~2. Gatani, Lo Re et Gaglio~\cite{GataniRG05} ont ensuite proposé une version répartie de l'algorithme d'Imase et Waxman~\cite{ImaseW91}. Ce dernier est une version \emph{on line} de la construction de l'arbre de Steiner. Dans cette approche \og dynamique \fg\/, les membres arrivent un par un dans le réseau. L'arbre de Steiner est donc calculé par rapport aux membres déjà en place. Dans l'algorithme d'Imase et Waxman,  un membre qui rejoint le réseau se connecte à l'arbre existant par le plus court chemin.  Par cette  approche, les auteurs obtiennent une $\log |S|$-approximation.

Lorsque je me suis  intéressée au problème de l'arbre de Steiner, seuls deux travaux auto-stabilisants existaient, \cite{KameiK02} et \cite{KameiK04}, par les mêmes auteurs: Kamei et Kakugawa. A ma connaissance, aucun autre publication n'a été produite dans le domaine depuis, sauf ma propre contribution, que je détaille dans la section suivante.  Kamei et Kakugawa considèrent dans \cite{KameiK02,KameiK04} un environnement dynamique. Le dynamisme étudié est toutefois assez contraint puisque le réseau est statique, et seuls les n{\oe}uds peuvent changer de statut, en devenant membres ou en cessant d'être membre. Dans les deux articles, les auteurs utilisent le modèle à registres partagés avec un adversaire centralisé non équitable. Dans leur premier article, ils proposent une version auto-stabilisante de l'algorithme dans~\cite{TakahashiM80}. Dans cette approche, l'existence d'un \MST\/ est supposée a priori, et seul le module d'élagage est fourni. Dans le second article \cite{KameiK04}, les auteurs proposent une approche en  quatre couches:
\begin{inparaenum}[\itshape (i\upshape)] 
\item Construction de fragments, i.e., chaque n{\oe}ud non membre se connecte au membre le plus proche par un plus court chemin, ce qui crée une forêt dont chaque sous-ensemble est appelé fragment.
\item Calcul du graphe réduit $H$, i.e., concaténation des arêtes entre les différents fragments.
\item Calcul d'un \MST\/ réduit: le \MST\/ de $H$.
\item Elagage de ce \MST.
\end{inparaenum}
\vspace*{-0,5cm}
\subsection{Contribution à la construction auto-stabilisante d'arbres de Steiner}

J'ai contribué à   l'approche auto-stabilisante pour le problème de l'arbre de Steiner  en collaboration de Maria Potop-Butucaru et Stéphane Rovedakis. Les résultats  de cette collaboration ont été publié dans ~\cite{BlinPR09b}. Dans les deux algorithmes que proposent Kamei et Kakugawa, il est supposé l'existence a priori d'un algorithme auto-stabilisant de construction de \MST. Nous avons conçu une solution qui ne repose pas sur un module de \MST.  Cette solution est basé sur l'algorithme \emph{on-line} d'Imase et Waxsman~\cite{ImaseW91}, dont il résulte une solution capable de supporter un  dynamisme plus important que celui de \cite{KameiK02,KameiK04}, à savoir l'arrivée et le départ de n{\oe}ud du réseau.

L'apport conceptuel de cet algorithme est donc double: 
\begin{itemize}
\item d'une part, il est le premier algorithme à ne pas reposer sur l'existence a priori d'un  algorithme auto-stabilisant pour le \MST; 
\item d'autre part, il est le premier à considérer un dynamisme important (départs et arrivées de n{\oe}uds). 
\end{itemize}
Ce double apport  nous a de plus permis de fournir des garanties sur la structure couvrante restante après une panne si l'algorithme avait eu le temps de converger (\og super stabilisation\fg). 

\paragraph{Brève description de l'algorithme.}

Notre travail se place dans le cadre d'un modèle par passage de messages, avec un adversaire faiblement équitable, et une atomicité envoie/réception. L'algorithme se décompose en quatre phases ordonnées et utilise un membre comme racine. Il s'exécute comme suit: 
\begin{inparaenum}[\itshape (i\upshape)] 
\item chaque n{\oe}ud met à jour sa distance à l'arbre de Steiner courant,
\item chaque n{\oe}ud souhaitant devenir membre envoie une requête de connexion,
\item connexion des nouveaux membres après accusé de réception des requêtes, 
\item mise à jour de l'arbre de Steiner courant, dont mise à jour de la distance de chaque n{\oe}ud de l'arbre à la racine de l'arbre. 
\end{inparaenum} 
L'algorithme utilise explicitement une racine et une variable gérant la distance entre chaque n{\oe}ud et la racine afin d'éliminer les cycles émanant d'une configuration initiale potentiellement erronée. Un arbre de plus courts chemins vers l'arbre de Steiner courant est maintenu pour tout n{\oe}ud du réseau. Les membres, ainsi que les n{\oe}uds du réseau impliqués dans l'arbre de Steiner, doivent être déclarés connectés. Si un n{\oe}ud connecté détecte une incohérence (problème de distance, de pointeur, etc.) il se déconnecte, et lance l'ordre de déconnexion dans son sous-arbre. Lorsque un membre est déconnecté, il lance une requête de connexion via l'arbre de plus courts chemins, et attend un accusé de réception pour réellement se connecter. Quand il est enfin  connecté, une mise à jour des distances par rapport au nouvel arbre de Steiner est initiée.

En procédant de cette manière nous obtenons, comme Imase et Waxsman~\cite{ImaseW91}, une $(\lceil \log |S| \rceil)$-approximation. La mémoire utilisée en chaque n{\oe}ud est de $O(\delta \log n)$ bits où $\delta$ est le degré de l'arbre couvrant courant. La convergence se fait en $O(D|S|)$ rondes, où $D$ est le diamètre du réseau.

\section{Arbre couvrant de degré minimum}

La construction d'un arbre couvrant de degré minimum a très peu été étudiée dans le domaine réparti. Pourtant, minimiser le degré d'un arbre est une contrainte naturelle. Elle peut par ailleurs se révéler d'importance pratique car les n{\oe}uds de fort degré favorisent la congestion des communications. Ils sont également les premiers  n{\oe}uds ciblés en cas d'attaque visant à déconnecter un réseau. Par aileurs, dans le monde du  pair-à-pair, il peut être intéressant pour les utilisateurs eux même d'être de faible degré. En effet, si un utilisateur possède une information très demandée, chaque lien (virtuel) de communication sera potentiellement utilisé pour fournir cette information, ce qui peut entrainer une diminution significative de sa propre bande passante.

En 2004 j'ai proposé avec Franck Butelle~\cite{BlinB04} le premier algorithme réparti pour la construction d'arbre couvrant de degré minimum. Lorsque je me suis intéressé à l'auto-stabilisation, c'est naturellement à la construction d'arbre couvrant de degré minimum que je me suis consacrée en premier lieu. De façon formelle, le problème est le suivant.

\begin{definition}[Arbre couvrant de degré  minimum]
Soit $G$ un graphe non orienté. On appelle arbre couvrant de degré minimum de $G$ tout arbre couvrant dont le degré\footnote{Le degré de l'arbre est le plus grand degré des n{\oe}uds} est minimum parmi tous les arbres couvrants de $G$.
\end{definition}

\subsection{Etat de l'art}

\begin{figure}[tb!]
\center{

\subfigure[\footnotesize{Degré maximum $u$}]{
\includegraphics[scale=0.35]{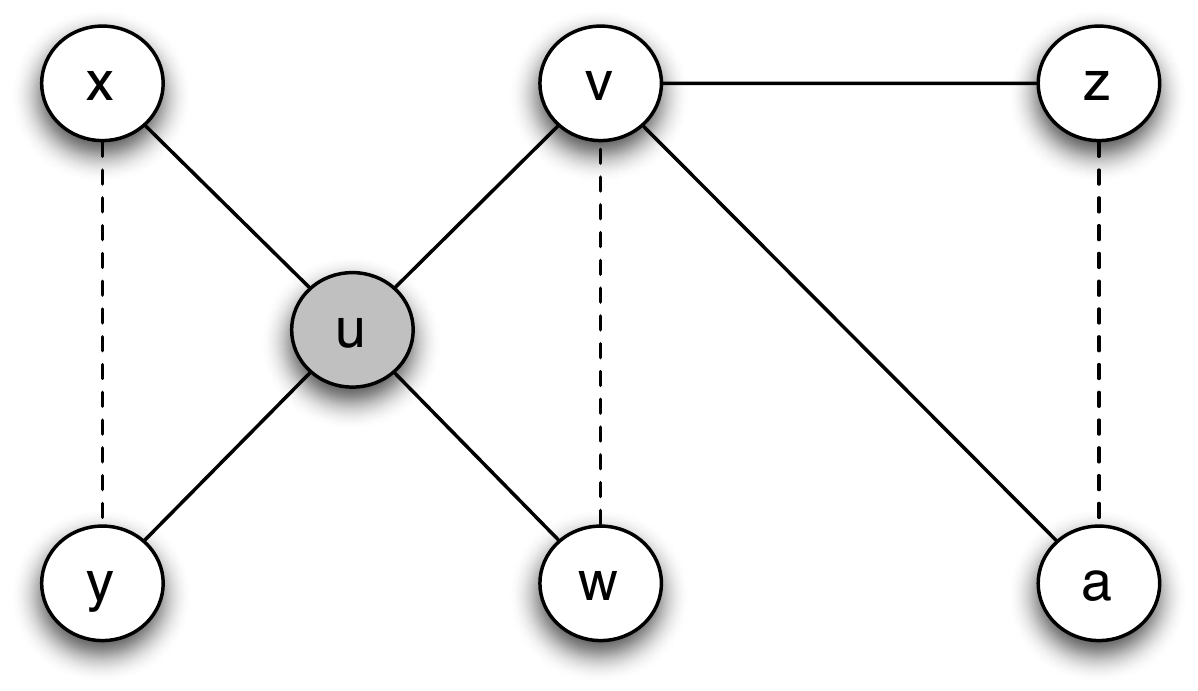}
\label{fig:MDST1}
}
\subfigure[\footnotesize{Degré maximum $u$ et $v$}]{
\includegraphics[scale=0.35]{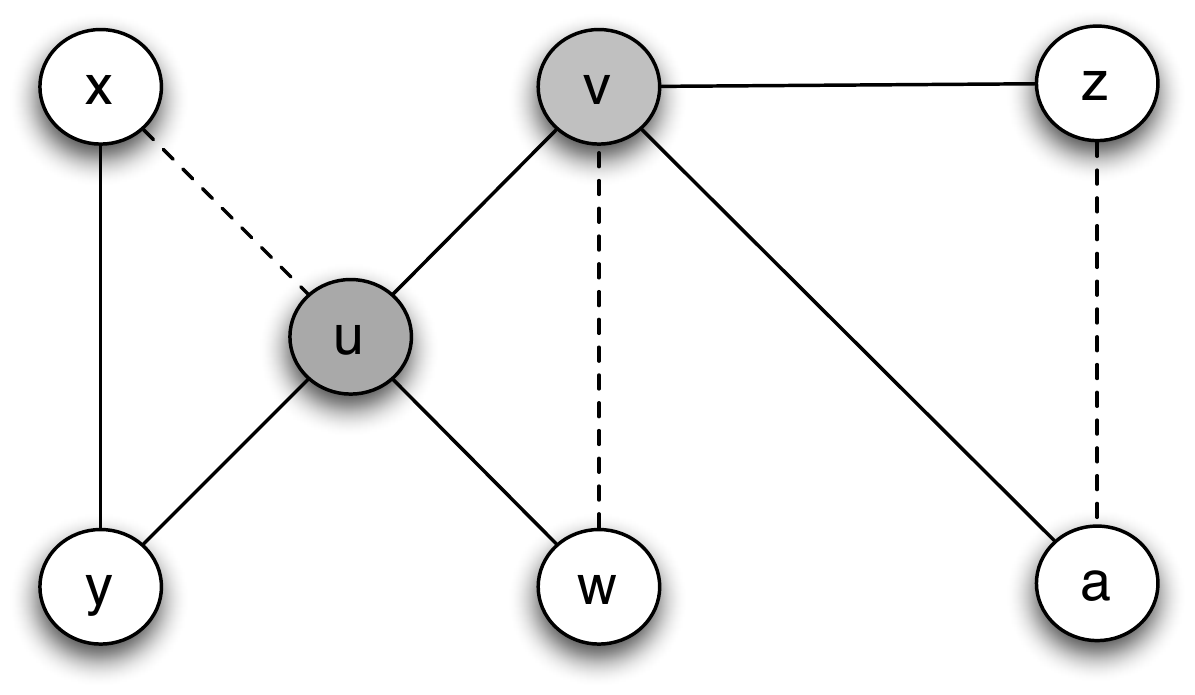}
\label{fig:ex2}
}
\subfigure[\footnotesize{Arbre optimal pour le degré}]{
\includegraphics[scale=0.35]{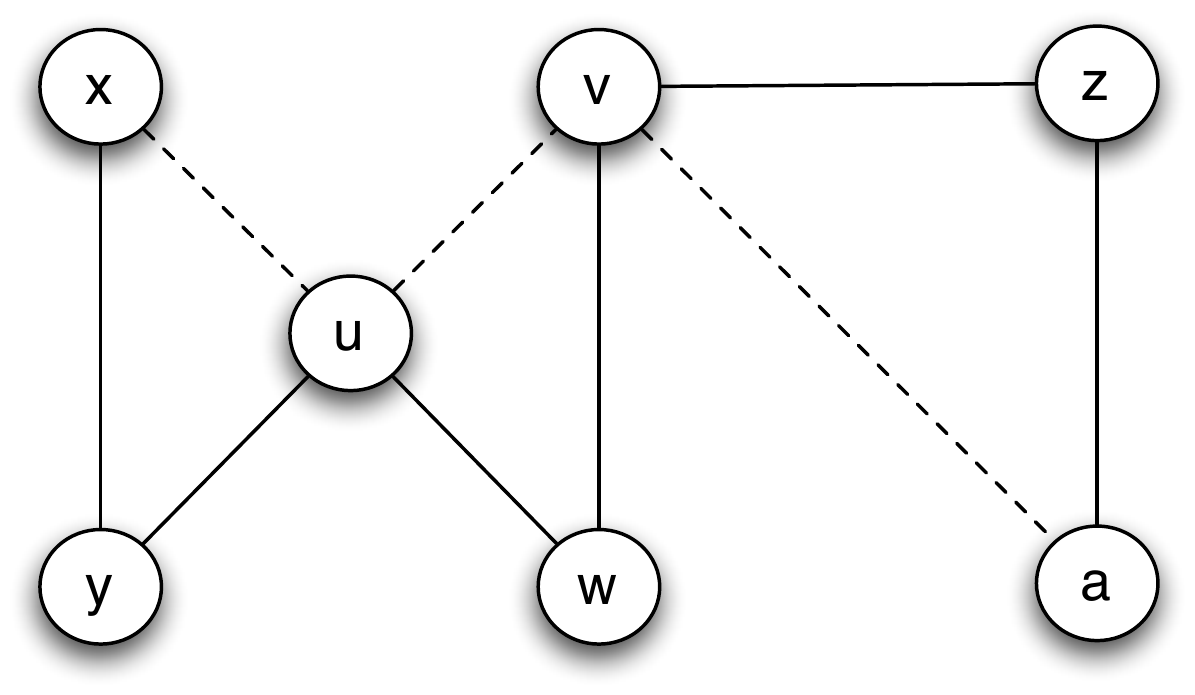}
\label{fig:ex2}
}

\caption{\small{Diminution des degrés de l'arbre couvrant.}} 
\label{fig:MDST}
}
\end{figure}

Le problème de l'arbre couvrant de degré minimum est connu comme étant  \emph{NP-difficile}, par réduction triviale du problème du chemin Hamiltonien. F\"urer et
Raghavachari~\cite{FurerR92,FurerR94} sont les premiers à s'être intéressés à ce  problème en séquentiel. Ils ont montré que le problème est facilement approximable en fournissant un algorithme calculant une solution de degré $\mbox{OPT}+1$ où $\mbox{OPT}$ est la valeur du degré minimum. L'algorithme est glouton, et fonctionne de la manière suivante. Au départ un arbre couvrant quelconque est construit. Puis, de façon itérative, on essaie de réduire le degré des n{\oe}uds de plus fort degré, jusqu'à que ce ne soit plus possible. Pour diminuer le degré $k$ d'un n{\oe}ud $u$, on identifie ses enfants, notés $u_1,...,u_k$, ainsi que les sous-arbres enracinés en ces enfants, noté $T_{u_1},...,T_{u_k}$, respectivement. Sans perte de généralité, considérons $v$ un descendant de $u_1$ tel que $\deg(v)<\deg(u)-2$. Si $v$ possède une arête $e$ qui ne fait pas parti de l'arbre, et dont l'extrémité $w$ est élément de $T_{u_j}$ avec $j\neq 1$,   alors l'algorithme échange $e$ avec l'arête $\{u,v\}$, ce qui a pour conséquence de diminuer le degré de $u$ tout en maintenant un arbre couvrant. La condition $\deg(v)<\deg(u)-2$ assure qu'après un échange, le nombre de n{\oe}uds de degré maximum aura diminué (voir l'exemple dans la Figure~\ref{fig:MDST}). Ce procédé est en fait récursif, car s'il n'existe pas de descendant $v$ de $u$ tel que $\deg(v)<\deg(u)-2$, il peut se faire  que le degré d'un des descendants de $u$ puisse être diminué afin que, par la suite, on puisse  diminuer le degré de $u$.  Cette opération est répétée jusqu'à ce qu'aucune amélioration puisse être effectuée. 

L'algorithme présenté en 2004, en collaboration avec Franck Butelle~\cite{BlinB04}, a été conçu pour  le modèle par passage de message ${\cal CONGEST}$. Le principe est le suivant. L'algorithme construit tout d'abord un arbre couvrant quelconque. Par la suite, les n{\oe}uds calculent (à partir des feuilles) le degré maximum de l'arbre couvrant courant. Une racine  $r$ est identifiée comme le noeud ayant un degré maximum (en cas d'égalité, il choisit celui d'identifiant maximum). Tous les enfants de $r$ effectuent un parcours en largeur  d'abord du graphe. Chaque parcours est marqué par l'identifiant de l'enfant ayant initié le parcours. Une arête $e$ ne faisant pas parti de l'arbre est candidate à l'échange avec une arête de l'arbre si et seulement si les deux conditions suivantes sont réunies:
\begin{inparaenum}[\itshape (i\upshape)] 
\item $e$ est traversée par deux parcours en largeur d'abord d'identifiants différents;
\item les deux  extrémités $u$ et $v$ de $e$ satisfont  $\deg(u)<\deg(r)-2$ et  $\deg(v)<\deg(r)-2$.
\end{inparaenum}
L'algorithme suit ensuite les grandes lignes de l'algorithme de F\"urer et
Raghavachari~\cite{FurerR92,FurerR94} pour les échanges, ainsi que pour la récursivité de la recherche des arêtes échangeables. Pour identifier les arêtes candidates à l'échange notre algorithme inonde le graphe de plusieurs parcours en largeur d'abord, le nombre de messages échangés est donc très important. Malheureusement,  à  l'heure actuelle, aucune étude n'a été faite sur la complexité distribuée de ce problème.

\begin{problem}
Dans le modèle ${\cal CONGEST}$, quel est le nombre minimum de messages à échanger et quel est le nombre minimum d'étapes  à effectuer pour la construction d'une approximation à $+1$ du degré de l'arbre couvrant de degré minimum~?
\end{problem}

\subsection{Un premier algorithme auto-stabilisant}

Je me suis consacré à la construction auto-stabilisante d'arbres couvrants de degré minimum en collaboration avec Maria Potop-Butucaru et Stéphane Rovedakis~\cite{BlinPR09a,BlinPR11}. Nous avons obtenu le premier (et pour l'instant le seul) algorithme auto-stabilisant pour ce problème. L'algorithme est décrit dans le même modèle que celui de la section~\ref{subsubsec:tapccenMST}, quoique dans un cadre semi-synchrone puisqu'il utilise un chronomètre. Il est basé sur la détection de cycles fondamentaux induits par des arêtes ne faisant pas partie de l'arbre couvrant courant. En effet, si l'on considère un n{\oe}ud $u$ de degré maximum dans l'arbre, diminuer le degré de $u$ requiert d'échanger une de ses arêtes incidentes, $f$,  avec une arête $e$ ne faisant pas partie de l'arbre couvrant courant, et créant un cycle fondamental dont $u$ est élément. Notons que cette approche est plus efficace que celle de~\cite{BlinB04} car elle permet de diminuer simultanément le degré des n{\oe}uds de plus grand degré, et de limiter l'échange d'information (communication au sein de cycles plutôt que par inondation).

L'apport conceptuel de cet algorithme est donc double: 
\begin{itemize}
\item d'une part, il est le premier algorithme auto-stabilisant pour la construction de degré minimum à utiliser une approche de construction par cycle élémentaire; 
\item d'autre part, il permet de diminuer le degré de tous les n{\oe}uds de degré maximum en parallèle. 
\end{itemize}
Ce double apport nous a permis d'obtenir un  temps de convergence $O(mn^2\log n)$ rondes pour un espace mémoire de $O(\log n)$~bits où $m$ est le nombre d'arêtes du réseau. 

\paragraph{Brève description de l'algorithme.}

Notre algorithme fonctionne en quatre étapes:

\begin{enumerate}
\item Construction et maintien d'un arbre couvrant.
\item Calcul du degré maximum de l'arbre couvrant courant.
\item Calcul des cycles élémentaires.
\item Reduction (si c'est possible) des degrés maximum.
\end{enumerate}

La principale difficulté en auto-stabilisation est de faire exécuter ces quatre étapes de façon indépendante, sans qu'une étape remette en cause l'intégrité (auto-stabilisation) d'une autre. Ces étapes sont décrites ci-après. 

\textbf{Construction et maintien d'un arbre couvrant.} Pour construire et maintenir un arbre couvrant, nous avons adapté à nos besoins l'algorithme auto-stabilisant de construction d'arbres couvrants en largeur d'abord proposé par Afek, Kutten et Yung~\cite{AfekKY91}. 
L'arbre construit est enraciné au n{\oe}ud ayant le plus petit identifiant. Par la suite, diminuer le degré de l'arbre couvrant ne changera pas l'identifiant de la racine, donc ne remettra pas en cause cette partie de l'algorithme.

\textbf{Calcul du degré maximum.} Pour que chaque n{\oe}ud sache  si son degré est maximum dans le graphe, nous utilisons la méthode classique de propagation d'information avec retour (ou PIF pour \og Propagation of Information with Feedback\fg). De nombreuses solutions  stabilisantes existent~\cite{BlinCV03,CournierDPV05}. Elles ont pour avantage de stabiliser instantanément. Cependant, garantir la stabilisation instantanée requiert des techniques complexes. Afin de faciliter l'analyse de notre algorithme, nous avons  proposé une solution auto-stabilisante plus simple pour le PIF.

\textbf{Identification des cycles fondamentaux\footnote{La technique présentée ici n'utilise pas les étiquettes informatives. Elle a en effet été développée avant que nous proposions les étiquettes informatives pour les cycles en auto-stabilisation.}.} Soit $e=\{u,v\}$ une arête ne faisant pas parti de l'arbre couvrant. Pour chercher son cycle fondamental $C_e$, le n{\oe}ud extrémité de $e$ d'identifiant minimum, par exemple $u$, lance un parcours en profondeur d'abord de l'arbre couvrant. Comme tout parcours en profondeur, il y a une phase de \og descente \fg\/ où  les n{\oe}uds sont visités pour la première fois, et une phase de \og remontée \fg\/ où le message revient sur des n{\oe}uds déjà visités. Le message qui effectue le parcours  stocke l'identifiant des n{\oe}uds et leur degré pendant la phase de descente, et efface ces données pendant la phase de remontée. Le parcours s'arrête quand il rencontre l'autre extrémité de $e$, c'est-à-dire $v$. Lorsque $v$ reçoit le message de parcours, celui-ci contient tous les identifiants et les degré des n{\oe}uds de l'unique chemin dans l'arbre entre $u$ et $v$. La taille de ce message peut donc atteindre $O(n \log n)$ bits. Cette idée est basé sur l'algorithme auto-stabilisant pour le \MST\/ de  Higham et Lyan~\cite{HighamL01}. L'algorithme résultant possède donc les mêmes défauts. En particulier, il nécessite un chronomètre pour déclencher périodiquement une recherche de cycles fondamentaux à partir des arêtes ne faisant pas partie de l'arbre couvrant. Il est donc semi-synchrone et non-silencieux. 

\textbf{Réduction des degrés.} Lorsque l'un des n{\oe}uds extrémités $v$ de l'arête $e$ ne faisant pas parti de l'arbre récupère les informations relatives au cycle fondamental $C_e$, l'algorithme détermine si $C_e$ contient un n{\oe}ud de degré maximum ou un n{\oe}ud bloquant. Dans le cas d'un n{\oe}ud de degré maximum,  si $u$ et $v$ ne sont pas des n{\oe}uds bloquants, alors $e$ est rajouté, et une des arête incidentes au n{\oe}ud de degré maximum  est supprimée. Cette suppression est effectuée à l'aide d'un message. L'important dans cette étape est de maintenir un arbre couvrant et de maintenir une orientation vers la racine, ce qui peut nécessiter une réorientation d'une partie de l'arbre couvrant avant la suppression de l'arête. Si $u$ et $v$ sont des n{\oe}uds bloquants, alors ils attendrons un message de recherche de cycle pour  signaler leur état, et être potentiellement débloqués par la suite.

\medskip

\paragraph{Complexité}
L'algorithme converge vers une configuration légitime  en $O(|E|n^2 \log n)$ rondes et utilise $O(\Delta \log n)$ bits de mémoire, ou $\Delta$ est le degré maximum du réseau. Le nombre important de rondes est dû à la convergence de chacune des étapes de notre algorithme. Il convient de noter que nous ne sommes pas dans le modèle ${\cal CONGEST}$, et que pour récolter les informations du cycle fondamental, nous utilisons des messages de taille $O(n \log n)$ bits. En revanche, l'information stockée sur la mémoire des n{\oe}uds est quant à elle de $O(\Delta \log n)$ bits. Enfin,  l'utilisation d'un chronomètre pour les arêtes ne faisant pas parti de l'arbre, rend cet algorithme semi-synchrone et non-silencieux.

\paragraph{Remarque.} 

Fort de  l'expérience acquise dans le domaine ces dernières années, un algorithme auto-stabilisant plus performant (dont silencieux) pour ce problème pourrait maintenant être proposé dans le modèle ${\cal CONGEST}$ avec une taille mémoire optimale de $O(\log n)$ bits. 

\section{Perspective: Arbre couvrant de poids et de degré minimum}

Cette section a pour objet d'ouvrir quelques perspectives en liaison avec l'optimisation d'arbres couvrants sous contrainte. Ayant traité séparément le problème de l'arbre couvrant de poids minimum et celui de l'arbre couvrant de degré minimum, il est naturel de s'intéresser maintenant à la combinaison des deux problèmes. Ce chapitre est ainsi consacré à ce problème bi-critère. Un bref état de l'art du problème de la construction d'arbres couvrants de poids minimum et de degré borné est présenté ci-après. Cet état de l'art me permettra de conclure par  un certain nombre de pistes de recherche.

En 2006, Goemans~\cite{Goemans06} émet la conjecture que l'approximation obtenue par F\"urer et Raghavachari~\cite{FurerR92} peut se généraliser aux graphes pondérés. Autrement dit, il conjecture que, parmi les \MST, on peut trouver en temps polynomial un \MST\/ de degré au plus $\Delta^*+1$, où $\Delta^*$ est le degré minimum de tout \MST.  Ce problème d'optimisation bi-critère est référencé dans la littérature par \emph{arbre couvrant de poids minimum et de degré borné} (en anglais \og Minimum Bounded Degree Spanning Trees\fg\/ ou MBDST). Sa définition formelle est la suivante. Soit $G$ un graphe. La solution cherchée est contrainte par un entier $B_v$ donné pour chacun des n{\oe}uds $v$ du graphe. MBDST  requiert de trouver un \MST\/ $T$ tel que, pour tout $v$, on ait $\deg_T(v) \leq B_v$. Soit $\mbox{OPT}$ le poids d'un tel \MST. Une $(\alpha,f(B_v))$-approximation de MBDST est un arbre $T$ dont le poids est au plus $\alpha \;\mbox{OPT}$, et tel que $\deg_T(v) \leq f(B_v)$. Par exemple le résultat de F\"urer et Raghavachari~\cite{FurerR92} peut est reformulé par une $(1,k+1)$ approximation dans le cas des graphes non pondérés (i.e., $B_v=k$ pour tout $v$).

Fischer propose, dans le rapport technique~\cite{Fischer93}, d'étendre la technique algorithmique de F\"urer et
Raghavachari~\cite{FurerR92} pour les graphes pondérés. Plus précisément, cet auteur cherche un \MST\/ de degré minimum. Pour cela, il introduit deux modifications à l'algorithme de~\cite{FurerR92}. D'une part, l'arbre initial n'est pas quelconque, mais est un \MST. D'autre part, les échanges d'arêtes se font entre arêtes de poids identiques.  Fischer annonce que ces modifications permettent d'obtenir en temps polynomial un \MST\/ dont les n{\oe}uds ont degré au plus $O(\Delta^*+\log n)$, où $\Delta^*$ est le degré minimum de tout \MST.  (La même année, Ravi et al.~\cite{RaviMRRH93} ont adapté leur travail sur l'arbre Steiner au problème MBDST pour obtenir une  $(O(\log n),O(B_v \log n))$-approximation). En 2000, Konemann et al.~\cite{KonemannR00} ont repris l'approche de Fischer~\cite{Fischer93} et en ont effectué une analyse plus détaillée. Dans leur article, la programmation linéaire est utilisée pour la première fois pour ce problème.  Les mêmes auteurs améliorent ensuite leurs techniques dans~\cite{KonemannR02,KonemannR05} pour obtenir une $(1,O(B_v + \log n))$-approximation.  Chaudhuri et al.~\cite{ChaudhuriRRT05,ChaudhuriRRT06} utilisent quant à eux  une méthode développée pour le problème du flot maximum, pour obtenir une $(1,O(B_v))$-approximation. Enfin, Singh et Lau~\cite{SinghL07} prouvent la conjecture de Goemans~\cite{Goemans06}, en obtenant une $(1,B_v+1)$-approximation, toujours sur la base de techniques de programmation linéaire. 

Dans un contexte réparti, seuls Lavault et Valencia-Pabon~\cite{LavaultV08} traitent à ma connaissance ce problème. Ils proposent une version répartie de l'algorithms Fischer~\cite{Fischer93}, garantissant ainsi la même approximation. Leur algorithme a une complexité temporelle de $O(\Delta^{2+\epsilon})$ étapes, où $\Delta$ est le degré du \MST\/ initial, et une complexité en nombre de messages échangés de $O(n^{3+\epsilon})$ bits. 

Cet ensemble de travaux sur le MBDST invitent à considérer les problèmes suivants. D'une part, partant du constat qu'il est difficile de donner des versions réparties d'algorithmes utilisant la programmation linéaire, nous souhaiterions aborder le problème de façon purement combinatoire:  
 
\begin{problem}
Développer une approche combinatoire pour  obtenir un algorithme polynomial calculant une $(1,\Delta^*+1)$-approximation pour le problème de l'arbre couvrant de poids minimum, et de degré borné, dans le cas des graphes pondérés.
\end{problem}

Bien sûr, tout algorithme combinatoire polynomial retournant une $(1,\Delta^*+o(\log n))$-approximation serait déjà intéressante. En fait, il serait déjà intéressant de proposer un algorithme réparti offrant la même approximation que Fischer~\cite{Fischer93} dans le modèle $\cal CONGEST$. 

\part{Entités autonomes}
\label{part2}
\chapter{Le nommage en présence de fautes internes}
\label{chap:corruptionRobot}

La seconde partie du document est consacrée à des \emph{entités autonomes} (agents logiciels mobiles, robots, etc.) se déplaçant dans un réseau. Les algorithmes sont exécutés non plus par les n{\oe}uds du réseau, mais par les entités autonomes. L'ensemble du réseau et des entités autonomes forme un système dans lequel le réseau joue le rôle d'environnement externe pour les entités autonomes. Par abus de langage, nous utilisons dans ce document le terme \emph{robots} pour désigner les entités autonomes. Ce chapitre est consacré à un modèle dans lequel les robots n'ont qu'une vision \og locale\fg\/ du système (chacun n'a accès qu'aux informations disponibles sur le n{\oe}ud sur lequel il se trouve). Le chapitre suivant sera consacré  à un modèle dans lequel les robots ont une vision  \og globale\fg\/  du système. 

La plupart des algorithmes auto-stabilisants pour les robots~\cite{Ghosh00,HermanM01,BeauquierHS01,DolevSW02} cherchent à se protéger de pannes \emph{externes}, autrement dit de fautes générées par l'environnement, mais pas par les robots eux-mêmes. Ce chapitre présente une nouvelle approche de l'auto-stabilisation pour les robots, à savoir la conception d'algorithmes auto-stabilisants pour des fautes internes \underline{et} externes, c'est-à-dire générées par les robots et par leur environnement.  Cette nouvelle approche a été étudiée en collaboration avec M.~Potop-Butucaru et S.~Tixeuil~\cite{BlinPT07}. 

La plupart des algorithmes conçus pour des robots utilisent les identifiants de ces robots. Dans ce chapitre, les pannes internes induisent une corruption de la mémoire des robots. Les identifiants des robots peuvent donc être corrompus. Par conséquent, la tâche consistant à attribuer des identifiants deux-à-deux distincts aux robots apparait  comme une brique de base essentielle à l'algorithmique pour les entités mobiles. Cette tâche est appelée le \emph{nommage} (\og naming \fg\/ en anglais). Dans le cadre étudié dans la Partie~\ref{partie:un} du document, c'est-à-dire les algorithmes auto-stabilisants pour les réseaux, l'existence d'identifiants deux-à-deux distincts attribués aux n{\oe}uds est équivalente à l'existence d'un unique leader. Ce chapitre est consacré à cette équivalence dans le cas des algorithmes auto-stabilisants pour les robots susceptibles de subir des fautes internes et externes. 

La première partie de ce chapitre est consacrée à la  formalisation d'un modèle pour l'étude de systèmes de robots sujets à des défaillances transitoires internes et externes. Dans un deuxième temps, le chapitre est consacrée à des résultats d'impossibilité pour le problème du nommage. Nous montrons que, dans le cas général, le nommage est impossible à résoudre de façon déterministe, mais qu'il l'est au moyen d'un algorithme probabiliste. Dans le cadre déterministe, nous montrons que le nommage est possible dans un arbre avec des liens de communication semi-bidirectionnels. Ces résultats complètent  les résultats d'impossibilité dans les réseaux répartis anonymes (voir~\cite{YamashitaK96ja,YamashitaK96jb}). De plus, nos algorithmes peuvent servir de brique de base pour résoudre d'autres problèmes, dont en particulier le regroupement --- problème connu pour avoir une solution uniquement si les robots ont un identifiant unique~\cite{DessmarkFKP06}.

\section{Un modèle local pour un système de robots}

Soit  $G=(V,E)$ un réseau anonyme. Les robots sont des machines de Turing qui se déplacent de n{\oe}uds en n{\oe}uds dans $G$ en traversant ses arêtes, et qui sont capables d'interagir avec leur environnement. On suppose un ensemble de $k>0$ robots. Durant l'exécution d'un algorithme, le nombre de robots ne change pas (i.e., les robots ne peuvent ni disparaitre ni apparaitre dans le réseau). Chaque robot possède un espace mémoire suffisant pour stocker au moins un identifiant, donc $\Omega(\log k)$ bits. Pour $u\in V$, les arêtes incidentes à $u$ sont étiquetées par des  numéros de port deux-à-deux distincts, entre 1 et $\deg(u)$. Chaque n{\oe}ud du réseau possède un \emph{tableau blanc} qui peut stocker un certain nombre d'information, sur lequel les robots peuvent lire et écrire. Les arêtes sont \emph{bidirectionnelles}, c'est-à-dire utilisables dans les deux sens. On considérera deux sous-cas selon qu'une arête peut être traversée par deux robots dans les deux sens simultanément, ou uniquement dans un sens à la fois. Dans le second cas, on dira que l'arête est \emph{semi-bidirectionnelle}. Une \emph{configuration} du système est définie par l'ensemble des n{\oe}uds occupés par les robots, l'état des robots, et l'information contenue dans tous les tableaux blancs. Les informations suivantes sont accessibles à un robot $\rob$ occupant un n{\oe}ud $u$ du réseau : 

\begin{itemize}
\item le numéro de port de l'arête par laquelle  $\rob$  est arrivé en $u$, et le degré de $u$;
\item l'état de chacun des robots présents sur le n{\oe}ud $u$ en même temps que $\rob$;
\item les données stockée sur le tableau blanc de $u$.
\end{itemize}

En fonction des informations ci-dessus, le robot change d'état, et décide possiblement de se déplacer. Le système est \emph{asynchrone}. L'adversaire qui modélise l'asynchronisme est distribué, faiblement équitable (voir Chapitre~\ref{chap:SS-ST}). Les n{\oe}uds contenant au moins un robot sont dits activables. A chaque étape atomique, l'adversaire doit choisir un sous-ensemble non vide $S\subseteq V$ de n{\oe}uds activables. (On dit que les n{\oe}uds de $S$ sont activés par l'adversaire). L'algorithme a ensuite la liberté de choisir quel robot est activé  sur chacun des n{\oe}uds de $S$. Autrement dit, en une étape atomique, tous les n{\oe}uds choisis par l'adversaire exécutent le code d'au moins un robot localisé sur chacun d'entre eux. Dans ce cadre, une ronde est définie par le temps minimum que mettent tous les n{\oe}uds activables à être activés par l'adversaire.

Ce chapitre étudie la résistance d'un système de robots aux fautes transitoires, internes et externes. Pour cela, nous supposons que chaque faute dans le système peut modifier le système de façon arbitraire, c'est-à-dire, plus précisément,
\begin{inparaenum}[\itshape (i\upshape)] 
\item la mémoire (i.e., l'état) des robots (faute interne), 
\item la localisation des robots (faute externe), et
\item le contenu des tableaux blancs (faute externe).
\end{inparaenum}
Notons que la structure du réseau est statique, et que, comme nous l'avons dit, il n'y a pas de modification du nombre de robots. 
Le modèle de fautes ci-dessus généralise le modèle utilisé dans~\cite{Ghosh00,HermanM01,BeauquierHS01,DolevSW02} qui ne considère que les fautes externes. 

\section{Les problèmes du nommage et de l'élection}

Comme il a été dit précédemment, une grande partie de la littérature sur les robots supposent que ces derniers ont des identifiants non corruptibles. Dans notre modèle, les robots peuvent avoir une mémoire  erronée après une faute du système, ce qui implique des valeurs d'identifiants potentiellement erronées. La capacité de redonner aux robots des identifiants deux-à-deux distincts est donc indispensable dans un système de robots avec fautes internes. 

Le problème du nommage est formalisé de la manière suivante: Soit $S$ un système composé de  $k$ robots dans un graphe $G$. Le système $S$ satisfait la spécification de  {\it nommage} si  les $k$ robots ont des identifiants entiers entre 1 et $k$ deux-à-deux distincts. Le problème de l'élection et  équivalent au problème du nommage, identifiant deux à deux distincts. Il se formalise de la manière suivante.Soit $S$ un système composé de  $k$ robots dans un graphe $G$. Le système $S$ satisfait la spécification d'{\it élection} si un unique robot est dans l'état \og leader \fg\/  et tous les autres robots sont dans l'état \og battu\fg.

\begin{theorem} \label{theo:equivnoelection}
Blin et al.~\cite{BlinPT07}. Les problèmes du nommage et de l'élection auto-stabilisants sont équivalents même en présence de fautes internes, c'est-à-dire que $k$ robots avec des identifiants deux-à-deux distincts peuvent élire un leader, et $k$ robots disposant d'un leader peuvent s'attribuer des identifiants deux-à-deux distincts. 
\end{theorem}

La preuve du théorème ci-dessous est dans~\cite{BlinPT07}. Intuitivement, pour résoudre l'élection, on procède de la façon suivantes. Chaque robot effectue un parcours Eulérien auto-stabilisant du graphe. A chaque arrivée sur un n{\oe}ud, il inscrit son identifiant sur le tableau blanc.  Après la stabilisation des parcours, tous les tableaux blancs ont la liste de tous les identifiants. Le robot avec l'identifiant maximum peut se déclarer leader. Réciproquement, le principe de l'algorithme est le suivant. Le leader $\rob_L$ suit un parcours Eulérien du graphe. Les autres robots procèdent de façon à rejoindre  $\rob_L$ en suivant l'information que ce dernier laisse sur les tableaux blancs. $\rob_L$ prend 1 comme identifiant. Lorsqu'un robot $\rob$ rejoint  $\rob_L$, $\rob$ reste avec $\rob_L$ et prend comme identifiant le nombre de robots actuellement avec $\rob_L$, incluant $\rob$ et  $\rob_L$.

Tout comme la plupart des résultats d'impossibilité dans le modèle discret (cf. Introduction), les résultats d'impossibilité relatifs au nommage et à l'élection sont dus à l'existence de symétries entre les robots impossibles à briser.
Considérons par exemple le cas où $G$ est un cycle. Supposons qu'après une défaillance du système, 
\begin{inparaenum}[\itshape (i\upshape)] 
\item chaque n{\oe}ud de $G$ contient un seul robot (i.e., $k=n$), 
\item les robots sont tous dans le même état (incluant le fait qu'ils ont le même identifiant), et  
\item les tableaux blancs son vides.
\end{inparaenum} 
Dans ce cas, l'adversaire pourra  activer tous les robots indéfiniment. En effet, à chaque activation, les robots effectuent la même action, et tous les robots garderont le même état, de même que tous les n{\oe}uds garderont le même tableau blanc. Par conséquent, résoudre le problème de nommage (ou de l'élection) de façon déterministe dans un cycle est impossible, même dans un environnement synchrone,  avec une mémoire infini pour les robots et les tableaux blancs.

Le présence d'arêtes bidirectionnelles est également un obstacle à la résolution du nommage (et de l'élection). Supposons en effet, la présence de deux robots avec la même information à l'extrémité d'une même arête, si les robots peuvent traverser en même temps dans les deux sens  cette  arête, les robots se croisent sans jamais briser la symétrie. 

\section{Algorithmes auto-stabilisants pour le nommage}

Dans cette section, nous décrivons tout d'abord un algorithme déterministe dans un cadre contournant les deux obstacles mis en évidence dans la section précédentes, c'est-à-dire la présence de cycles et d'arêtes bidirectionnelles. Nous nous restreignons donc aux arbres dont les arêtes sont semi-bidirectionnelles (i.e., non utilisables dans les deux sens en même temps). Dans un second temps,  nous décrivons un algorithme \emph{probabiliste} réalisant le nommage dans tout réseau (connexe) avec  arêtes  bidirectionnelles.

\subsection{Algorithme déterministe}

Soit $k$ robots placés de façon arbitraire dans un arbre. Nous supposons que les  tableaux blancs peuvent stocker  $\Omega(k (\log k +\log \Delta))$ bits, où $\Delta$ est le degré maximum du graphe. Chaque robot $\rob$ a un identifiant entier $\id_{\rob}\in[1,k]$. Cet entier peut être corrompu. Sans perte de généralité, on suppose que la corruption d'un identifiant préserve toutefois l'appartenance à $[1,k]$. 

Succinctement, l'algorithme fonctionne de la manière suivante. Chaque n{\oe}ud stocke dans son tableau blanc jusqu'à $k$ paires  (identifiant, numéro de port). L'écriture sur chaque tableau blanc se fait en ordre FIFO afin d'ordonner les écritures. Lorsqu'une paire est écrite dans un tableau contenant déjà $k$ paires, la plus ancienne paire est détruite. Chaque robot effectue un parcours Eulérien de l'arbre. Quand un robot $\rob$ arrive sur un \nd\/ $u$, il vérifie si son identifiant $\id_{\rob}$ est présent dans une des paires stockées sur le tableau blanc de $u$, noté $\tb_u$. Si cet identifiant n'est pas présent, alors le robot inscrit la paire $(\id_{\rob},p)$ sur $\tb_u$ où $p$ est le numéro de port par lequel $\rob$ partira pour continuer son parcours Eulérien. Si l'identifiant de $\rob$ est  présent dans une paire sur $\tb_u$, deux cas doivent  être considérés. S'il y a déjà un robot $\rob'$ localisé en $u$ avec le même identifiant que $\rob$, alors $\rob$ est activé et prend un nouvel identifiant, le plus petit identifiant non présent sur le tableau. Le robot $\rob$ continue ensuite son parcours Eulérien, en notant la paire $(\id_{\rob},p)$ convenable sur $\tb_u$. Enfin, si $\rob$ est le seul robot sur $u$ avec identifiant $\id_{\rob}$, il teste si la dernière arête $e$ associée  à son identifiant est  l'arête par laquelle il est entré sur $u$. Si oui, alors cela est cohérent avec un parcours Eulérien et $\rob$ continue ce parcours.  Si non, alors $\rob$ sort de $u$ par l'arête $e$ afin de rencontrer le robot portant le même identifiant que lui, s'il existe, et provoquer ainsi le changement d'identifiant de l'un des deux.

Afin de prouver la correction de l'algorithme, il convient de noter deux  remarques importantes. D'une part, puisque le réseau est un arbre et que les arêtes sont semi-bidirectionnelles, deux robots possédant le même identifiant se retrouveront sur un même \nd\/ en un nombre fini d'étapes. D'autre part, l'espace mémoire de chaque tableau étant borné, est l'écriture étant FIFO, si un tableau possède des informations erronées, celles-ci finiront par disparaitre en étant recouvertes par des informations correctes.

Pour ce qui est de la complexité de l'algorithme, on peut montrer que l'algorithme converge en $O(kn)$ rondes. Ce temps de convergence découle du fait que deux robots portant le même identifiant mettrons dans le pire des cas $O(n)$ rondes pour se rencontrer.

\subsection{Algorithme probabiliste}

L'approche probabiliste permet de considérer le cadre général de graphes arbitraires avec liens bidirectionnels. De fait, il est très simple d'obtenir une solution, sans même utiliser de tableaux blancs sur les \nd s. Chaque robot se déplace suivant une marche aléatoire uniforme. Lorsque plusieurs robots se rencontrent,  ils s'ignorent  s'ils ont des identifiants différents. Deux robots ayant le même identifiant se rencontrant sur un n{\oe}ud choisissent chacun un nouvel identifiant de manière aléatoire uniforme entre 1 et $k$. 

Pour prouver la convergence, nous considérons le cas d'une configuration initiale dans laquelle deux robots ont le même identifiant. Il est connu~\cite{TetaliW91} que la marche aléatoire non biaisée implique que  les deux robots se rencontreront en au plus $O(n^3)$ rondes. Lorsque deux robots ayant le même identifiant se rencontrent, les robots ont chacun une probabilité au moins $\frac{1}{k}$ de choisir un identifiant utilisé par aucun autre robot. L'algorithme probabiliste donc a un temps de stabilisation de $O(kn^3)$.

\section{Perspectives}

Les travaux présentés dans ce chapitre sont les premiers à considérer la conception d'algorithmes auto-stabilisants pour des robots susceptibles de subir des pannes internes, en plus des pannes externes usuellement traitées dans la littérature. Nous avons montré qu'il était possible sous ces hypothèses de réaliser le nommage et l'élection dans les arbres en déterministe, et dans tous les graphes en probabiliste. 

La restriction aux arbres, dans le cas déterministe, est motivée par les symétries pouvant être induites par la présence de cycles.
Une piste de recherche évidente consiste donc à considérer le cas des graphes avec cycles mais suffisamment \og asymétriques\fg\/ pour permettre le nommage.  

Par ailleurs, le nommage et l'élection ne sont intéressants qu'en tant que briques élémentaires pour la réalisation de tâches plus élaborées, comme le rendez-vous ou la recherche d'intrus. Concevoir des algorithmes basés sur ces briques élémentaires demande de composer avec soin des algorithmes auto-stabilisants pour des robots. Si la composition d'algorithmes auto-stabilisants pour les réseaux est maintenant bien comprise, il n'en va pas nécessairement de même dans le cadre de l'algorithmique pour entités mobiles. L'étude de la composition d'algorithmes auto-stabilisants pour entités mobiles est à ma connaissance un problème ouvert.

\chapter{Auto-organisation dans un modèle à vision globale}
\label{chap:perpetuelle}

Le chapitre précédent était consacré à un modèle dans lequel les robots n'avaient qu'une vision \og locale\fg\/ du système (chacun n'a accès qu'aux informations disponibles sur le n{\oe}ud sur lequel il se trouve). Ce chapitre est consacré  à un modèle dans lequel les robots ont une vision  \og globale\fg\/  du système. Le chapitre considère le modèle CORDA (asynchronisme)  dans un modèle discret (réseau). Il a pour objectif d'identifier les hypothèses minimales nécessaires  à la réalisation de tâches complexes de façon auto-stabilisante. Mes contributions dans ce cadre ont été réalisées en collaboration de A.~Milani, M.~Potop-Butucaru et de S.~Tixeuil~\cite{BlinMPT10}. 

La première section de ce chapitre est consacrée à un bref état de l'art des algorithmes conçus pour le modèle CORDA discret. Un modèle minimale est ensuite formalisé, dans la section suivante. La tâche algorithmique utilisée pour la compréhension de ce modèle est  l'\emph{exploration perpétuelle}, définie comme suit. La position initiale des robots est arbitraire --- elle peut résulter par exemple d'une faute du système. Partant de cette position initiale, les robots doivent agir de façon à ce que chaque n{\oe}ud du réseau soit  visité infiniment souvent par chacun des robots. La  section~\ref{sec:impoexplo} est ainsi consacrée à établir des bornes inférieures et supérieures sur le nombre de robots pouvant réaliser l'exploration perpétuelle dans l'anneau. La section~\ref{sec:exploperpet} résume quant à elle une de nos contributions principales, à savoir deux algorithmes d'exploration utilisant respectivement un nombre minimal et maximal de robots. La dernière section liste quelques perspectives sur le modèle CORDA et sur le problème de l'exploration perpétuelle. 

La contribution de ce chapitre à l'auto-organisation d'un système de robots est donc double, 
\begin{itemize}
\item d'une part, la mise en évidence d'un modèle \og minimaliste \fg\/ n'ajoutant aucune hypothèse non inhérente à l'esprit d'un modèle observation-calcul-déplacement tel que le modèle CORDA; 
\item d'autre part, la conception d'algorithmes pour les robots créant et maintenant  des asymétries entre les positions de ces robots permettant de donner une \og direction \fg\/ à l'exploration en l'absence de références extérieurs (numéro de ports, identifiant des n{\oe}uds, sens de direction, etc.). 
\end{itemize}

\section{Etat de l'art des algorithmes dans le modèle CORDA discret}

Une grande partie de la littérature sur l'algorithmique dédiée à la coordination de robots distribués considère que les robots évoluent dans un espace euclidien continu à deux dimensions. Les sujets essentiellement traités dans ce domaine sont la formation de patterns (cercle, carré, etc.), le regroupement, l'éparpillement, le rendez-vous, etc. Nous renvoyons à~\cite{SuzukiY99,CieliebakFPS03,AsahiroFSY08,DefagoS08,FlocchiniPSW08,YamashitaS10} pour des exemples de résolutions de tels problèmes. Dans cadre continu, le modèle suppose que les robots utilisent des capteurs visuels possédant une parfaite précision qui permettent ainsi de localiser la position des autres robots. Les robots sont également supposés capables de se déplacer. Les déplacements sont souvent supposés idéaux, c'est-à-dire sans déviation par rapport à une destination fixée. 

Ce modèle est critiquable car la technologie actuelle permet difficilement de le mettre en oeuvre. Les capteurs visuels et les déplacements des robots sont en effet loin d'être parfaits. La tendance ces dernières années a donc été de déplacer le cadre d'étude du modèle  continu au modèle discret. 

Dans le modèle discret, l'espace est divisé en un nombre fini d'\emph{emplacements} modélisant une pièce, une zone de couverture d'une antenne, un accès à un bâtiment, etc. La structure de l'espace est idéalement représentée par un graphe où les n{\oe}uds du graphe représentent les emplacements qui peuvent être détectés par les robots, et où les arêtes représentent la possibilité pour un robot de se déplacer d'un emplacement à un autre. Ainsi, le modèle discret facilite à la fois le problème de la détection et du déplacement.  Au lieu de devoir détecter  la position exacte d'un autre robot, il est en effet plus aisé pour un robot de détecter si un emplacement est vide ou s'il contient un ou plusieurs autres robots. En outre, un robot peut plus facilement rejoindre un emplacement que rejoindre des coordonnées géographiques exactes. Le modèle discret permet également de simplifier les algorithmes en raisonnant sur des structures finies (\emph{i.e}, des graphes) plutôt que sur des structures infinies (le plan ou l'espace 3D). Il y a cependant un prix à payer à cette simplification. Comme l'ont noté la plupart des articles relatifs à l'algorithmique pour les entités mobiles dans le modèle discret~\cite{KlasingMP08, KlasingKN08, FlocchiniIPS07, FlocchiniIPS08, DevismesPT09}, le modèle discret est livrée avec le coût de la \emph{symétrie} qui n'existe pas dans le modèle continu où les robots possèdent leur propre système de localisation (e.g., GPS et boussole). Ce chapitre est consacré à l'étude du modèle discret, et en particulier à la conception d'algorithmes auto-stabilisant s'exécutant correctement en dépit des symétries potentielles. 

Un des premiers modèles pour l'étude de l'auto-organisation de robots distribués, appellé \emph{SYm},  a été proposé par Suzuki et Yamashita~\cite{SuzukiY06}. Dans le modèle SYm, les robots n'ont pas d'état (ils n'ont donc pas de mémoire du passé), et fonctionne par cycle élémentaire synchrone. Un cycle élémentaire est constitué des trois actions atomiques synchrones suivantes: observation, calcul, déplacement  (\og Look-Compute-Move\fg\/ en anglais). Autrement dit, à chaque cycle, chaque robot observe tout d'abord les positions des autres robots (les robots sont supposés capables de \og voir\fg\/ les positions de tous les autres robots), puis il calcule le déplacement qu'il doit effectuer, et enfin il se déplace selon ce déplacement.  Le modèle \emph{CORDA}~\cite{FlocchiniPSW99,Prencipe01} peut se définir comme la variante asynchrone du modèle SYm. Il reprend toutes les hypothèses du modèle SYm mais il suppose que les robots sont asynchrones. La littérature traitant de l'auto-organisation de robots dans un modèle du type \og observation-calcul-déplacement \fg\/  utilise presque systématiquement une variante de SYm ou de CORDA. Le modèle peut en particulier supposer la présence ou non d'identifiants distincts affectés aux robots, la capacité de stockage d'information par les robots et/ou les n{\oe}uds du réseau dans lequel ils se déplacent, la présence ou non d'un sens de la direction attribué au robots, la possibilité ou non d'empiler des robots, etc. Ce sont autant d'hypothèses qui influent sur la capacité d'observer, de calculer et de se déplacer. 

A ma connaissance, tous les articles traitant du modèle CORDA discret enrichissent ce modèle en supposant des hypothèses supplémentaires, comme par exemple la présence d'identifiants distincts, ou d'un sens de direction.

Dans les modèles SYm et CORDA, l'ensemble des positions des robots détermine une \emph{configuration} du système. Une  \emph{photographie instantanée} (\og snapshot \fg\/ en anglais), ou simplement \emph{photo}, du système à l'instant $t$ par un robot est la configuration du système à cet instant $t$.  La phase d'observation exécuté par un robot consiste à prendre une \emph{photo} de la position des autres robots. Durant sa phase de calcul, chaque robot calcule son action en fonction de la dernière photo prise. Enfin, dans sa phase déplacement, chaque robot se déplace en traversant \emph{une seule} arête incidente au n{\oe}ud courant. Cette arête est déterminée durant la phase de calcul. En présence de numéro de ports, ou d'un sens de direction, établir la correspondance entre une arête identifiée sur la photo et l'arête réelle à emprunter est direct. Nous verrons dans la section suivante que l'absence de numéro de port et de sens de direction n'empêche pas d'établir cette correspondance. 

Dans le cadre du modèle CORDA discret, les deux problèmes les plus étudiés sont sans doute le \emph{regroupement}~\cite{KlasingMP08,KlasingKN08} et l'\emph{exploration}, dans ses versions avec arrêt~\cite{FlocchiniIPS07,FlocchiniIPS08,DevismesPT09,ChalopinFMS10,FlocchiniIPS10} et perpétuelle~\cite{BaldoniBMR08}.
La tâche de regroupement demande aux robots de se réunir en un n{\oe}ud du réseau~; la tâche d'exploration demande aux robots de visiter chaque n{\oe}ud du réseau. Ce chapitre se focalise uniquement sur la tâche d'exploration. 

Dans l'exploration avec arrêt, le fait que les robots doivent s'arrêter après avoir exploré tous les n{\oe}uds du réseau requiert de leur part de se \og souvenir \fg\/ quelle partie du réseau a été explorée. Les robots doivent donc être capable de distinguer les différentes étapes de l'exploration (n{\oe}ud exploré ou pas, arête traversée ou pas, etc.) bien qu'ils n'aient pas de mémoire persistante. La symétrie des configurations est le principal problème rencontré dans le modèle discret. C'est pourquoi la plupart des articles du domaine se sont dans un premier temps restreints à l'étude de réseaux particuliers tels que les arbres~\cite{FlocchiniIPS10} et les cycles~\cite{FlocchiniIPS07,KlasingMP08,KlasingKN08,DevismesPT09,FlocchiniIPS10}. Dans~\cite{ChalopinFMS10}, les auteurs considèrent les réseaux quelconques mais supposent que les positions initiales des robots sont asymétriques. 

Une technique classique pour éviter la présence de symétrie est d'utiliser un grand nombre de robots pour créer des groupes de robots de taille différentes et donc asymétriques.  Une mesure de complexité souvent considérée est donc le nombre minimum de robots requis pour explorer un réseau donné. Pour les arbres à $n$ \nd s, $\Omega(n)$ robots sont nécessaires~\cite{FlocchiniIPS08} pour l'exploration avec arrêt, même si le degré maximum est $4$. En revanche, pour les arbres de degré maximum~3, un nombre de robots exponentiellement plus faible, $O(\log n/\log\log n)$, est suffisant. Dans un cycle de $n$ n{\oe}uds, l'exploration avec arrêt par $k$ robots est infaisable si $k|n$, mais faisable si $\mbox{pgcd}(n,k)=1$ avec $k\geq 17$~\cite{FlocchiniIPS07}. En conséquence, le nombre de robots nécessaire et suffisant pour cette tâche est $\Theta(\log n)$ dans le cycle. Enfin, dans \cite{ChalopinFMS10}, les auteurs proposent un algorithme d'exploration avec arrêt  dans un réseau quelconque avec des arêtes étiquetées par des numéro de ports, pour un nombre impair de robots $k\geq 4$. 
 
Dans \cite{BaldoniBMR08}, les auteurs résolvent le problème de l'exploration perpétuelle dans une grille partielle anonyme (c'est-à-dire une grille anonyme à laquelle un ensemble de n{\oe}uds et d'arêtes ont été supprimés). Cet article introduit la contrainte d'\emph{excusivité} mentionnée précédemment, qui stipule qu'au plus un robot  peut occuper le même \nd, ou traverser la même arête. Un certain nombre de travaux utilisent des \og tours \fg\/ de robots pour casser la symétrie (voir \cite{FlocchiniIPS07}). La contrainte d'exclusivité interdit ce type de stratégies. Nos contributions personnelles considèrent également la contrainte d'exclusivité mais, contrairement à~\cite{BaldoniBMR08},  les robots n'ont pas de sens de la direction.

\begin{figure}[tb]
\centering
\subfigure[$(\rob_2,\free_2,\rob_1,\free_z)$]{
\includegraphics[width=.22\textwidth]{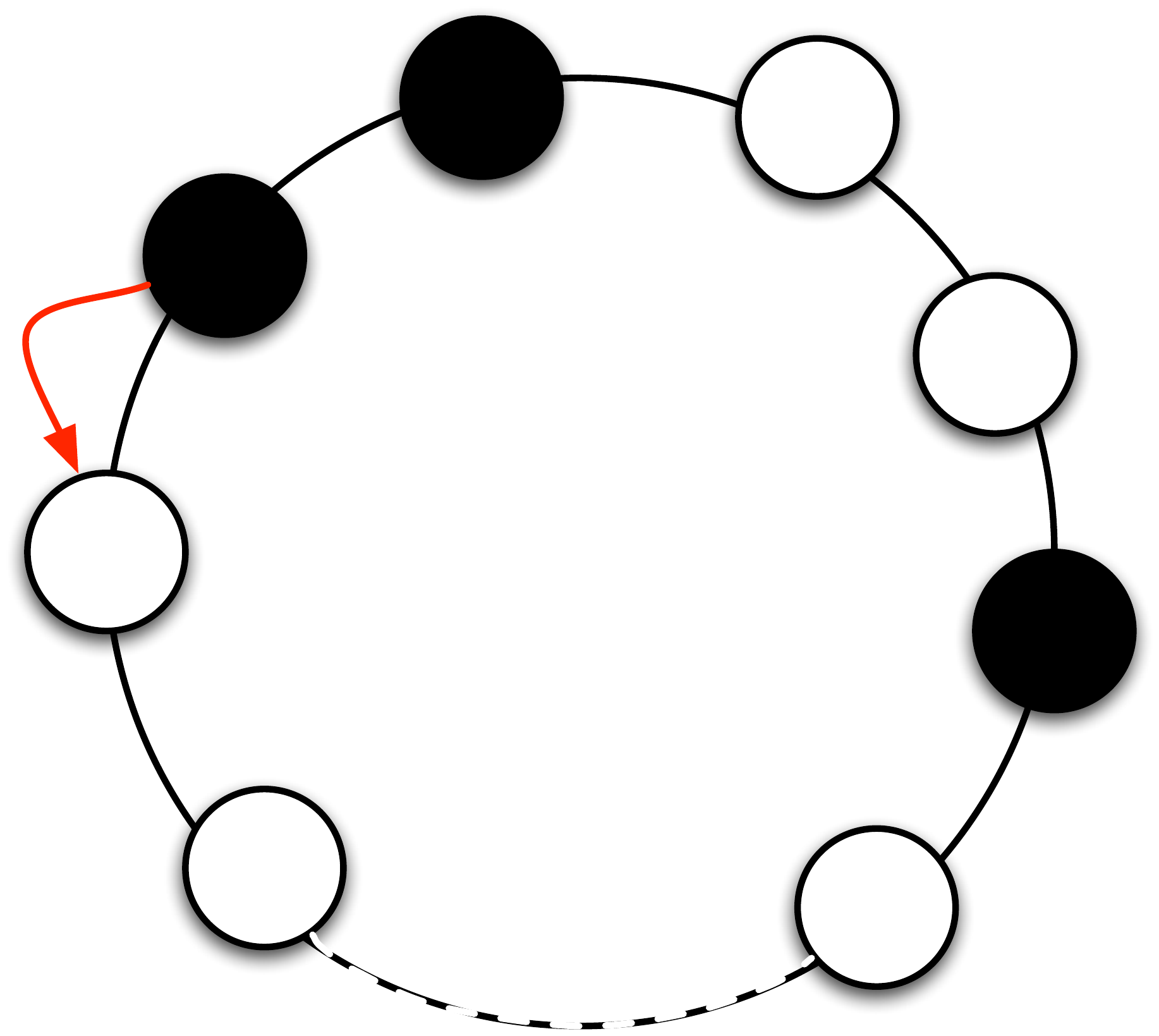}
\label{fig:AlgoA}
}
\subfigure[$(\rob_1,\free_1,\rob_1,\free_2,\rob_1,\free_{z-1})$]{
\includegraphics[width=.22\textwidth]{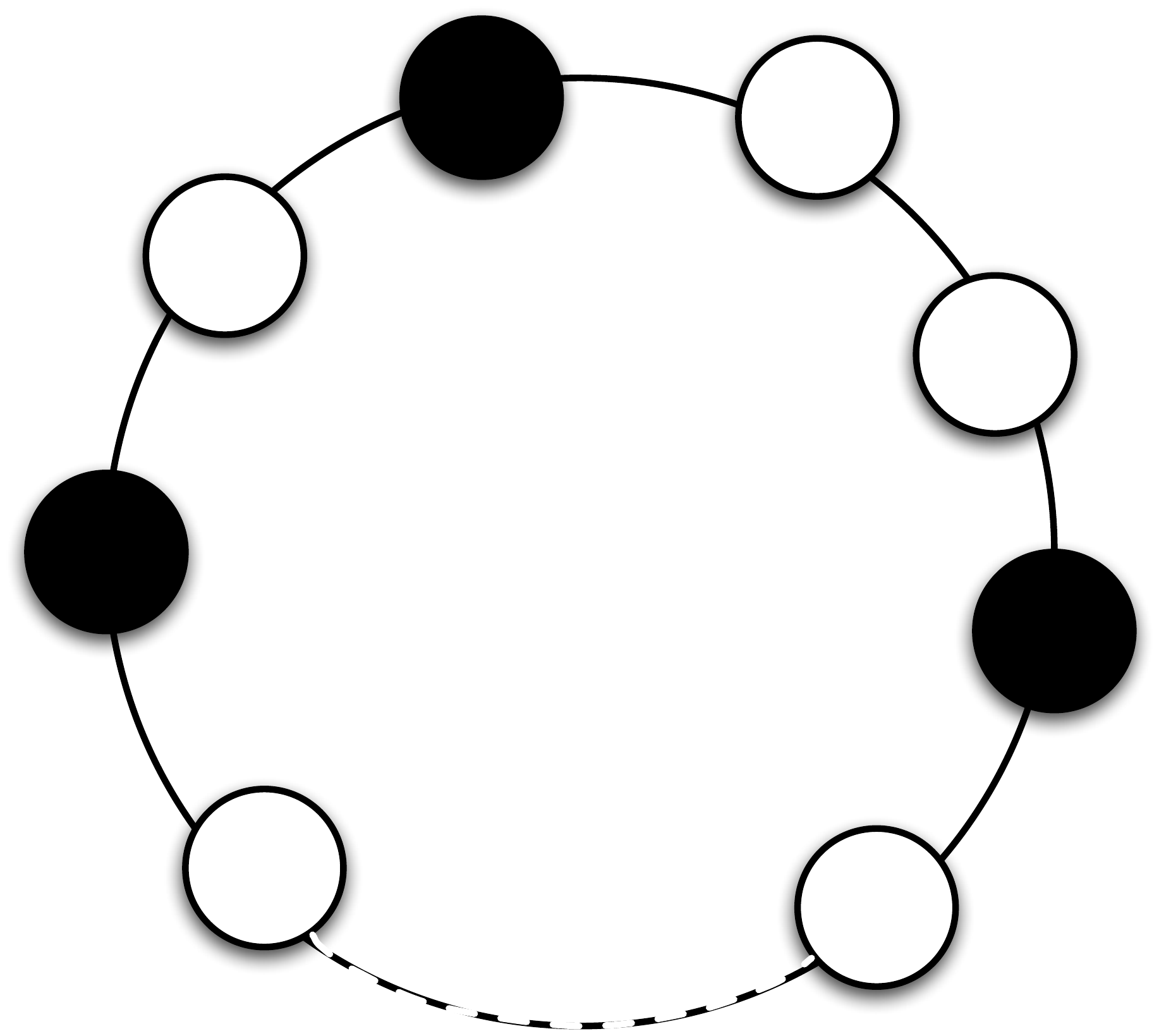}
\label{fig:AlgoB}
}
\caption{Photos}
\vspace*{-0,2cm}
\end{figure}

\section{Un modèle global minimaliste pour un système de robots}
Ce chapitre se focalise sur le modèle CORDA dans sa version élémentaire, sans hypothèse supplémentaire. Les robots n'ont donc pas d'état, sont totalement asynchrones, ont une vision globale du système (graphe et robots), et sont capable de calculer et de déplacer. En revanche, il ne sont munis d'aucune information supplémentaire. En particulier, ils sont anonymes, ne possèdent pas de moyen de communication direct, et n'ont pas de sens de direction (ils ne peuvent donc pas distinguer  la droite de la gauche, ou le nord du sud). Egalement, les n{\oe}uds sont anonymes, et les arêtes ne possèdent pas de numéro de port. En terme de déplacement, il ne peut y avoir qu'au plus un  robot par n{\oe}ud, et une arête ne peut être traversée que par un seul robot à la fois (pas de croisement, i.e., arête semi-bidirectionnelle). Un algorithme ne respectant ces dernières spécifications relatives au nombre de robots par n{\oe}uds et aux croisements le long des arêtes entraine une \emph{collision}, et sera considéré incorrecte. Dans de telles conditions minimalistes, les robots ne collaborent que par l'intermédiaire de leurs positions qui dictent leurs actions à chacun. 
Spécifions précisément le modèle CORDA discret pour un anneau et pour une chaîne, selon les principes établis dans~\cite{BlinMPT10}. Les robots sont \emph{asynchrones}, \emph{anonymes}, \emph{silencieux} (ils ne possèdent pas de moyen de communication direct), sont \emph{sans état} (pas de mémoire du passé, \og oblivious \fg\/ en anglais), et n'ont \emph{aucun sens de la direction}. L'asynchronisme est modélisé par un adversaire qui décide quel robot, ou quel sous-ensemble de robots, est activé parmi les robots activables par l'algorithme. Les robots sont soumis à la contrainte d'exclusivité. 

Afin de simplifier notre propos, nous avons besoin de formaliser la notion de photo présentée précédemment. Une \emph{photo} $\C$  (pour \og snapshot\fg) impliquant $k$ robots dans un anneau à $n$ \nd s,  est une séquence non orientée (circulaire dans le cas du cycle) de symboles $\rob$ et $\free$ indexés par des entiers: $\rob_i$ signifie que $i$ \nd s consécutifs sont occupés par des robots, et $\free_j$ signifie que $j$ \nd s consécutifs sont non occupés. Par exemple, la photo $\C=(\rob_{i_1},\free_{j_1},\dots,\rob_{i_\ell},\free_{j_\ell})$ décrit le cas où $k$ robots sont divisés en $\ell$ groupes, et, pour $m=1,\dots,\ell$ le $m$-ème groupe de robots occupe $i_m$ \nd s consécutifs dans l'anneau, et les $m$-ème et $(m+1)$-ème groupes de robots sont séparés par $j_m\geq 1$ \nd s libres.  
Le résultat d'une observation par un robot $\rob$ est une photo $\C=(\rob_{i_1},\free_{j_1},\dots,\rob_{i_\ell},\free_{j_\ell})$. Le calcul effectué par ce robot résulte en une autre photo $\C'$ à atteindre par un unique déplacement effectué par $\rob$ ou par un autre robot. Un algorithme sera donc défini par un ensemble de transitions entre photos $\C\rightarrow \C'$ spécifiant la configuration $\C'$ image de $\C$, pour chaque $\C$. Par exemple, la transition \vspace*{-0,2cm} $$(\rob_2,\free_2,\rob_1,\free_{n-5})\rightarrow (\rob_1,\free_1,\rob_1,\free_2,\rob_1,\free_{n-6})$$ stipule que le robot du groupe de deux robots à côté de $n-5$ n{\oe}uds libres (voir Figure~\ref{fig:AlgoA}) doit se déplacer d'un cran vers ces $n-5$ n{\oe}uds libres (voir Figure~\ref{fig:AlgoB}). 

\section{Résultat d'impossibilités}
\label{sec:impoexplo}

Les résultats d'impossibilités que nous avons obtenus dans~\cite{BlinMPT10} sont résumés dans cette section. Ils résultent de la mise en évidence de cas de symétrie qu'il est impossible de briser dans le modèle CORDA discret minimaliste que nous utilisons. Soit \rob\/ un robot sur une chaîne, et soit $u$ un \nd\/ à l'extrémité de cette chaîne. Soit $v$ le n\oe ud voisin de $u$. Tout   algorithme d'exploration doit spécifier à \rob\/ localisé sur  $v$ d'aller en $u$ car sinon $u$ ne serait jamais exploré. Il en résulte que si \rob\/ est initialement placé en $v$ alors le fait que le robot soit sans état implique que l'adversaire pourra systématiquement ordonner le déplacement de \rob\/ de $v$ vers $u$ et de $u$ vers $v$, indéfiniment, et le reste de la chaîne ne sera jamais explorée. (Notons que cette impossibilité n'est pas vérifiée dans le cas où le robot aurait un sens de direction, ou dans le cas de l'existence de numéros de port. En effet, dans les deux cas, la fonction de transition est de la forme  $(\C,d) \rightarrow  (\C',d')$ où $d$ et $d'$ sont soit des numéros de port, soit des directions.) L'exploration perpétuelle d'une chaîne par $k>1$ robots est impossible, simplement parce que les robots n'ont aucun moyen de se croiser du fait de la contrainte d'exclusivité. 

\begin{figure}[tb]
\centering
\subfigure[$(\rob_1,\free_z)$]{
\includegraphics[width=.18\textwidth]{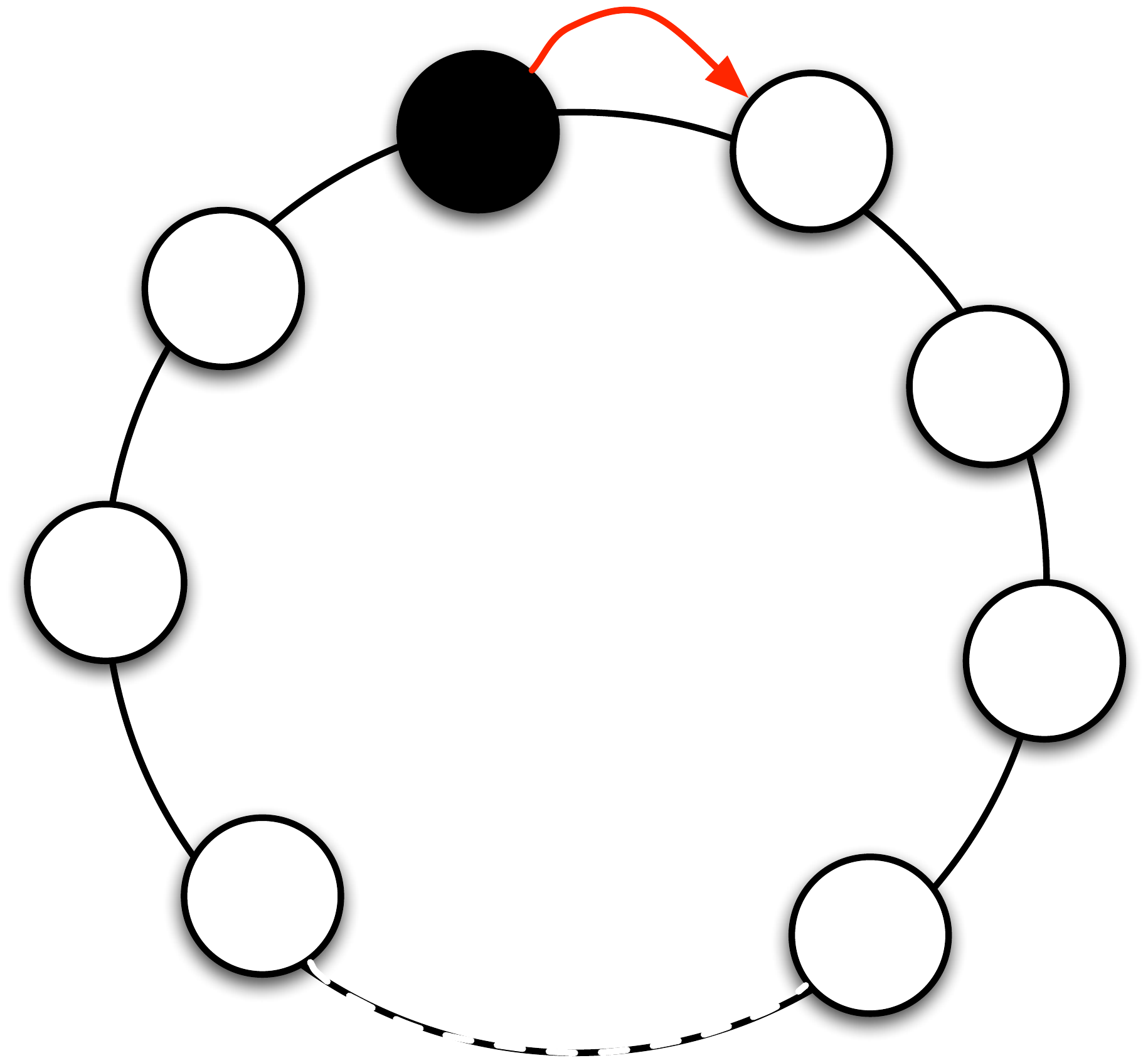}
\label{fig1-1}
}
\subfigure[$(\rob_1,\free_z)$]{
\includegraphics[width=.18\textwidth]{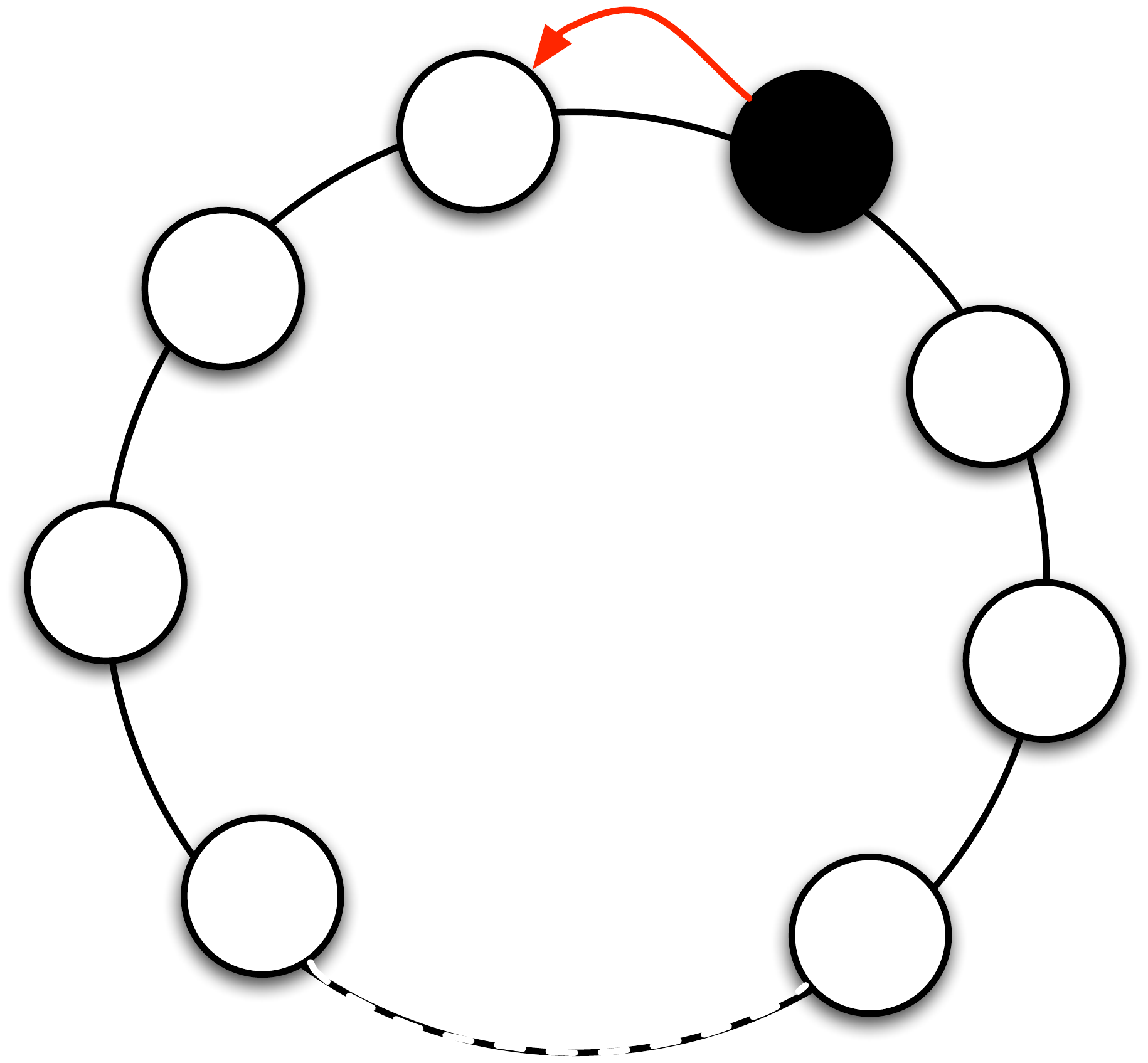}
\label{fig1-2}
}
\subfigure[]{
\includegraphics[width=.16\textwidth]{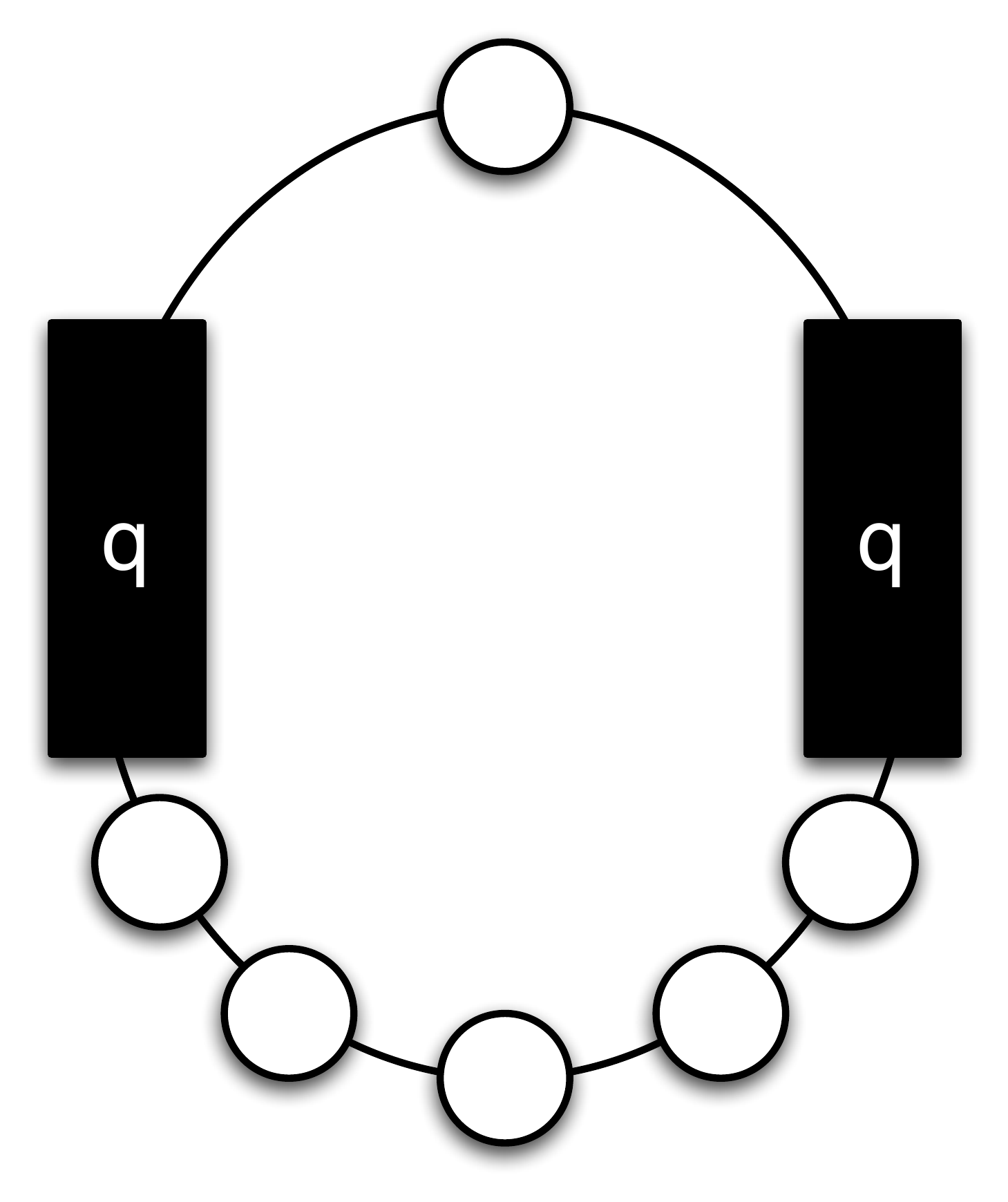}
\label{fig:even1}
}
\subfigure[]{
\includegraphics[width=.16\textwidth]{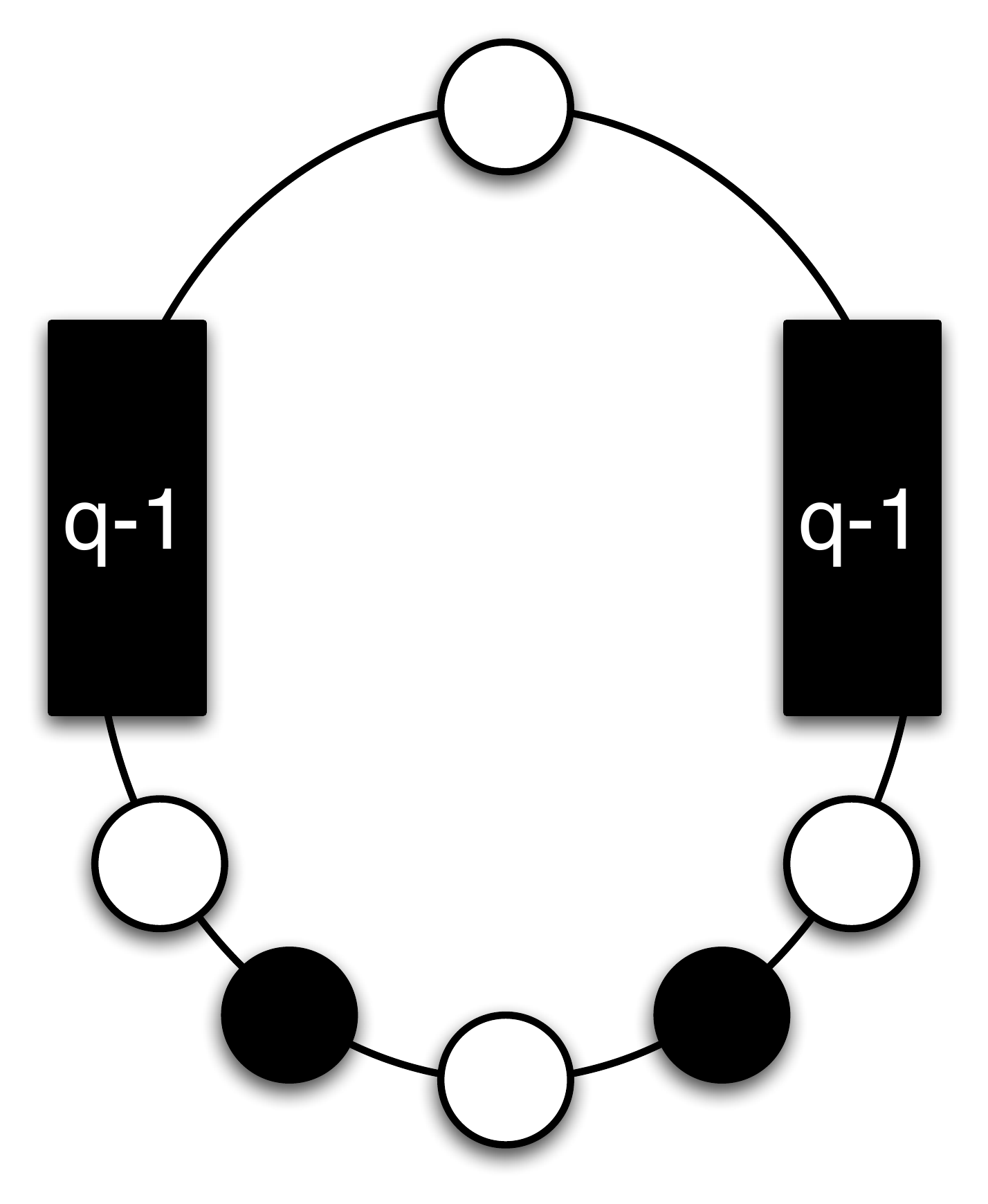}
\label{fig:even2}
}
\subfigure[]{
\includegraphics[width=.16\textwidth]{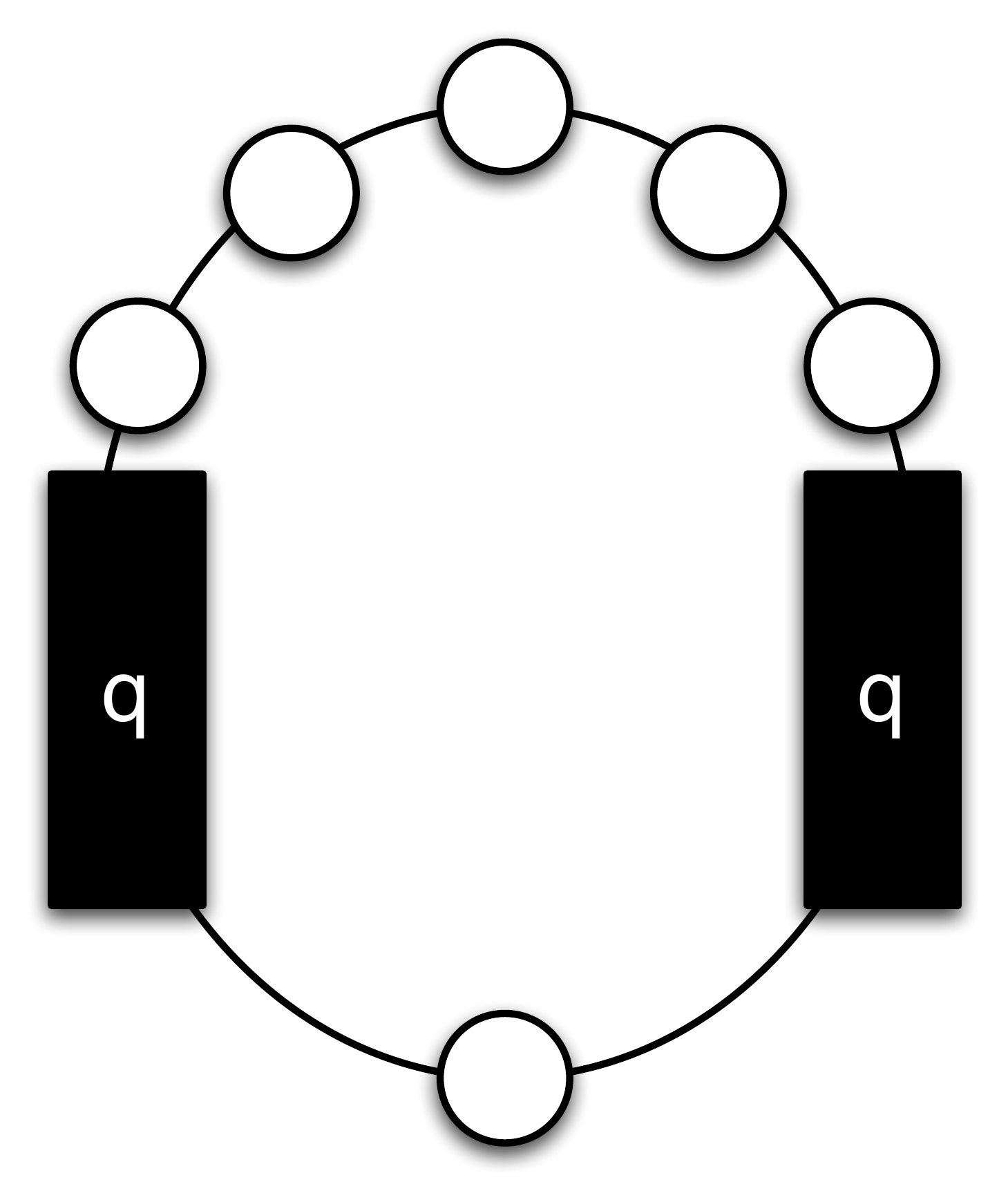}
\label{fig:even3}
}
\caption{Cas d'impossibilité avec un robot, ou avec un nombre pair de robots}
\end{figure}

Partant du fait que l'exploration perpétuelle est impossible dans une chaîne, il est naturel des s'intéresser  par la suite à l'exploration perpétuelle dans un anneau et aux cas d'impossibilité dans cette topologie. L'impossibilité d'explorer l'anneau avec un unique robot est illustrée sur les figures~\ref{fig1-1} et \ref{fig1-2}. L'impossibilité d'explorer l'anneau avec un nombre pair de robots est illustrée sur les figures~\ref{fig:even1}-\ref{fig:even3}. De même, il est simple de montrer qu'explorer l'anneau de $n$ n{\oe}uds avec $k$ robots, $n-4\leq k\leq n$, est impossible. Nous avons montré dans~\cite{BlinMPT10} qu'en revanche, $k=3$ et $k=n-5$ sont des valeurs universelles, c'est-à-dire que pour $n$ assez grand, et non multiple de $k$, l'exploration perpétuelle de l'anneau à $n$ n{\oe}uds est possible avec $k$ robots, quelle que soit la configuration initiale. 

\begin{figure}[tb!]
\centering
\subfigure[font=scriptsize][Pellicule $G_{7,3}$ pour 7 n{\oe}uds et 3 robots]{
\includegraphics[width=.45\textwidth]{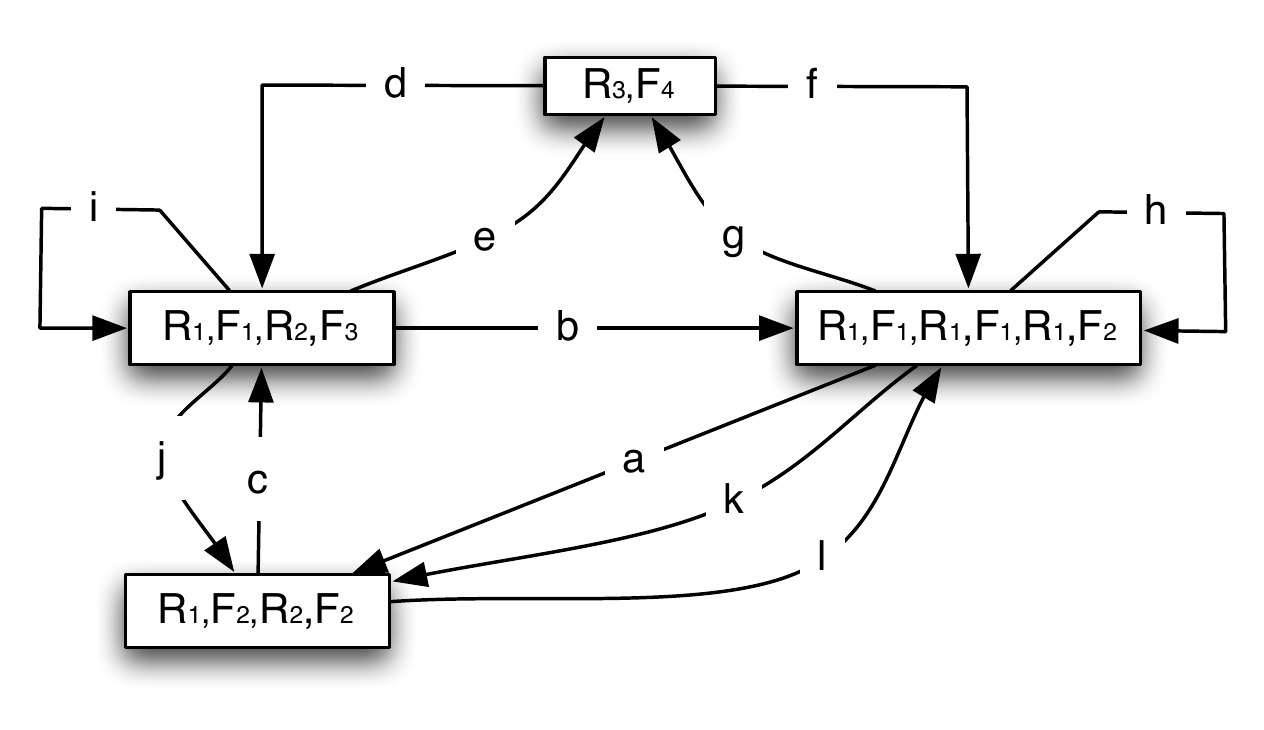}
\label{graphe71}
}
\subfigure[font=scriptsize][Réduction de la pellicule $G_{7,3}$]{
\includegraphics[width=.45\textwidth]{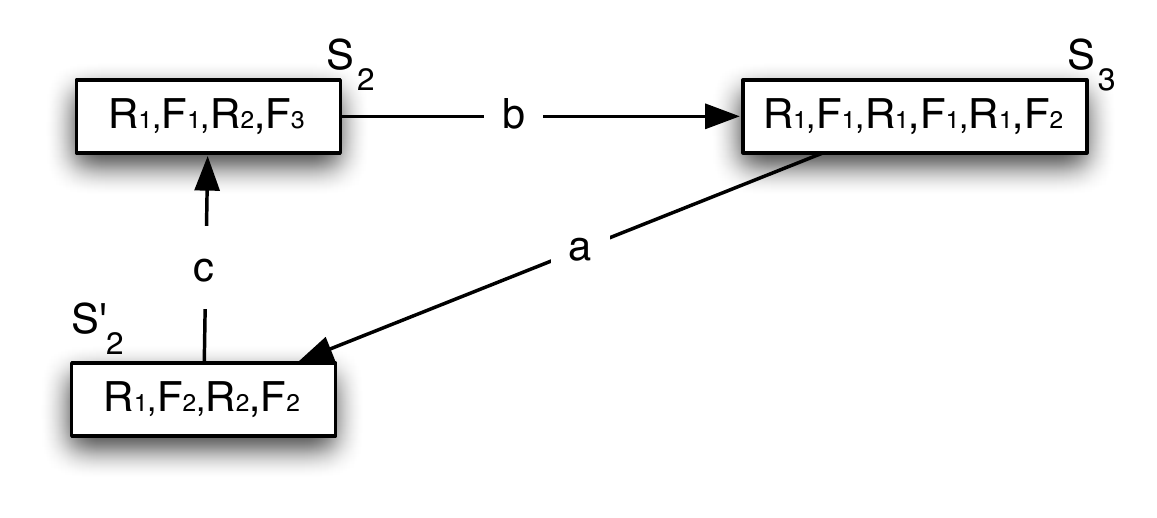}
\label{graphe72}
}
\caption{Pellicules et réductions}
\end{figure}

\begin{figure}[t!]
\centering
\includegraphics[width=.5
\textwidth]{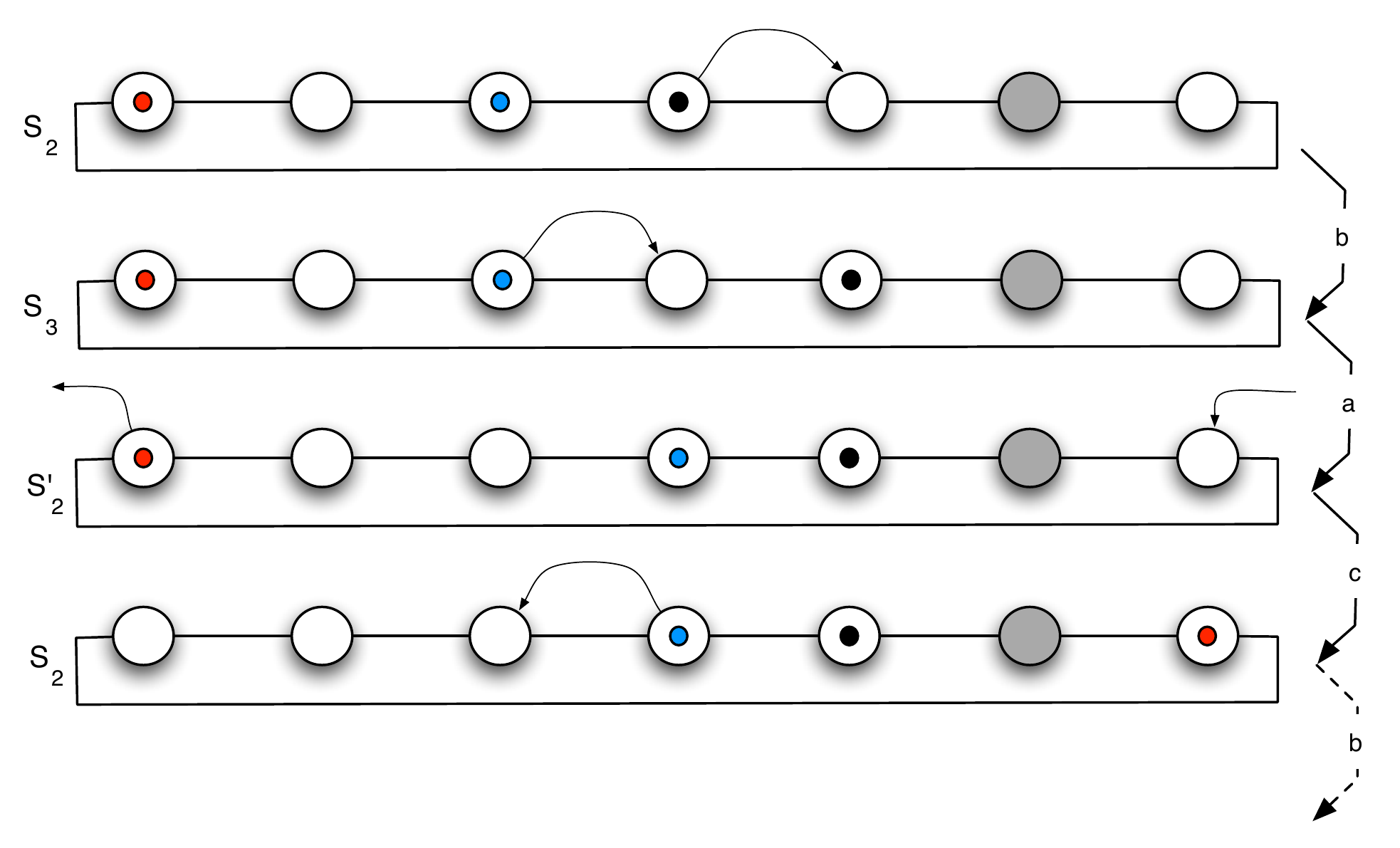}
\caption{Dans la pellicule $G_{7,3}$, le n{\oe}ud gris n'est jamais visité}
\label{graphe73}
\end{figure}

Déterminer la valeur minimale de $n$ pour laquelle 3 robots peuvent effectuer l'exploration perpétuelle de l'anneau à $n$ n{\oe}uds requiert une étude de cas spécifique. Nous montrons que cette  valeur est~10. En utilisant le résultat de  Flocchini et al.~\cite{FlocchiniIPS07} qui stipule en particulier qu'il est impossible d'explorer un anneau de $n$ n{\oe}uds avec un nombre de robots $k$ divisant $n$ nous pouvons diminuer le nombre de cas à traiter à $n=4, 5, 7, 8$. Nous avons introduit la notion de \emph{pellicule}. Une pellicule représente  toutes les photos possibles pour un nombre de n{\oe}uds et de robots fixés, ainsi que tous les mouvements possibles entre ces photos (voir la figure~\ref{graphe71} pour un anneau avec 7 n{\oe}uds et 3 robots). Nous avons prouvé qu'une pellicule  peut-être \emph{réduite} à un sous-ensemble de photos particulières en supprimant les photos qui sont trivialement un obstacle à l'exploration perpétuelle (voir la figure~\ref{graphe72} pour un anneau avec 7 n{\oe}uds et 3 robots). Par exemple, les photos permettant à l'adversaire de forcer le robot à effectuer un \og ping-pong \fg\/ perpétuel entre deux  n{\oe}uds peuvent être trivialement supprimées. Une fois cette réduction faite, il reste à analyser la sous-pellicule restante. Cela est effectué en simulant les déplacements induits par les photos de cette sous-pellicule.  Ainsi dans le cas d'un anneau à sept n{\oe}uds, on montre qu'il n'existe pas d'algorithme déterministe permettant la visite perpétuelle par trois robots (voir figure~\ref{graphe73}).

On procède de même dans le cas des anneaux à 4, 5, ou 8 n{\oe}uds pour 3~robots. Dans le cas de 4 ou 5 n{\oe}uds, la réduction résulte en une sous-pellicule vide. Dans le cas de 8 n{\oe}uds la réduction résulte en la même sous-pellicule que celle obtenue pour 7 n{\oe}uds. Il est donc impossible de faire l'exploration d'un anneau de moins de dix n{\oe}uds avec exactement trois robots. 

\section{Algorithme d'exploration perpétuelle}
\label{sec:exploperpet}

Dans cette section, nous montrons tout d'abord qu'il existe un algorithme d'exploration perpétuelle pour trois robots dans un anneau de $n$ n{\oe}uds, avec $n\geq10$ et $n$ différent d'un multiple de trois.  Nous décrivons ensuite un algorithme déterministe permettant de faire l'exploration perpétuelle avec $n-5$ robots dans un anneau de $n$ n{\oe}uds, où $n>10$ et $k$ est impair. 

\begin{figure}[t!]
\centering
\subfigure[Photo $\C^\circ_2$]{
\includegraphics[width=.18\textwidth]{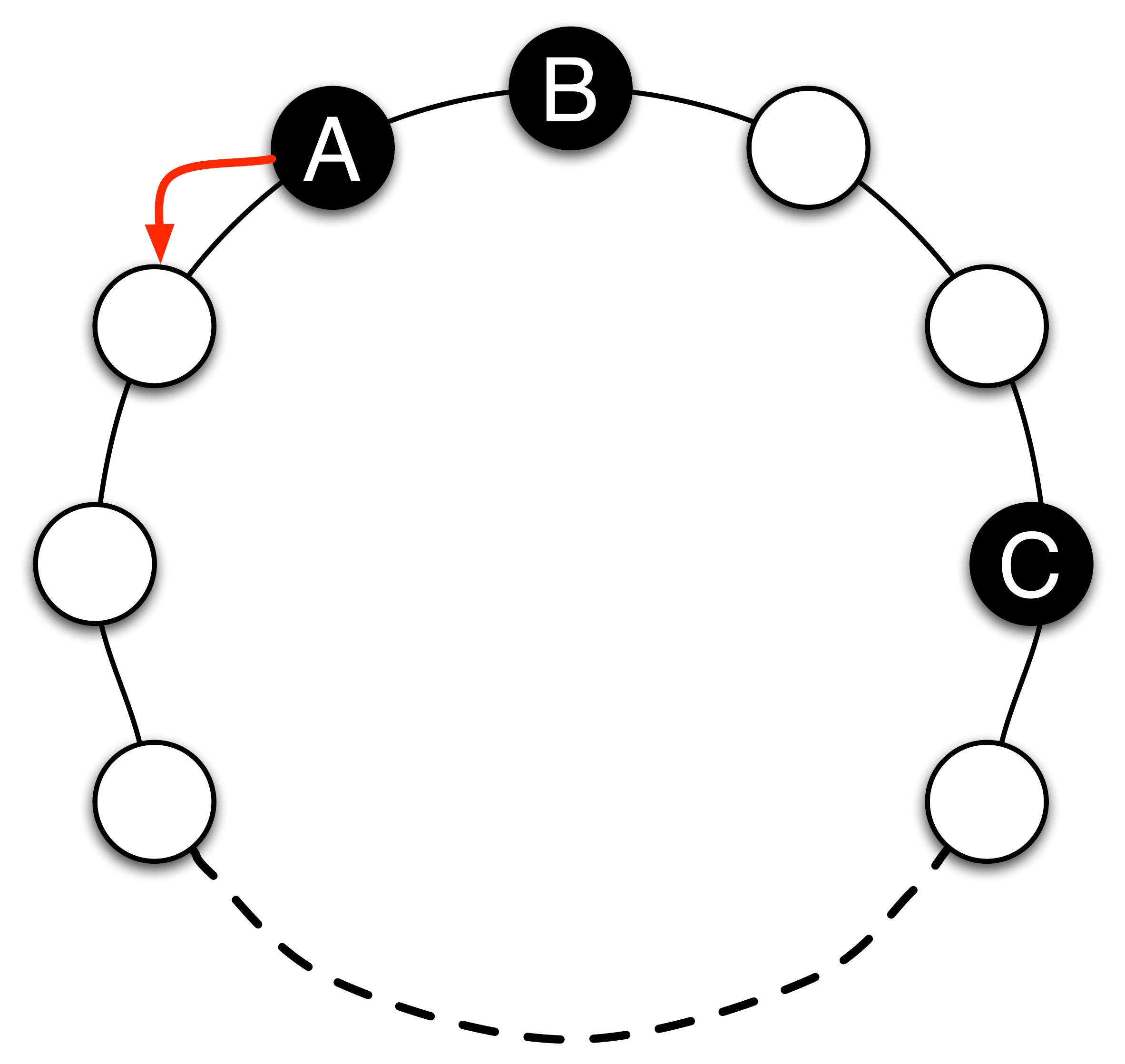}
\label{fig:AlgoMin1}
}
\subfigure[Photo $\C^\circ_3$]{
\includegraphics[width=.18\textwidth]{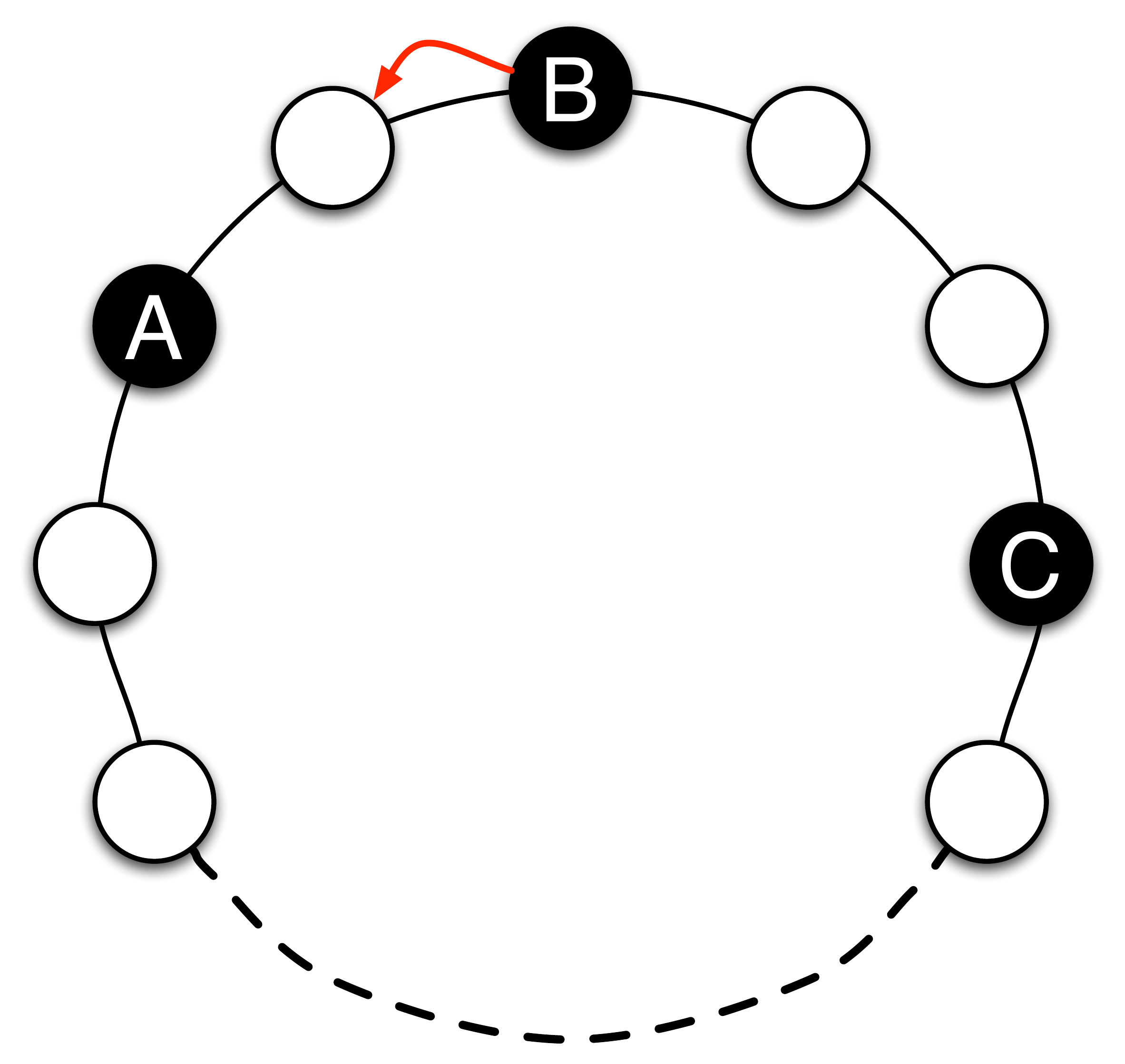}
\label{fig:AlgoMin2}
}
\subfigure[Photo $ \overline{\C^\circ_2}$]{
\includegraphics[width=.18\textwidth]{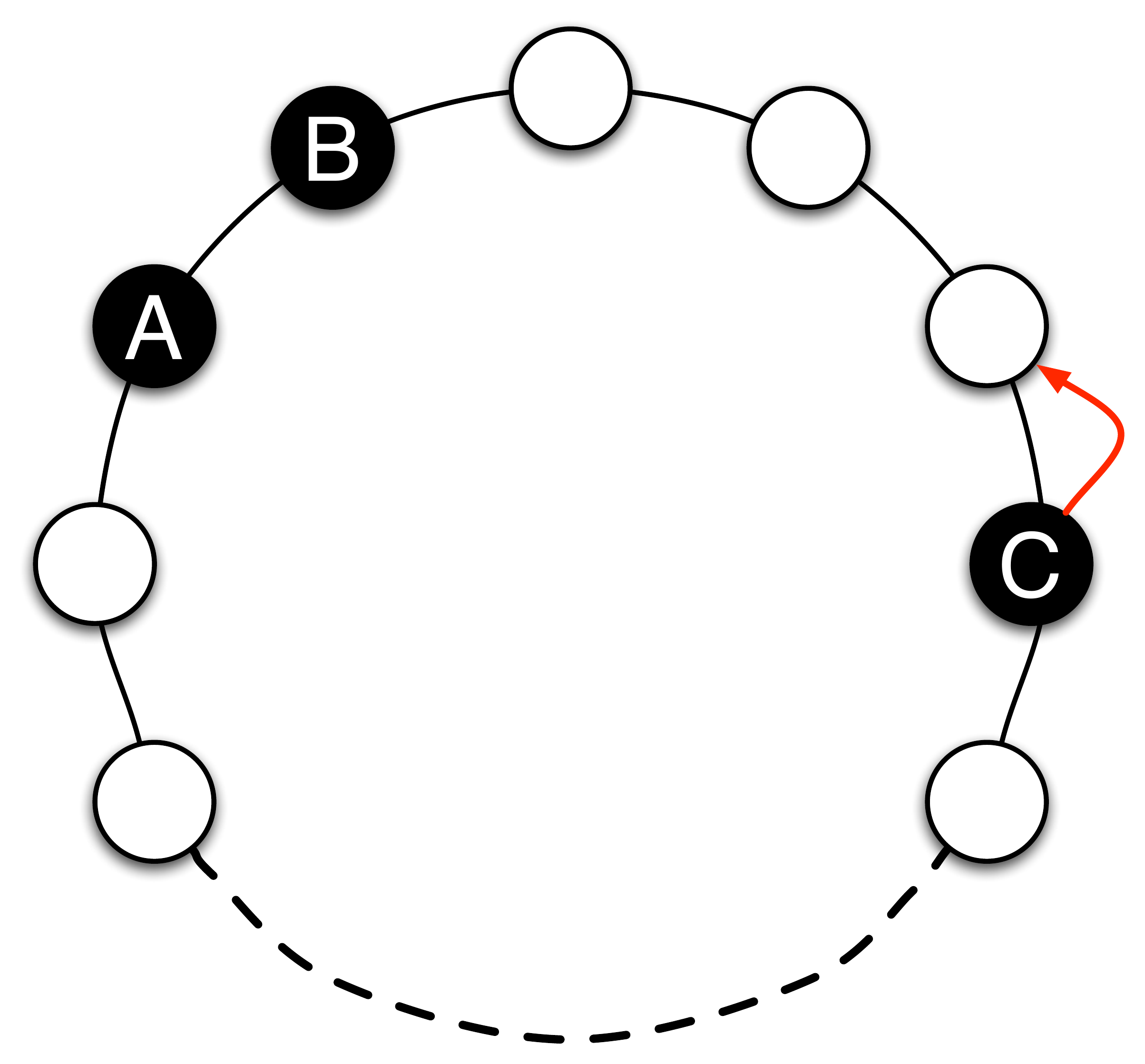}
\label{fig:AlgoMin3}
}
\subfigure[Photo $\C^\circ_2$]{
\includegraphics[width=.18\textwidth]{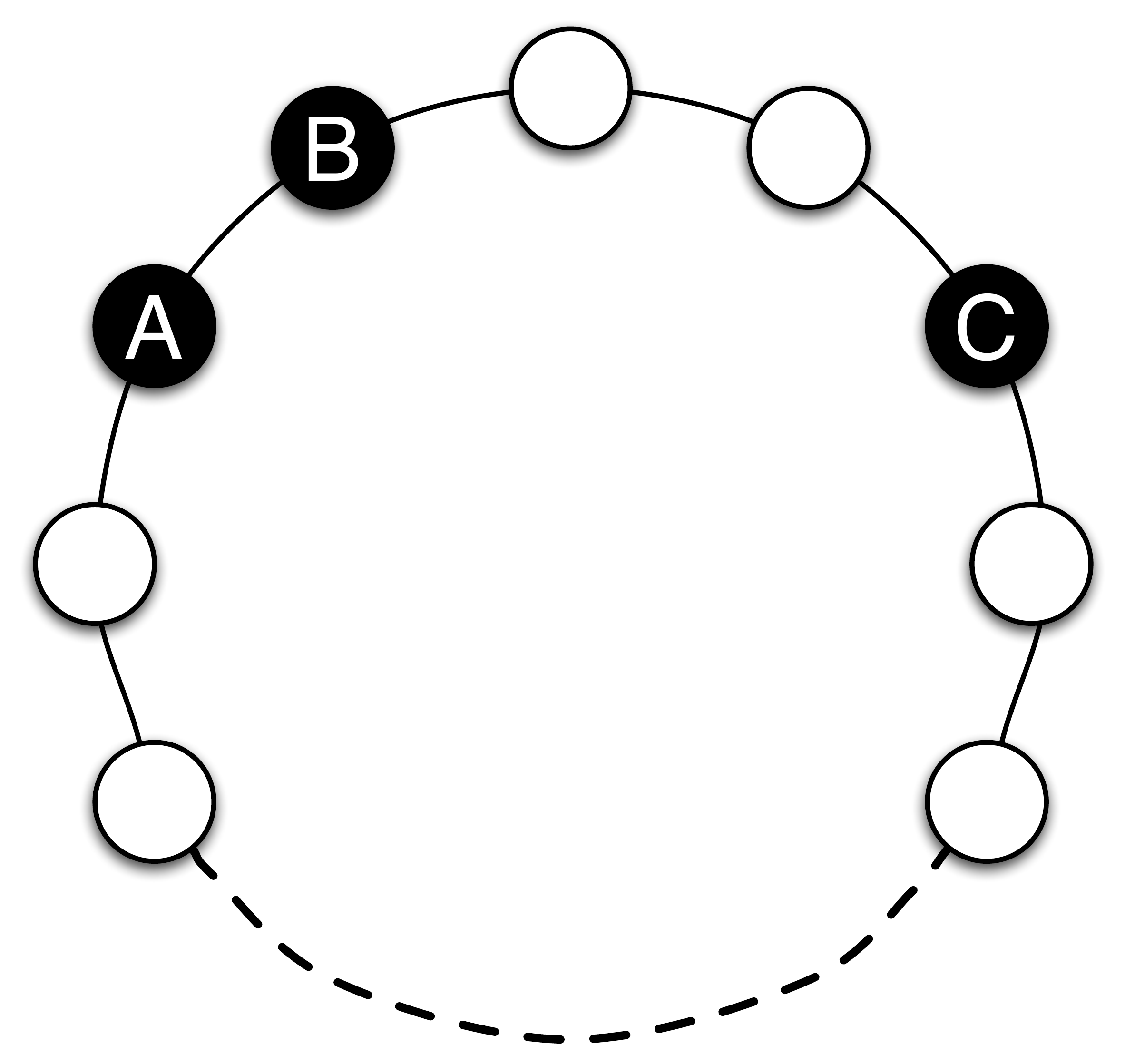}
\label{fig:AlgoMin4}
}
\caption{Exploration perpétuelle avec $3$ robots}
\label{fig:AlgoMin}
\end{figure}

\subsection{Algorithme utilisant un nombre minimum de robots}
\label{subsection:algo-min}

Notre algorithme traite différemment deux types de photos: les photos du régime permanent, et les photos du régime transitoire. Les premières sont appelées photos \emph{permanentes}, et les secondes photos \emph{transitoires}. 

Les photos permanentes présentent une asymétrie des positions des robots qui permet de donner une \og direction \fg\/ à l'exploration (sens des aiguilles d'une montre, ou sens inverse des aiguilles d'une montre). Cette asymétrie est créée à la fois par les groupes de robots (placés sur des n{\oe}uds consécutifs) et par les groupes de n{\oe}uds libres. Notre algorithme assure que la même asymétrie sera maintenue tout au long de son exécution dans le régime permanent. Dans ce régime, l'algorithme préserve une formation en deux groupes de robots,  l'un constitué par un robot et l'autre par deux robots. Le robot  seul, noté $\rob_C$ (abusivement, car les robots n'ont pas d'identifiant), sera toujours séparé par au moins deux n{\oe}uds libres des autres robots. Dans l'algorithme,  $\rob_C$ avance dans la direction indiquée par un plus grand nombre de n{\oe}uds libres. Les deux autres robots,  $\rob_B$ et  $\rob_A$, avancent vers dans la direction indiquée par un plus petit nombre de n{\oe}uds libres. Leur objectif est de rejoindre le robot  $\rob_C$ (voir figure~\ref{fig:AlgoMin}). Plus formellement, notre algorithme d'exploration perpétuelle en régime permanent est décrit ci-dessous. Il a l'avantage de ne rendre activable qu'un seul robot à chaque étape, ce qui force le choix de l'adversaire. Dans cet algorithme paramètré par $z$, on suppose que $z\neq \{0,1,2,3\}$.

\begin{center}
\begin{tabular}{ll}\hline
     \multicolumn{2}{l}{\textbf{Algorithme d'exploration perpétuelle avec un nombre minimum de robots } } \\\hline
 $\C^\circ_2=(\rob_2,\free_2,\rob_1,\free_z)$ &  $\rightarrow \C^\circ_3=(\rob_1,\free_1,\rob_1,\free_2,\rob_1,\free_{z-1})$\\
$\C^\circ_3=(\rob_1,\free_1,\rob_1,\free_2,\rob_1,\free_z)$&  $ \rightarrow  \overline{\C^\circ_2}=(\rob_2,\free_3,\rob_1,\free_z)$\\
$ \overline{\C^\circ_2}=(\rob_2,\free_3,\rob_1,\free_z)$&$\rightarrow \C^\circ_2=(\rob_2,\free_2,\rob_1,\free_{z+1})$\\
      \hline
\end{tabular}
\label{alg:phase1}
\end{center}

\begin{figure}[bt!]
\centering
\includegraphics[width=0.8\textwidth]{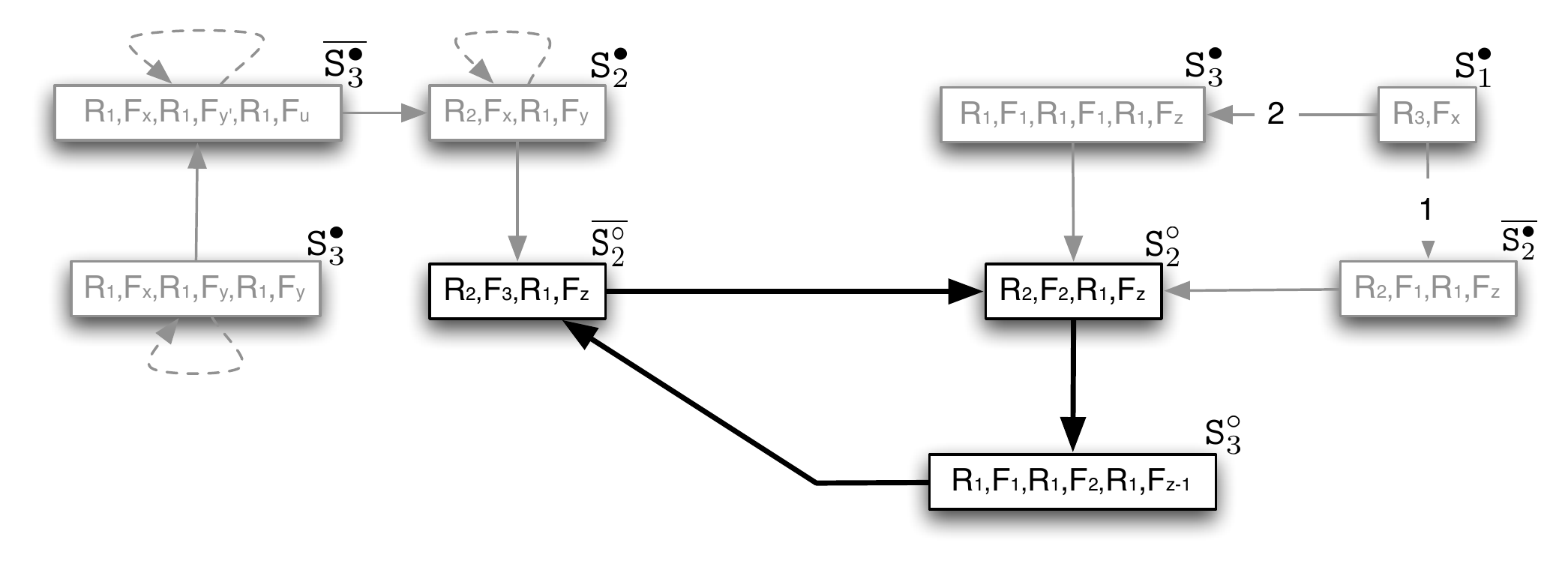}
\caption{\small Pellicule avec $3$ robots représentant les photos permanentes et transitoires, ainsi que leur convergence}
\label{fig:PG3z}
\end{figure}

Les photos transitoires nécessitent un traitement spécifique.  Le nombre de robots étant limité à $3$, le nombre de photos transitoires est limité à $5$. Grâce à la pellicule $G_{n,3}$, nous avons pu construire un algorithme de convergence pour passer du régime transitoire au régime permanent. Cet algorithme est décrit ci-dessous (voir aussi la figure~\ref{fig:PG3z}) . Notons que, dans le cas  $\C^\bullet_1$, l'adversaire a le choix d'activer un ou deux robots, ce qui est délicat à traiter.

\begin{flushleft}
{\small
\begin{tabular}{lll}\hline
\multicolumn{3}{l}{\textbf{Algorithme de convergence avec un nombre minimum de robots }.} \\
\hline
 $\C^\bullet_2=(\rob_2,\free_y,\rob_1,\free_z)$&$\rightarrow \C^\bullet_2=(\rob_2,\free_{y-1},\rob_1,\free_{z+1})$& {\footnotesize avec} $y<z, (y,z)\not \in \{1,2,3\}$ \\
 $\C^\bullet_3=(\rob_1,\free_x,\rob_1,\free_y,\rob_1,\free_y)$& $  \rightarrow \overline{\C^\bullet_3}=(\rob_1,\free_x,\rob_1,\free_{y-1},\rob_1,\free_{y+1})$& {\footnotesize avec} $x \neq y \neq 0$\\
 $\overline{\C^\bullet_3}= (\rob_1,\free_x,\rob_1,\free_y,\rob_1,\free_z)$&$ \rightarrow \overline{\C^\bullet_3}=(\rob_1,\free_{x-1},\rob_1,\free_{y},\rob_1,\free_{z+1})$&avec $x < y < z$\\
$\C^\bullet_1=(\rob_3,\free_z)$&$ \rightarrow \overline{\C^\bullet_2}=(\rob_2,\free_1,\rob_1,\free_{z-1})$& {\footnotesize quand} $1$ robot activé \\
                  & $\rightarrow$ $\C^\bullet_3=(\rob_1,\free_1,\rob_1,\free_1,\rob_1,\free_{z-2})$& {\footnotesize quand} $2$ robots sont activés\\ 
$\overline{\C^\bullet_2}=(\rob_2,\free_1,\rob_1,\free_z)$&$ \rightarrow \C^\circ_2=(\rob_2,\free_2,\rob_1,\free_{z-1})$&\\
      \hline
\end{tabular}
\label{alg:phase2}
}
\end{flushleft}

\subsection{Algorithme utilisant un nombre maximum de robots}
\label{subsection:algo-max}

Dans cette partie, nous décrivons succinctement l'algorithme d'exploration perpétuelle par un nombre maximum de robots, à savoir $k=n-5$ robots pour $k$ impair, $k>3$ et $n\mod k \neq 0$. Comme pour le nombre minimum de robots,  l'algorithme présente  un régime transitoire et un régime permanent. Malheureusement, la phase transitoire implique un nombre de photos (transitoires)  très importants. De surcroît, le nombre de photos transitoires pour lesquelles l'adversaire peut choisir d'activer plus d'un robot est également très important. Le manque de place nous empêche de décrire ici l'algorithme de passage du régime transitoire au régime permanent. 

L'idée principale de l'algorithme en régime permanent est de créer la même asymétrie que dans l'exploration perpétuelle avec 3~robots. Toutefois, dans le cas présent, les rôles des n{\oe}uds libres et des robots sont inversés.  Ainsi, l'algorithme maintient deux ou trois groupes de n{\oe}uds libres, et deux ou trois groupes de robots (voir la figure~\ref {fig:AlgoMax}). Par exemple, considérons une photo $\C_1$ indiquant deux groupes de robots, l'un constitué de deux robots, et l'autre constitué de $n-7$ robots. Ces deux groupes sont séparés d'un côté par trois n{\oe}uds libres, et de l'autre par deux n{\oe}uds libres. Le robot appartenant au plus grand groupe de robots rejoint alors le groupe de deux robots à travers les trois n{\oe}uds libres. Ensuite, le robot appartenant maintenant au groupe de trois robots rejoint le grand groupe de robots à travers les deux n{\oe}uds libres. La configuration obtenue sera identique à celle de la photo $\C_1$. Le processus peut donc être répété indéfiniment. 

\begin{figure}[b!]
\centering
\subfigure[{\footnotesize } $\C^\circ_2$]{
\includegraphics[width=.13\textwidth]{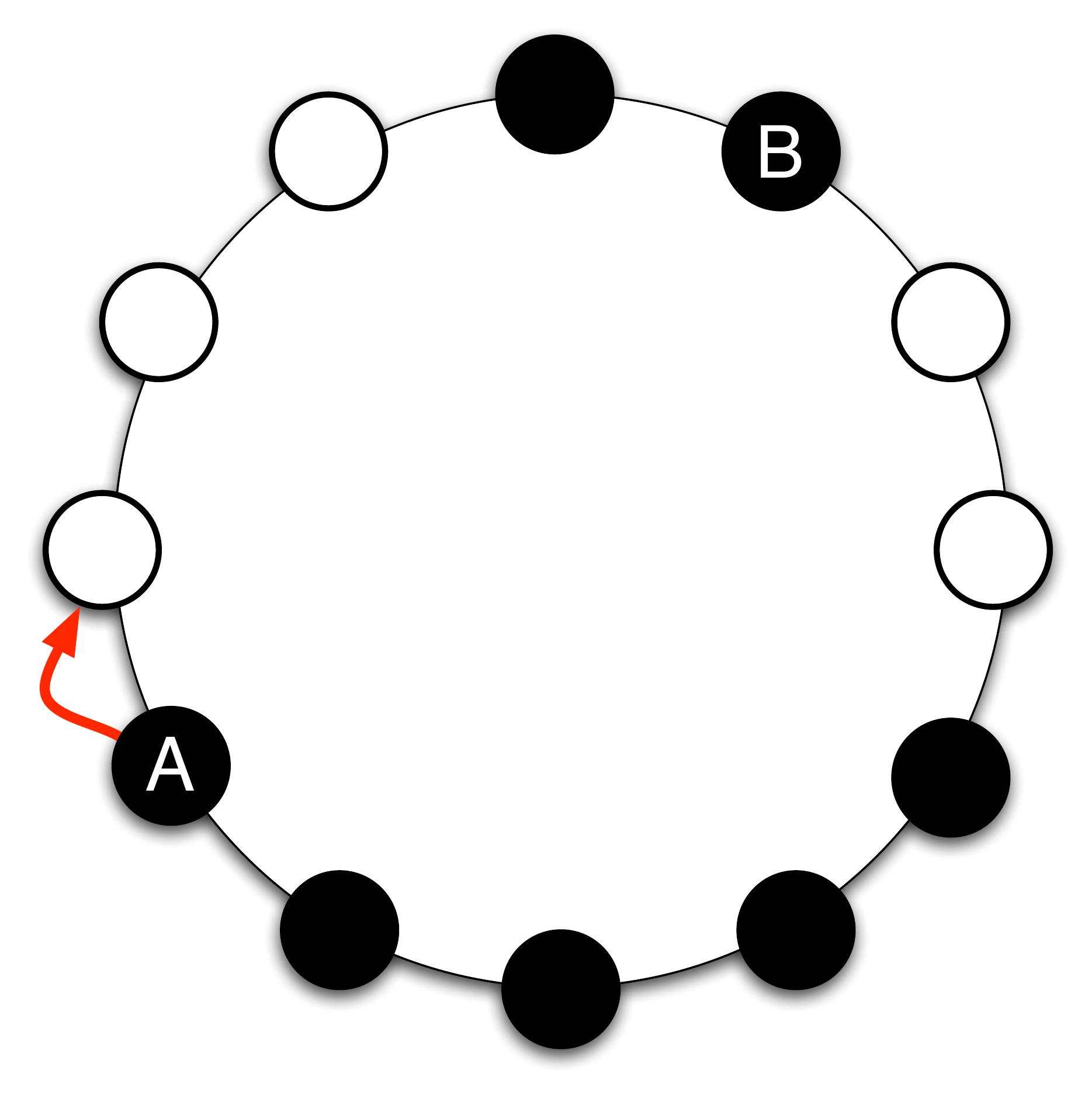}
\label{fig:AlgoMax1}
}
\subfigure[ $\C^\circ_3$]{
\includegraphics[width=.13\textwidth]{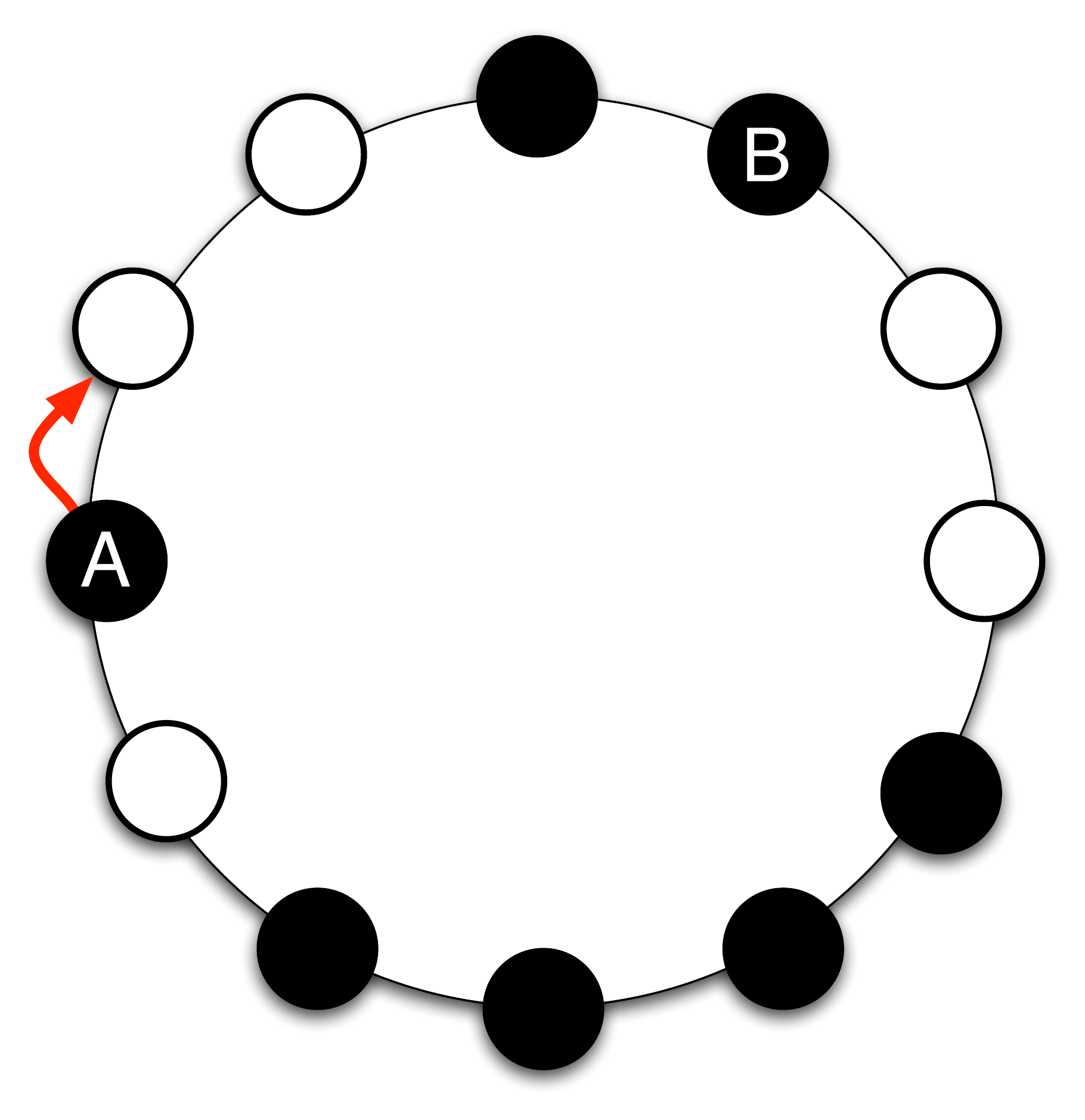}
\label{fig:AlgoMax2}
}
\subfigure[  $\overline{\C^\circ_3}$]{
\includegraphics[width=.13\textwidth]{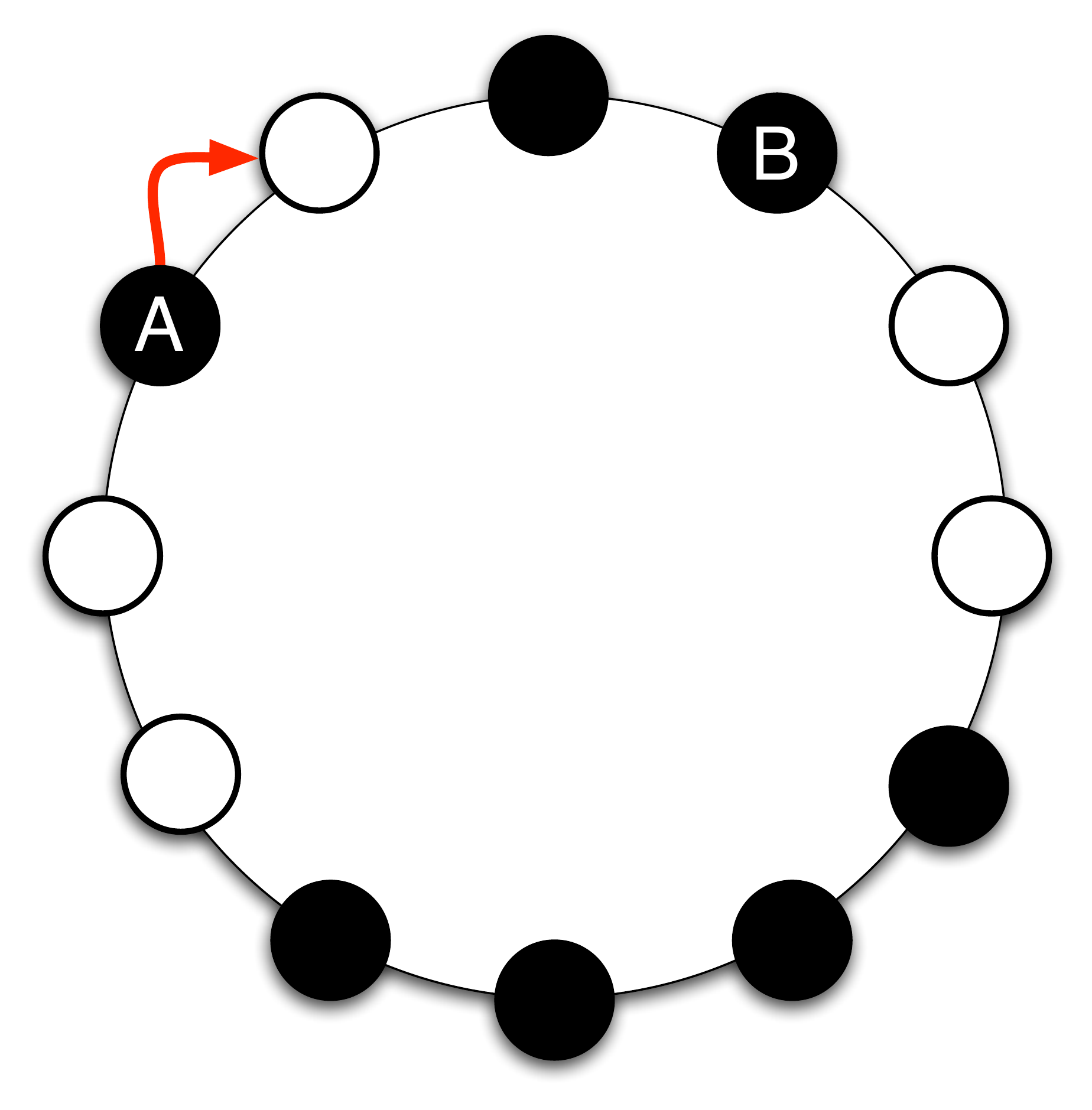}
\label{fig:AlgoMax3}
}
\subfigure[ $\overline{\C^\circ_2}$]{
\includegraphics[width=.13\textwidth]{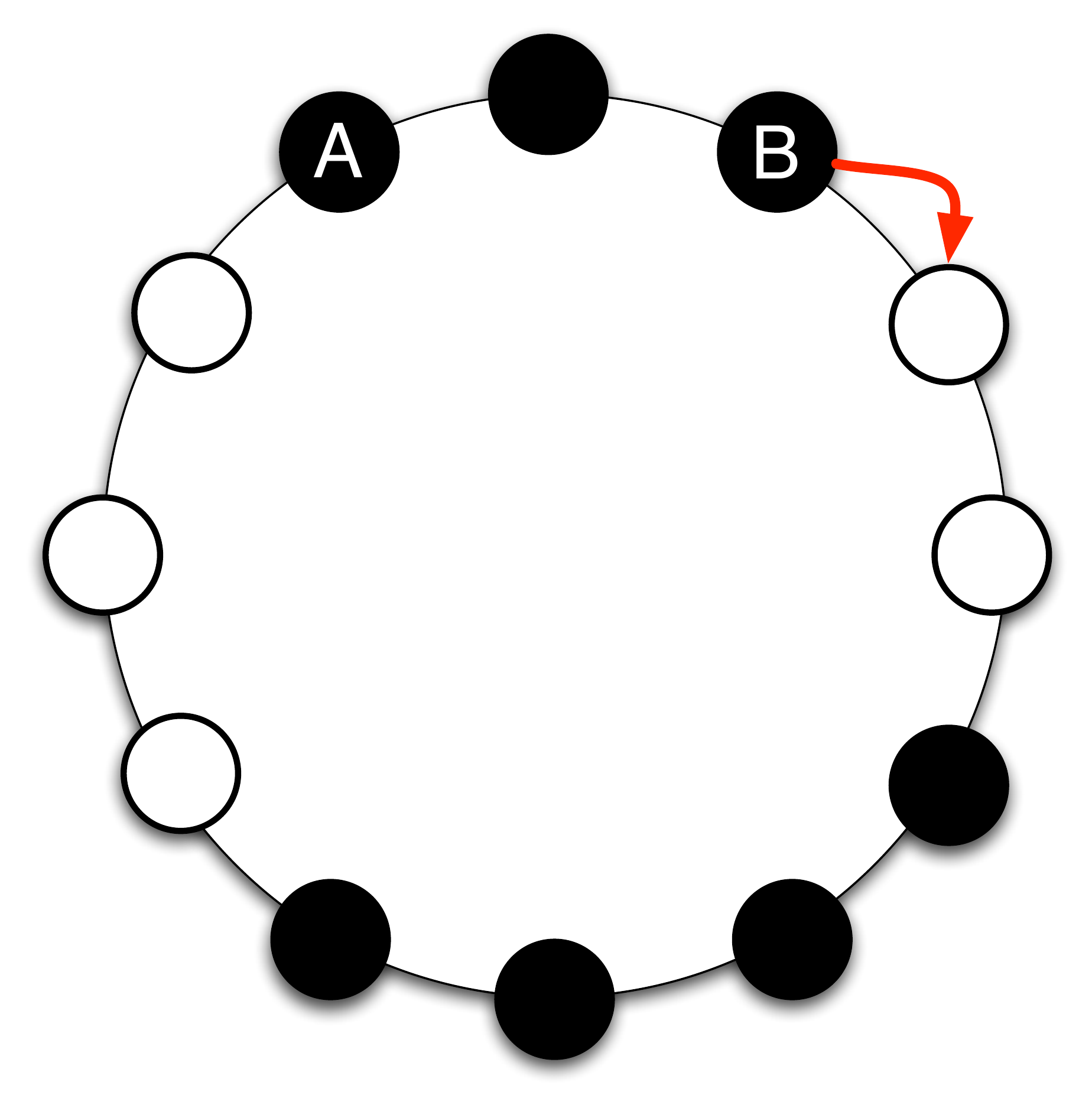}
\label{fig:AlgoMax}4
}
\subfigure[ $\wideparen{\C^\circ_3}$]{
\includegraphics[width=.13\textwidth]{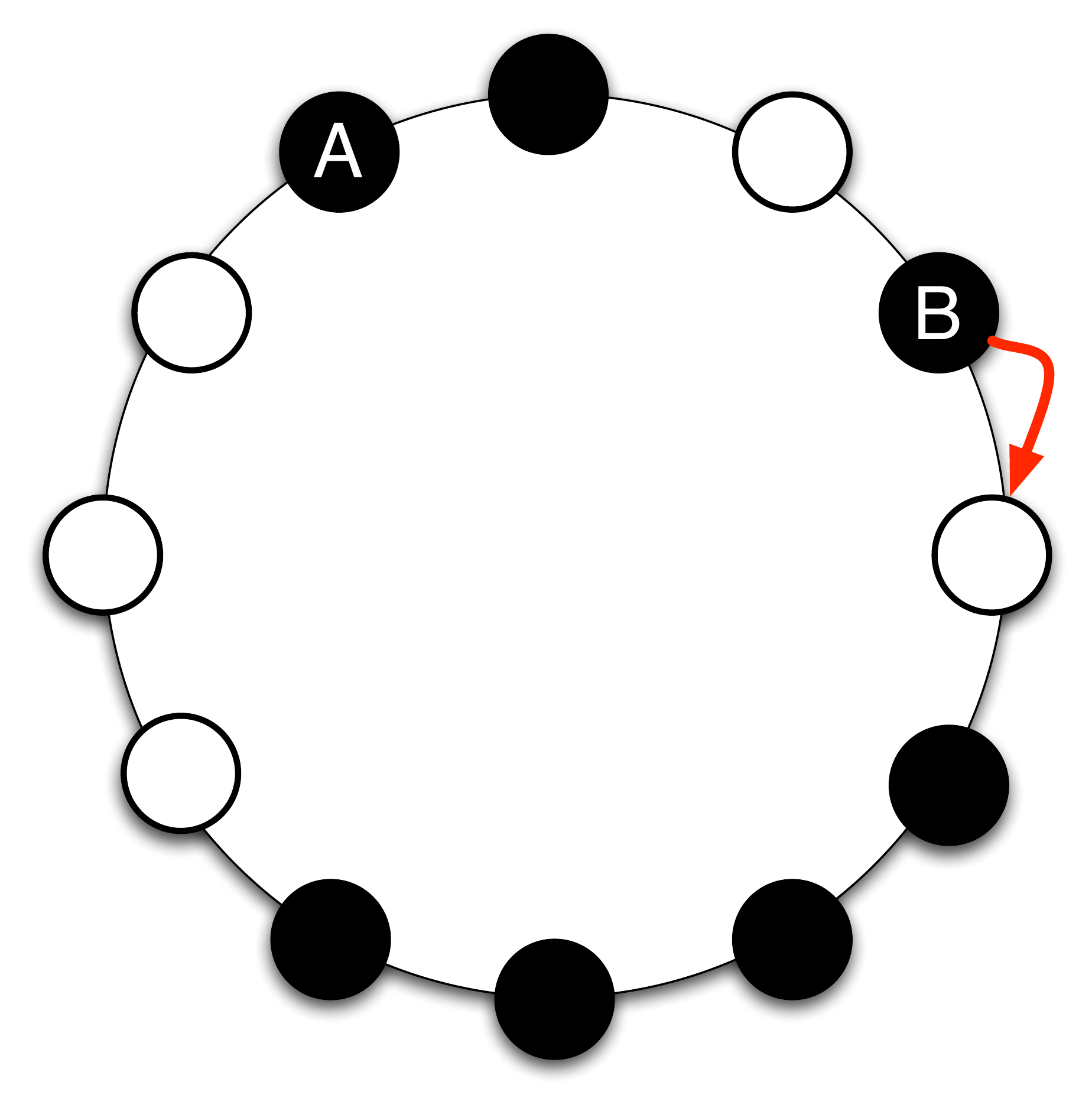}
\label{fig:AlgoMax5}
}
\subfigure[ $\C^\circ_2$]{
\includegraphics[width=.13\textwidth]{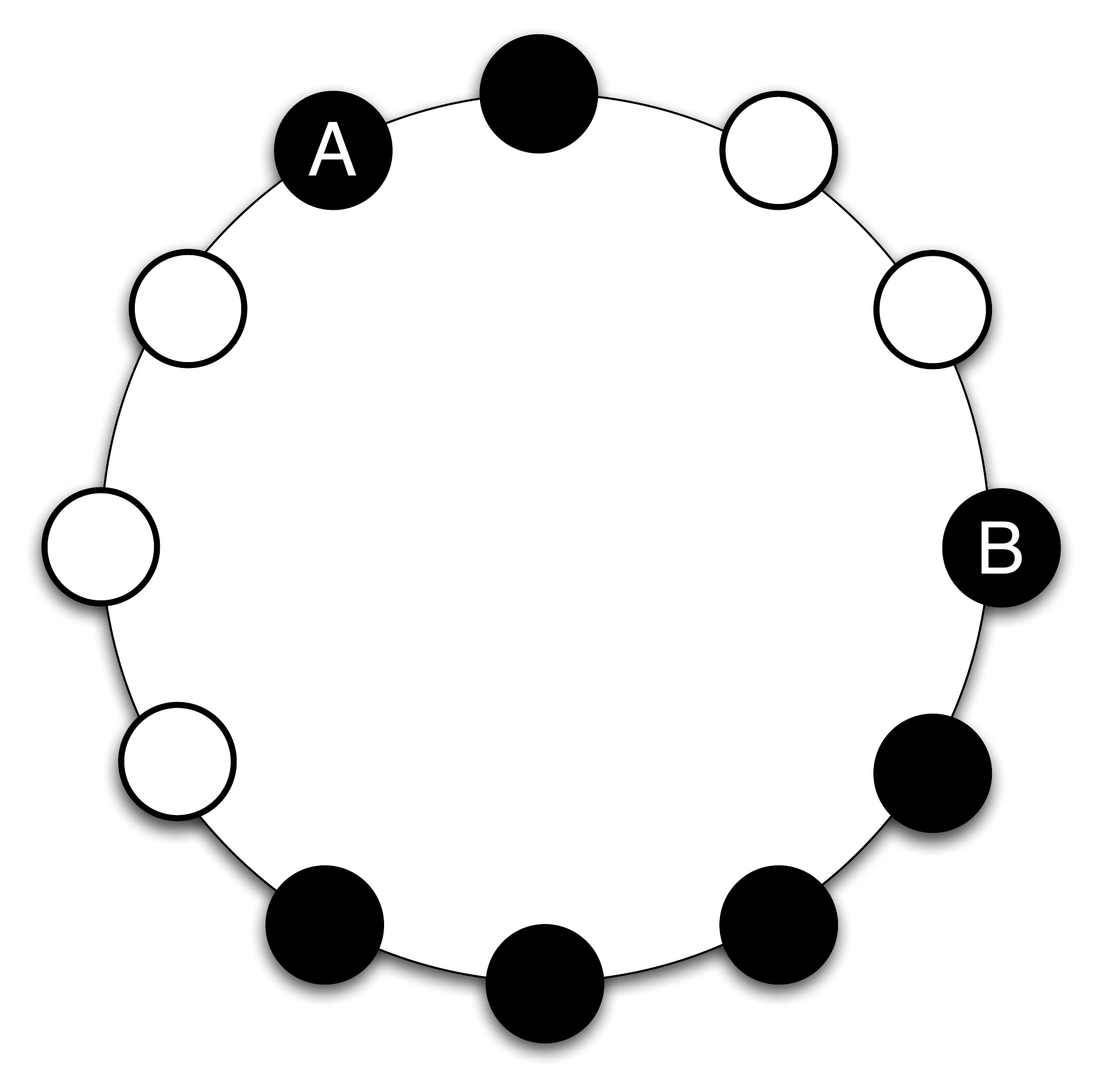}
\label{fig:AlgoMax6}
}
\caption{Exploration perpétuelle utilisant un nombre maximum de robots}
\label{fig:AlgoMax}
\end{figure}


\section{Perspectives}

Les travaux présentés dans ce chapitre sont les premiers à considérer la conception d'algorithmes auto-stabilisants pour les robots dans le modèle CORDA sans hypothèse supplémentaire. Nous avons montré qu'il est possible, sous certaines conditions élémentaires liant la taille de l'anneau au nombre de robots, de réaliser l'exploration perpétuelle dans tout anneau, de façon déterministe.  Le choix de l'anneau est motivé par le fait que, malgré sa simplicité, son étude permet de mettre en évidence des techniques déjà sophistiquées. Il n'en reste pas moins qu'une piste de recherche évidente consiste  à considérer des familles de réseaux plus complexes, comme les grilles et les tores, voire des réseaux quelconques. Par ailleurs, il serait évidemment intéressant d'étudier le \emph{regroupement} dans le modèle CORDA discret avec contrainte d'exclusivité. Notons néanmoins qu'il n'est même pas clair comment définir ce problème dans ce cadre. On pourrait imaginer le regroupement sur un sous-réseau connexe du réseau initial, mais d'autres définitions sont possibles, potentiellement incluant la formation de formes spécifiques (chemin, anneau, etc.). 

\part{Conclusions et perspectives}

\chapter{Perspectives de recherche}
\label{chap:Conclusion}

Ce document fournit un résumé de mes principales contributions récentes à l'auto-stabilisation, aussi bien dans le cadre des réseaux que dans celui des entités mobiles. Chacun des chapitres a listé un certain nombre de problèmes ouverts et de directions de recherche spécifiques à chacune des thématiques abordées dans le chapitre. Dans cette dernière partie du document, je développe des perspectives de recherche générales, à longs termes, autour \emph{des compromis entre l'espace utilisé par les n{\oe}uds, le temps de convergence de l'algorithme, et la qualité de la solution retournée par l'algorithme}. 

Plusieurs paramètres peuvent en effet être pris en compte pour mesurer l'efficacité d'un algorithme auto-stabilisant, dont en particulier le temps de convergence et la complexité mémoire. L'importance du temps de convergence vient de la nécessité évidente pour un système de retourner le plus rapidement possible dans un état valide après une panne. La nécessité de  minimiser la mémoire vient, d'une part, de l'importance grandissante de réseaux tel que les réseaux de capteurs qui ont des espaces mémoires restreints et, d'autre part, de l'intérêt de minimiser l'échange d'information et le stockage d'information afin de limiter la corruption.

La minimisation de la mémoire peut se concevoir au détriment d'autres critères, dont en particulier le temps de convergence, et la qualité de la solution espérée. Mes perspectives de recherche s'organisent autour de deux axes~: 
\begin{itemize}
\item compromis mémoire - temps de convergence; 
\item compromis mémoire - qualité de la solution. 
\end{itemize}
Ces deux axes sont bien évidemment complémentaires, et peuvent avoir à être imbriqués. Dans un but de simplicité de la présentation, ils sont toutefois décrits ci-dessous de façon indépendantes. 

\section{Compromis mémoire - temps de convergence}

Nous avons vu par exemple dans le chapitre~\ref{chap:MST} que, pour le cas de la construction d'un arbre couvrant de poids minimum, un certain compromis espace-temps peut être mis en évidence. Nous avons en effet conçu un algorithme en temps de convergence $O(n^2)$ utilisant une mémoire $O(\log^2  n)$ bits  en chaque n{\oe}ud, mais nous avons montré qu'au prix d'une augmentation du temps en $O(n^3)$, il est possible de se limiter à une mémoire $O(\log n)$ bits par n{\oe}ud. C'est précisément ce type de compromis que nous cherchons à mettre en évidence\footnote{Notons qu'un meilleurs compromis a été trouvé recemment~\cite{KormanKM11}.}.

Nous comptons  aborder le compromis mémoire - temps de convergence selon deux approches. D'une part, nous allons considérer un grand nombre de problèmes dans le cadre de l'optimisation de structures couvrantes, afin d'étudier si le compromis  mis en évidence pour le problème de l'arbre couvrant de poids minimum peut s'observer dans d'autres cadres. Le problème sur lequel nous comptons focaliser nos efforts est celui de la construction de \emph{spanners}, c'est-à-dire de graphes partiels couvrants. Les spanners sont principalement caractérisés par leur nombre d'arêtes et leur facteur d'élongation. Ce dernier paramètre est défini par le maximum, pris sur toutes les paires de n{\oe}uds $(u,v)$, du rapport entre la distance $\mbox{dist}_G(u,v)$ entre ces deux n{\oe}uds dans le réseau $G$ et la distance  $\mbox{dist}_S(u,v)$ entre ces mêmes n{\oe}uds dans le spanner $S$~:
\[\mbox{élongation}=\max_{u,v}\frac{\mbox{dist}_S(u,v)}{\mbox{dist}_G(u,v)}~.\]
La littérature sur les spanners a pour objectif le meilleur compromis entre nombre d'arêtes et élongation, selon des approches centralisées ou réparties. Dans~\cite{AlthoferDDJS93}, un algorithme réparti est ainsi proposé, construisant pour tout $k\geq 1$, un spanner d'élongation $2k-1$ avec un nombre d'arêtes $O(n^{1+1/k})$. Nous avons pour but de reprendre cette approche, mais dans un cadre auto-stabilisant. Est-il possible de concevoir des algorithmes auto-stabilisants offrant les mêmes performances que celles ci-dessus~? Quelle est l'espace mémoire requis pour ce type d'algorithmes (s'ils existent)~? Peut-on mettre en évidence des compromis espace - élongation - nombre d'arêtes~? Ce sont autant de questions que nous comptons aborder dans l'avenir. 

D'autre part, nous avons également pour souhait la mise en évidence de bornes inférieures. L'établissement de bornes inférieures non-triviales est un des défis de l'informatique (cf. P versus NP). Le cadre du réparti et de l'auto-stabilisation ne simplifie pas forcément la difficulté de la tâche, mais certaines restrictions, comme imposer aux algorithmes de satisfaire certaines contraintes de terminaison (par exemple d'être silencieux), semble permettre l'obtention de bornes non-triviales (voir~\cite{DolevGS99}).  

\section{Compromis mémoire - qualité de la solution}

Nous avons également pour objectif l'étude du compromis entre l'espace mémoire utilisé par un algorithme et le rapport d'approximation qu'il garantit pour un problème d'optimisation donné. Ces dernières années, différents types de compromis ont fait l'objet de recherches intensives. La théorie des \emph{algorithmes d'approximation} est basée sur un tel compromis. Celle-ci a été développée autour de l'idée que, pour certains problèmes d'optimisation NP-difficiles, il est possible de produire de bonnes solutions approchées en temps de calcul polynomial.  Nous nous proposons d'étudier le compromis entre l'espace mémoire utilisé et le rapport d'approximation dans le cas des algorithmes auto-stabilisants, dont les n{\oe}uds sont restreints à utiliser un espace limité.

Dans le cas de la construction d'arbres, la plupart des algorithmes répartis de la littérature se focalisent sur des algorithmes dont les caractéristiques sont à un facteur d'approximation $\rho$ de l'optimal, où $\rho$ est  proche du \emph{meilleur} facteur connu pour un algorithme séquentiel. C'est, par exemple, le cas de la construction d'arbres de Steiner ($\rho=2$), ou d'arbres de degré minimum ($\mbox{OPT}+1$). Cette approche, satisfaisante du point de vue des performances en terme d'optimisation, peut se révéler très coûteuse en mémoire dans un cadre auto-stabilisant. Pour optimiser la mémoire, il pourrait se révéler plus efficace de relaxer quelque peu le facteur d'approximation. C'est cette voie que je me propose d'étudier dans l'avenir.

\chapter*{Research perspectives (in English)}
\addcontentsline{toc}{chapter}{Research perspectives (in English)}
\chaptermark{Research perspectives (in English)}
\label{chap:AConclusionEnglish}

This document has summarized my recent contributions in the field of self-stabilization, within the networking framework as well as within the framework of computing with mobile entities. Each chapter has listed open problems, and some specific research directions related to the topics addressed in the chapter. In this last chapter of the document, I am going to develop general long term research perspectives organized around the study of \emph{tradeoffs between the memory space used by nodes, the convergence time of the algorithm, and the quality of the solution returned by the algorithm}. 

Several parameters can be taken into account for measuring the efficiency of a self-stabilizing algorithm, among which the  convergence time and the memory space play an important role. The importance of the convergence time  comes from the evident necessity for a system to return to a valid state after a fault, as quickly as possible. The importance of minimizing the memory space comes from, on one hand, the growing importance of networks, such as sensor networks, which involve computing facilities subject to space constraints, and, on the other hand, the minimization of the amount of information exchange and storage, in order to limit the probability of information corruption.

Minimizing the memory space can be achieved, though potentially to the detriment of other criteria, among which the  convergence time, and the quality of the returned solution. My research perspectives thus get organized around two main subjects:
\begin{itemize} 
\item tradeoff between memory size and convergence time;
\item tradeoff between memory size and  quality of  solutions. 
\end{itemize}
These two subjects are obviously complementary, and thus must not be treated independently from each other. Nevertheless, for the sake of simplifying the presentation, they are described below as two independent topics.

\section*{Tradeoff between memory size and convergence time}
\addcontentsline{toc}{section}{Tradeoff between memory size and convergence time}

As we have seen in Chapter~\ref{chap:MST}, in the case of MST construction, some space-time tradeoffs can be identified. We have indeed conceived an algorithm whose convergence time is $O(n^2)$, with a memory space of $O(\log^2 n)$ bits at every nodes, and we have shown that, to the prize of increasing the convergence time to $O(n^3)$, it is possible to reduce the memory space to $O(\log n)$ bits per node. This is precisely this kind of tradeoffs that are aiming at studying in the future\footnote{Note that a better tradeoff has recently been  identified in~\cite{KormanKM11}.}.

We plan to tackle tradeoffs between memory space and convergence time according to two approaches. First, we are going to consider a large number of problems within the framework of optimizing spanning structures, and we will analyze whether the kind of tradeoffs brought to light for MST construction can be observed in different frameworks. One of the problems on which we plan to focus our efforts is the construction of \emph{spanners}  (i.e., spanning subgraphs). Spanners are essentially characterized by their number of  edges and by their stretch factor. This latter parameter is defined by the maximum, taken over all pairs $(u,v)$ of nodes, of the ratio between the distance $\mbox{dist}_G(u,v)$ between these two nodes in the network $G$, and the distance $\mbox{dist}_S(u,v)$ between the same two nodes in the spanner~$S$:
\[\mbox{stretch}=\max_{u,v}\frac{\mbox{dist}_S(u,v)}{\mbox{dist}_G(u,v)}~.\]
The literature on spanners mostly focuses on the best tradeoff between the number of edges and the stretch, according to  centralized or distributed approaches. In~\cite{AlthoferDDJS93}, a distributed algorithm is proposed which, for any $k\geq $1, constructs spanners of stretch $2k-1$ with a number of edges $O(n^{1+1 / k})$. We aim at revisiting distributed spanner construction, in the framework of self-stabilization. Is it possible to conceive  self-stabilizing algorithms offering the same performances as those described above? What would be the memory  space requirement of such algorithms (if they exist)? Can we bring to light tradeoffs between memory space, stretch, and number of edges? These questions are typical of the ones we plan to tackle in the future.

Our second approach aims at identifying lower bounds. Establishing lower bounds is one of challenges of computer science (as exemplified by the P versus NP question). The framework of  distributed computing, and/or self-stabilization does not necessarily simplify the difficulty of the task. However, restrictions such as imposing the algorithms to satisfy certain termination constraints  (for example to be silent), have been proved to be helpful for deriving lower bounds (see, e.g.,~\cite{DolevGS99}).

\section*{Tradeoff between memory size and  quality of  solutions}
\addcontentsline{toc}{section}{Tradeoff between memory size and  quality of  solutions}

One of our objectives is also to study tradeoffs between, on the one hand, the memory space used by the algorithm, and, on the other hand,  the quality of the solution provided by the algorithm, in the framework of optimization problems. In this framework, various types of tradeoffs have been the object of extensive researches these last years. The theory of \emph{approximation algorithms} is precisely based on that sort of tradeoffs. It was developed around the idea that, for many  NP-hard optimization problems, it is possible to compute ``good'' solutions (though not necessarily optimal) in polynomial time. We are aiming at studying the tradeoff between memory space and approximation ratio in the case of self-stabilizing  algorithms (in a context in  which nodes are restricted to use a limited space).

In the case of spanning tree construction, most of the distributed algorithms  in the literature are based on sequential approximation algorithms with approximation factor $\rho$ close to the best known approximation factor. That is, for example, the case of Steiner tree construction  ($\rho=2$), and minimum-degree spanning tree ($\mbox{OPT}+1$).  In the context of self-stabilization, this approach, which is satisfying from the point of view of optimization, can be very expensive as far as memory is concerned. For optimizing memory, one may consider relaxing the quality of the approximation factor. That is this approach that   I suggest studying in the future.

\newpage
\bibliographystyle{plain}
\addcontentsline{toc}{part}{Bibliographie}
\bibliography{biblioLB,biblio}

\begin{thebibliography}{100}

\bibitem{AfekB98}
Yehuda Afek and Anat Bremler-Barr.
\newblock Self-stabilizing unidirectional network algorithms by power supply.
\newblock {\em Chicago J. Theor. Comput. Sci.}, 1998, 1998.

\bibitem{AfekKY91}
Yehuda Afek, Shay Kutten, and Moti Yung.
\newblock Memory-efficient self stabilizing protocols for general networks.
\newblock In {\em Proceedings of the 4th international workshop on Distributed
  algorithms}, pages 15--28. Springer, 1991.

\bibitem{AggarwalK93}
Sudhanshu Aggarwal and Shay Kutten.
\newblock Time optimal self-stabilizing spanning tree algorithms.
\newblock In {\em 13th Conference on Foundations of Software Technology and
  Theoretical Computer Science,(FSTTCS 1993)}, volume 761 of {\em Lecture Notes
  in Computer Science}, pages 400--410. Springer, 1993.

\bibitem{AlthoferDDJS93}
Ingo Alth{\"o}fer, Gautam Das, David~P. Dobkin, Deborah Joseph, and Jos{\'e}
  Soares.
\newblock On sparse spanners of weighted graphs.
\newblock {\em Discrete {\&} Computational Geometry}, 9:81--100, 1993.

\bibitem{AroraG90}
Anish Arora and Mohamed~G. Gouda.
\newblock Distributed reset (extended abstract).
\newblock In {\em 10th Conference Foundations of Software Technology and
  Theoretical Computer Science, (FSTTCS 1990)}, volume 472 of {\em Lecture
  Notes in Computer Science}, pages 316--331. Springer, 1990.

\bibitem{AroraG94}
Anish Arora and Mohamed~G. Gouda.
\newblock Distributed reset.
\newblock {\em IEEE Trans. Computers}, 43(9):1026--1038, 1994.

\bibitem{AsahiroFSY08}
Fujita S.-Suzuki I. Yamashita~M. Asahiro, Y.
\newblock A self-stabilizing marching algorithm for a group of oblivious
  robots.
\newblock In {\em 12th International Conference on Principles of Distributed
  Systems, (OPODIS 2008)}, volume 5401 of {\em Lecture Notes in Computer
  Science}, pages 125--144. Springer, 2008.

\bibitem{Awerbuch87}
Baruch Awerbuch.
\newblock Optimal distributed algorithms for minimum weight spanning tree,
  counting, leader election and related problems.
\newblock In {\em 19th Annual ACM Symposium on Theory of Computing, (STOC
  1987)}, pages 230--240, 1987.

\bibitem{AwerbuchGPV90}
Baruch Awerbuch, Oded Goldreich, David Peleg, and Ronen Vainish.
\newblock A trade-off between information and communication in broadcast
  protocols.
\newblock {\em J. ACM}, 37(2):238--256, 1990.

\bibitem{AwerbuchKMPV93}
Baruch Awerbuch, Shay Kutten, Yishay Mansour, Boaz Patt-Shamir, and George
  Varghese.
\newblock Time optimal self-stabilizing synchronization.
\newblock In ACM, editor, {\em 25th Annual ACM Symposium on Theory of
  Computing, (STOC 1993)}, pages 652--661, 1993.

\bibitem{BaldoniBMR08}
Roberto Baldoni, Fran\c{c}ois Bonnet, Alessia Milani, and Michel Raynal.
\newblock On the solvability of anonymous partial grids exploration by mobile
  robots.
\newblock In {\em 12th International Conference on Principles of Distributed
  Systems, (OPODIS 2008)}, volume 5401 of {\em Lecture Notes in Computer
  Science}, pages 428--445. Springer, 2008.

\bibitem{BarriereFFS03}
Lali Barri{\`e}re, Paola Flocchini, Pierre Fraigniaud, and Nicola Santoro.
\newblock Can we elect if we cannot compare?
\newblock In {\em Proceedings of the Fifteenth Annual ACM Symposium on
  Parallelism in Algorithms and Architectures, (SPAA 2003)}, pages 324--332.
  ACM, 2003.

\bibitem{BeauquierHS01}
J.~Beauquier, T.~Herault, and E.~Schiller.
\newblock {Easy Stabilization with an Agent}.
\newblock {\em 5th Workshop on Self-Stabilizing Systems (WSS)}, 2194:35--51,
  2001.

\bibitem{BermanR94}
Piotr Berman and Viswanathan Ramaiyer.
\newblock Improved approximations for the steiner tree problem.
\newblock {\em J. Algorithms}, 17(3):381--408, 1994.

\bibitem{BlinB04}
L{\'e}lia Blin and Franck Butelle.
\newblock The first approximated distributed algorithm for the minimum degree
  spanning tree problem on general graphs.
\newblock {\em Int. J. Found. Comput. Sci.}, 15(3):507--516, 2004.

\bibitem{BlinCV03}
L{\'e}lia Blin, Alain Cournier, and Vincent Villain.
\newblock An improved snap-stabilizing pif algorithm.
\newblock In {\em 6th International Symposium on Self-Stabilizing Systems, (SSS
  2003)}, volume 2704 of {\em Lecture Notes in Computer Science}, pages
  199--214. Springer, 2003.

\bibitem{BlinDPR10}
L{\'e}lia Blin, Shlomi Dolev, Maria~Gradinariu Potop-Butucaru, and Stephane
  Rovedakis.
\newblock Fast self-stabilizing minimum spanning tree construction - using
  compact nearest common ancestor labeling scheme.
\newblock In {\em 24th International Symposium on Distributed Computing, (DISC
  2010)}, volume 6343 of {\em Lecture Notes in Computer Science}, pages
  480--494. Springer, 2010.

\bibitem{BlinMPT10}
L{\'e}lia Blin, Alessia Milani, Maria Potop-Butucaru, and S{\'e}bastien
  Tixeuil.
\newblock Exclusive perpetual ring exploration without chirality.
\newblock In {\em 24th International Symposium on Distributed Computing, (DISC
  2010)}, volume 6343 of {\em Lecture Notes in Computer Science}, pages
  312--327. Springer, 2010.

\bibitem{BlinPRT09}
L{\'e}lia Blin, Maria Potop-Butucaru, Stephane Rovedakis, and S{\'e}bastien
  Tixeuil.
\newblock A new self-stabilizing minimum spanning tree construction with
  loop-free property.
\newblock In {\em 23rd International Symposium on Distributed Computing, (DISC
  2009)}, volume 5805 of {\em Lecture Notes in Computer Science}, pages
  407--422. Springer, 2009.

\bibitem{BlinPR09a}
L{\'e}lia Blin, Maria~Gradinariu Potop-Butucaru, and Stephane Rovedakis.
\newblock Self-stabilizing minimum-degree spanning tree within one from the
  optimal degree.
\newblock In {\em 23rd IEEE International Symposium on Parallel and Distributed
  Processing, (IPDPS 2009)}, pages 1--11. IEEE, 2009.

\bibitem{BlinPR09b}
L{\'e}lia Blin, Maria~Gradinariu Potop-Butucaru, and Stephane Rovedakis.
\newblock A superstabilizing log({\it })-approximation algorithm for dynamic
  steiner trees.
\newblock In {\em 11th International Symposium on Stabilization, Safety, and
  Security of Distributed Systems, (SSS 2009)}, volume 5873 of {\em Lecture
  Notes in Computer Science}, pages 133--148. Springer, 2009.

\bibitem{BlinPR11}
L{\'e}lia Blin, Maria~Gradinariu Potop-Butucaru, and Stephane Rovedakis.
\newblock Self-stabilizing minimum degree spanning tree within one from the
  optimal degree.
\newblock {\em J. Parallel Distrib. Comput.}, 71(3):438--449, 2011.

\bibitem{BlinPRT10}
L{\'e}lia Blin, Maria~Gradinariu Potop-Butucaru, Stephane Rovedakis, and
  S{\'e}bastien Tixeuil.
\newblock Loop-free super-stabilizing spanning tree construction.
\newblock In {\em 12th International Symposium on Stabilization, Safety, and
  Security of Distributed Systems, (SSS 2010)}, volume 6366 of {\em Lecture
  Notes in Computer Science}, pages 50--64. Springer, 2010.

\bibitem{BlinPT07}
L{\'e}lia Blin, Maria~Gradinariu Potop-Butucaru, and S{\'e}bastien Tixeuil.
\newblock On the self-stabilization of mobile robots in graphs.
\newblock In {\em 11th International Conference on Principles of Distributed
  Systems, (OPODIS 2007)}, volume 4878 of {\em Lecture Notes in Computer
  Science}, pages 301--314. Springer, 2007.

\bibitem{Boruvka26}
Otakar Bo{\r{r}}uvka.
\newblock O jistém problému minimáln\'im.
\newblock {\em Czech, German summary. Pr\'ace mor. p\u{r}\'irodov\u{e}d. spol.
  v Brn\u{e}}, III 3:37--58, 1926.

\bibitem{BurmanK07}
Janna Burman and Shay Kutten.
\newblock Time optimal asynchronous self-stabilizing spanning tree.
\newblock In {\em 21st International Symposium on Distributed Computing, (DISC
  2007)}, volume 4731 of {\em Lecture Notes in Computer Science}, pages
  92--107. Springer, 2007.

\bibitem{ButelleLB95}
Franck Butelle, Christian Lavault, and Marc Bui.
\newblock A uniform self-stabilizing minimum diameter tree algorithm (extended
  abstract).
\newblock In {\em 9th International Workshop on Distributed Algorithms, (WDAG
  1995)}, volume 972 of {\em Lecture Notes in Computer Science}, pages
  257--272. Springer, 1995.

\bibitem{ByrkaGRS10}
Jaroslaw Byrka, Fabrizio Grandoni, Thomas Rothvo{\ss}, and Laura Sanit{\`a}.
\newblock An improved lp-based approximation for steiner tree.
\newblock In {\em 42nd ACM Symposium on Theory of Computing,(STOC 2010)}, pages
  583--592. ACM, 2010.

\bibitem{ChalopinGMO06}
J{\'e}r{\'e}mie Chalopin, Emmanuel Godard, Yves M{\'e}tivier, and Rodrigue
  Ossamy.
\newblock Mobile agent algorithms versus message passing algorithms.
\newblock In {\em Principles of Distributed Systems, 10th International
  Conference, (OPODIS 2006)}, volume 4305 of {\em Lecture Notes in Computer
  Science}, pages 187--201. Springer, 2006.

\bibitem{ChalopinFMS10}
Jérémie Chalopin, Paola Flocchini, Bernard Mans, and Nicola Santoro.
\newblock Network exploration by silent and oblivious robots.
\newblock In {\em 36th International workshop on Graph Theory Concepts in
  Computer Science, (WG 2010)}, Lecture Notes in Computer Science, pages
  208--219. Springer, 2010.

\bibitem{ChaudhuriRRT05}
Kamalika Chaudhuri, Satish Rao, Samantha Riesenfeld, and Kunal Talwar.
\newblock What would edmonds do? augmenting paths and witnesses for
  degree-bounded msts.
\newblock In {\em 8th International Workshop on Approximation Algorithms for
  Combinatorial Optimization Problems, (APPROX 2005) and 9th
  InternationalWorkshop on Randomization and Computation,(RANDOM 2005)}, volume
  3624 of {\em Lecture Notes in Computer Science}, pages 26--39. Springer,
  2005.

\bibitem{ChaudhuriRRT06}
Kamalika Chaudhuri, Satish Rao, Samantha Riesenfeld, and Kunal Talwar.
\newblock A push-relabel algorithm for approximating degree bounded msts.
\newblock In {\em 33rd International Colloquium in Automata, Languages and
  Programming, (ICALP 2006)}, volume 4051 of {\em Lecture Notes in Computer
  Science}, pages 191--201. Springer, 2006.

\bibitem{ChenHK93}
Gen-Huey Chen, Michael~E. Houle, and Ming-Ter Kuo.
\newblock The steiner problem in distributed computing systems.
\newblock {\em Inf. Sci.}, 74(1-2):73--96, 1993.

\bibitem{CieliebakFPS03}
Flocchini-P.-Prencipe G. Santoro~N. Cieliebak, M.
\newblock Solving the robots gathering problem.
\newblock In {\em ICALP 2003}, volume 2719 of {\em Lecture Notes in Computer
  Science}, pages 1181--1196. Springer, 2003.

\bibitem{CobbG02}
Jorge~Arturo Cobb and Mohamed~G. Gouda.
\newblock Stabilization of general loop-free routing.
\newblock {\em J. Parallel Distrib. Comput.}, 62(5):922--944, 2002.

\bibitem{CollinD94}
Zeev Collin and Shlomi Dolev.
\newblock Self-stabilizing depth-first search.
\newblock {\em Inf. Process. Lett.}, 49(6):297--301, 1994.

\bibitem{CournierDPV02}
Alain Cournier, Ajoy~Kumar Datta, Franck Petit, and Vincent Villain.
\newblock Snap-stabilizing pif algorithm in arbitrary networks.
\newblock In {\em The 22nd International Conference on Distributed Computing
  Systems}, pages 199--, 2002.

\bibitem{CournierDPV05}
Alain Cournier, Ajoy~Kumar Datta, Franck Petit, and Vincent Villain.
\newblock Optimal snap-stabilizing pif algorithms in un-oriented trees.
\newblock {\em J. High Speed Networks}, 14(2):185--200, 2005.

\bibitem{Dalal79}
Yogen~K. Dalal.
\newblock A distributed algorithm for constructing minimal spanning trees.
\newblock Technical Report 111, Stanford University California, June 1979.

\bibitem{Dalal87}
Yogen~K. Dalal.
\newblock A distributed algorithm for constructing minimal spanning trees.
\newblock {\em IEEE Trans. Software Eng.}, 13(3):398--405, 1987.

\bibitem{DanturiNT09}
Praveen Danturi, Mikhail Nesterenko, and S{\'e}bastien Tixeuil.
\newblock Self-stabilizing philosophers with generic conflicts.
\newblock {\em TAAS}, 4(1), 2009.

\bibitem{DattaGT00}
Ajoy~Kumar Datta, Maria Gradinariu, and S{\'e}bastien Tixeuil.
\newblock Self-stabilizing mutual exclusion using unfair distributed scheduler.
\newblock In {\em 14th International Parallel {\&} Distributed Processing
  Symposium, (IPDPS'00)}, pages 465--. IEEE Computer Society, 2000.

\bibitem{DattaJPV00}
Ajoy~Kumar Datta, Colette Johnen, Franck Petit, and Vincent Villain.
\newblock Self-stabilizing depth-first token circulation in arbitrary rooted
  networks.
\newblock {\em Distributed Computing}, 13(4):207--218, 2000.

\bibitem{DattaLV08}
Ajoy~Kumar Datta, Lawrence~L. Larmore, and Priyanka Vemula.
\newblock Self-stabilizing leader election in optimal space.
\newblock In {\em 10th International Symposium on Stabilization, Safety, and
  Security of Distributed Systems, (SSS 2008)}, volume 5340 of {\em Lecture
  Notes in Computer Science}, pages 109--123. Springer, 2008.

\bibitem{DessmarkFKP06}
Anders Dessmark, Pierre Fraigniaud, Dariusz~R. Kowalski, and Andrzej Pelc.
\newblock Deterministic rendezvous in graphs.
\newblock {\em Algorithmica}, 46(1):69--96, 2006.

\bibitem{Devisme06}
Stephane Devisme.
\newblock {\em Quelques Contributions à la Stabilisation Instantanée.}
\newblock PhD thesis, Université de Picardie Jules Verne, 2006.

\bibitem{DevismesPT09}
St{\'e}phane Devismes, Franck Petit, and S{\'e}bastien Tixeuil.
\newblock Optimal probabilistic ring exploration by semi-synchronous oblivious
  robots.
\newblock In {\em 16th International Colloquium on Structural Information and
  Communication Complexity, (SIROCCO 2009)}, volume 5869 of {\em Lecture Notes
  in Computer Science}, pages 195--208. Springer, 2009.

\bibitem{Dijkstra74}
Edsger~W. Dijkstra.
\newblock Self-stabilizing systems in spite of distributed control.
\newblock {\em Communications of the ACM}, 17(11):643--644, 1974.

\bibitem{DolevSW02}
S.~Dolev, E.~Schiller, and J.~Welch.
\newblock {Random walk for self-stabilizing group communication in ad-hoc
  networks}.
\newblock {\em Reliable Distributed Systems, 2002. Proceedings. 21st IEEE
  Symposium on}, pages 70--79, 2002.

\bibitem{Dolev93}
Shlomi Dolev.
\newblock Optimal time self stabilization in dynamic systems (preliminary
  version).
\newblock In {\em 7th International Workshop on Distributed Algorithms, (WDAG
  1993)}, volume 725 of {\em Lecture Notes in Computer Science}, pages
  160--173. Springer, 1993.

\bibitem{Dolev00}
Shlomi Dolev.
\newblock {\em Self-Stabilization}.
\newblock MIT Press, 2000.

\bibitem{DolevGS96}
Shlomi Dolev, Mohamed~G. Gouda, and Marco Schneider.
\newblock Memory requirements for silent stabilization (extended abstract).
\newblock In {\em 15th Annual ACM Symposium on Principles of Distributed
  Computing, (PODC 1996)}, pages 27--34. ACM, 1996.

\bibitem{DolevGS99}
Shlomi Dolev, Mohamed~G. Gouda, and Marco Schneider.
\newblock Memory requirements for silent stabilization.
\newblock {\em Acta Informatica}, 36:447--462, 1999.

\bibitem{DolevIM90}
Shlomi Dolev, Amos Israeli, and Shlomo Moran.
\newblock Self-stabilization of dynamic systems assuming only read/write
  atomicity.
\newblock In {\em 9th Annual ACM Symposium on Principles of Distributed
  Computing, (PODC 1990)}, pages 103--117. ACM, 1990.

\bibitem{DolevIM93}
Shlomi Dolev, Amos Israeli, and Shlomo Moran.
\newblock Self-stabilization of dynamic systems assuming only read/write
  atomicity.
\newblock {\em Distributed Computing}, 7(1):3--16, 1993.

\bibitem{DolevIM97}
Shlomi Dolev, Amos Israeli, and Shlomo Moran.
\newblock Uniform dynamic self-stabilizing leader election.
\newblock {\em IEEE Trans. Parallel Distrib. Syst.}, 8(4):424--440, 1997.

\bibitem{DefagoS08}
Souissi-S. Défago, X.
\newblock Non-uniform circle formation algorithm for oblivious mobile robots
  with convergence toward uniformity.
\newblock {\em Theor. Comput. Sci.}, 396(1-3):97--112, 2008.

\bibitem{Fischer93}
Ted Fischer.
\newblock Optimizing the degree of minimum weight spanning trees.
\newblock Technical Report TR 93-1338, Dept. of Computer Science, Cornell
  University, Ithaca, NY 14853, 1993.

\bibitem{FlocchiniIPS07}
Paola Flocchini, David Ilcinkas, Andrzej Pelc, and Nicola Santoro.
\newblock Computing without communicating: Ring exploration by asynchronous
  oblivious robots.
\newblock In {\em 11th International Conference on Principles of Distributed
  Systems, (OPODIS 2007)}, volume 4878 of {\em Lecture Notes in Computer
  Science}, pages 105--118. Springer, 2007.

\bibitem{FlocchiniIPS08}
Paola Flocchini, David Ilcinkas, Andrzej Pelc, and Nicola Santoro.
\newblock Remembering without memory: Tree exploration by asynchronous
  oblivious robots.
\newblock In {\em 15th International Colloquium on Structural Information and
  Communication Complexity, (SIROCCO 2008)}, volume 5058 of {\em Lecture Notes
  in Computer Science}, pages 33--47. Springer, 2008.

\bibitem{FlocchiniIPS10}
Paola Flocchini, David Ilcinkas, Andrzej Pelc, and Nicola Santoro.
\newblock Remembering without memory: Tree exploration by asynchronous
  oblivious robots.
\newblock {\em Theor. Comput. Sci.}, 411(14-15):1583--1598, 2008.

\bibitem{FlocchiniPSW99}
Paola Flocchini, Giuseppe Prencipe, Nicola Santoro, and Peter Widmayer.
\newblock Hard tasks for weak robots: The role of common knowledge in pattern
  formation by autonomous mobile robots.
\newblock In {\em 10th International Symposium onAlgorithms and Computation,
  (ISAAC99)}, volume 1741 of {\em Lecture Notes in Computer Science}, pages
  93--102. Springer, 1999.

\bibitem{FlocchiniPSW08}
Paola Flocchini, Giuseppe Prencipe, Nicola Santoro, and Peter Widmayer.
\newblock Arbitrary pattern information by asynchronous anonymous oblivious
  robots.
\newblock {\em Theor. Comput. Sci.}, 407(1-3):412--447, 2008.

\bibitem{FredericksonL84}
Greg~N. Frederickson and Nancy~A. Lynch.
\newblock The impact of synchronous communication on the problem of electing a
  leader in a ring.
\newblock In {\em Sixteenth Annual ACM Symposium on Theory of Computing, (STOC
  1984)}, pages 493--503. ACM, 1984.

\bibitem{FredmanW90}
Michael~L. Fredman and Dan~E. Willard.
\newblock Trans-dichotomous algorithms for minimum spanning trees and shortest
  paths.
\newblock In {\em 31st Annual Symposium on Foundations of Computer Science,
  (FOCS 1990)}, volume~II, pages 719--725. IEEE, 1990.

\bibitem{FurerR92}
Martin F{\"u}rer and Balaji Raghavachari.
\newblock Approximating the minimum degree spanning tree to within one from the
  optimal degree.
\newblock In {\em Third Annual ACM/SIGACT-SIAM Symposium on Discrete
  Algorithms, (SODA 92)}, pages 317--324. ACM/SIAM, 1992.

\bibitem{FurerR94}
Martin F{\"u}rer and Balaji Raghavachari.
\newblock Approximating the minimum-degree steiner tree to within one of
  optimal.
\newblock {\em J. Algorithms}, 17(3):409--423, 1994.

\bibitem{GafniB81}
Eli~M. Gafni and P.~Bertsekas.
\newblock Distributed algorithms for generating loop-free routes in networks
  with frequently changing topology.
\newblock {\em IEEE Transactions on Communications}, 29(3):11--18, 1981.

\bibitem{GallagerHS83}
Robert~G. Gallager, Pierre~A. Humblet, and Philip~M. Spira.
\newblock A distributed algorithm for minimum-weight spanning trees.
\newblock {\em ACM Trans. Program. Lang. Syst.}, 5(1):66--77, 1983.

\bibitem{GarayKP98}
Juan~A. Garay, Shay Kutten, and David Peleg.
\newblock A sublinear time distributed algorithm for minimum-weight spanning
  trees.
\newblock {\em SIAM J. Comput.}, 27(1):302--316, 1998.

\bibitem{Aceves93}
J.~J. Garcia-Luna-Aceves.
\newblock Loop-free routing using diffusing computations.
\newblock {\em IEEE/ACM Trans. Netw.}, 1(1):130--141, 1993.

\bibitem{Gartner03}
Felix~C. G{\"a}rtner.
\newblock A survey of self-stabilizing spanning-tree construction algorithms.
\newblock Technical report, EPFL, October 2003.

\bibitem{GataniRG05}
Luca Gatani, Giuseppe~Lo Re, and Salvatore Gaglio.
\newblock A dynamic distributed algorithm for multicast path setup.
\newblock In {\em 11th International Euro-Par Parallel Processing
  Conference,(Euro-Par 2005)}, volume 3648 of {\em Lecture Notes in Computer
  Science}, pages 595--605. Springer, 2005.

\bibitem{Ghosh00}
S.~Ghosh.
\newblock {Agents, distributed algorithms, and stabilization}.
\newblock {\em Computing and Combinatorics (COCOON 2000), Springer LNCS}, pages
  242--251, 2000.

\bibitem{Goemans06}
Michel~X. Goemans.
\newblock Minimum bounded degree spanning trees.
\newblock In {\em 47th Annual IEEE Symposium on Foundations of Computer Science
  (FOCS 2006)}, pages 273--282. IEEE Computer Society, 2006.

\bibitem{GoudaS99}
Mohamed~G. Gouda and Marco Schneider.
\newblock Stabilization of maximal metric trees.
\newblock In {\em Workshop on Self-stabilizing Systems, (WSS 1999)}, pages
  10--17. IEEE Computer Society, 1999.

\bibitem{GuptaS99}
Sandeep K.~S. Gupta and Pradip~K. Srimani.
\newblock Using self-stabilization to design adaptive multicast protocol for
  mobile ad hoc networks.
\newblock In {\em Workshop on Mobile Networks and Computing, (DIMACS 99)}.
  Rutgers University, 1999.

\bibitem{GuptaS03}
Sandeep K.~S. Gupta and Pradip~K. Srimani.
\newblock Self-stabilizing multicast protocols for ad hoc networks.
\newblock {\em J. Parallel Distrib. Comput.}, 63(1):87--96, 2003.

\bibitem{HarelT84}
Dov Harel and Robert~Endre Tarjan.
\newblock Fast algorithms for finding nearest common ancestors.
\newblock {\em SIAM J. Comput.}, 13(2):338--355, 1984.

\bibitem{HermanM01}
T.~Herman and T.~Masuzawa.
\newblock {Self-Stabilizing Agent Traversal}.
\newblock {\em WSS01 Proceedings of the Fifth International Workshop on
  Self-Stabilizing Systems, Springer LNCS}, 2194:152--166, 2001.

\bibitem{HighamL01}
Lisa Higham and Zhiying Liang.
\newblock Self-stabilizing minimum spanning tree construction on
  message-passing networks.
\newblock In {\em 15th International Conference on Distributed Computing, (DISC
  2001)}, volume 2180 of {\em Lecture Notes in Computer Science}, pages
  194--208. Springer, 2001.

\bibitem{HuangC92}
Shing-Tsaan Huang and Nian-Shing Chen.
\newblock A self-stabilizing algorithm for constructing breadth-first trees.
\newblock {\em Inf. Process. Lett.}, 41(2):109--117, 1992.

\bibitem{HuangC93}
Shing-Tsaan Huang and Nian-Shing Chen.
\newblock Self-stabilizing depth-first token circulation on networks.
\newblock {\em Distributed Computing}, 7(1):61--66, 1993.

\bibitem{HuangW97}
Shing-Tsaan Huang and Lih-Chyau Wuu.
\newblock Self-stabilizing token circulation in uniform networks.
\newblock {\em Distributed Computing}, 10(4):181--187, 1997.

\bibitem{HuangL02}
Tetz~C. Huang and Ji-Cherng Lin.
\newblock A self-stabilizing algorithm for the shortest path problem in a
  distributed system.
\newblock {\em Computers $\&$ Mathematics with Applications}, 43(1-2):103 --
  109, 2002.

\bibitem{ImaseW91}
Makoto Imase and Bernard~M. Waxman.
\newblock Dynamic steiner tree problem.
\newblock {\em SIAM J. Discrete Math.}, 4(3):369--384, 1991.

\bibitem{Jain98}
Kamal Jain.
\newblock Factor 2 approximation algorithm for the generalized steiner network
  problem.
\newblock In {\em 39th Annual Symposium on Foundations of Computer Science,
  (FOCS 1998)}, pages 448--457. IEEE, 1998.

\bibitem{JohnenT03}
Colette Johnen and S{\'e}bastien Tixeuil.
\newblock Route preserving stabilization.
\newblock In {\em Self-Stabilizing Systems}, pages 184--198, 2003.

\bibitem{KakugawaY02}
Hirotsugu Kakugawa and Masafumi Yamashita.
\newblock Uniform and self-stabilizing fair mutual exclusion on unidirectional
  rings under unfair distributed daemon.
\newblock {\em J. Parallel Distrib. Comput.}, 62(5):885--898, 2002.

\bibitem{KameiK02}
S.~Kamei and H.~Kakugawa.
\newblock A self-stabilizing algorithm for the steiner tree problem.
\newblock In {\em SRDS}, pages 396--, 2002.

\bibitem{KameiK04}
S.~Kamei and H.~Kakugawa.
\newblock A self-stabilizing algorithm for the steiner tree problem.
\newblock {\em IEICE TRANSACTIONS on Information and System},
  E87-D(2):299--307, 2004.

\bibitem{KargerKT95}
David~R. Karger, Philip~N. Klein, and Robert~Endre Tarjan.
\newblock A randomized linear-time algorithm to find minimum spanning trees.
\newblock {\em J. ACM}, 42(2):321--328, 1995.

\bibitem{KlasingKN08}
Ralf Klasing, Adrian Kosowski, and Alfredo Navarra.
\newblock Taking advantage of symmetries: Gathering of asynchronous oblivious
  robots on a ring.
\newblock In {\em 12th International Conference on Principles of Distributed
  Systems, (OPODIS 2008)}, volume 5401 of {\em Lecture Notes in Computer
  Science}, pages 446--462. Springer, 2008.

\bibitem{KlasingMP08}
Ralf Klasing, Euripides Markou, and Andrzej Pelc.
\newblock Gathering asynchronous oblivious mobile robots in a ring.
\newblock {\em Theor. Comput. Sci.}, 390(1):27--39, 2008.

\bibitem{KonemannR00}
Jochen K{\"o}nemann and R.~Ravi.
\newblock A matter of degree: improved approximation algorithms for
  degree-bounded minimum spanning trees.
\newblock In {\em Thirty-Second Annual ACM Symposium on Theory of Computing,
  (STOC 2000)}, pages 537--546, 2000.

\bibitem{KonemannR02}
Jochen K{\"o}nemann and R.~Ravi.
\newblock A matter of degree: Improved approximation algorithms for
  degree-bounded minimum spanning trees.
\newblock {\em SIAM J. Comput.}, 31(6):1783--1793, 2002.

\bibitem{KonemannR05}
Jochen K{\"o}nemann and R.~Ravi.
\newblock Primal-dual meets local search: Approximating msts with nonuniform
  degree bounds.
\newblock {\em SIAM J. Comput.}, 34(3):763--773, 2005.

\bibitem{KormanK07}
Amos Korman and Shay Kutten.
\newblock Distributed verification of minimum spanning tree.
\newblock {\em Distributed Computing}, 20(4):253--266, 2007.

\bibitem{KormanKM11}
Amos Korman, Shay Kutten, and Toshimitsu Masuzawa.
\newblock Fast and compact self stabilizing verification, computation, and
  fault detection of an mst.
\newblock In {\em 30th Annual ACM Symposium on Principles of Distributed
  Computing, (PODC 2011)}, pages 311--320. ACM, 2011.

\bibitem{KormanKP10}
Amos Korman, Shay Kutten, and David Peleg.
\newblock Proof labeling schemes.
\newblock {\em Distributed Computing}, 22:215--233, 2010.

\bibitem{Kurskal64}
J.B Kurskal.
\newblock Multi-dimensional scaling by optimizing goodness of fit to a
  non-metric hypothesis.
\newblock {\em Psychometrika 29}, pages 1--27, March 1964.

\bibitem{KuttenP98}
Shay Kutten and David Peleg.
\newblock Fast distributed construction of small {\it }-dominating sets and
  applications.
\newblock {\em J. Algorithms}, 28(1):40--66, 1998.

\bibitem{LavaultV08}
Christian Lavault and Mario Valencia-Pabon.
\newblock A distributed approximation algorithm for the minimum degree minimum
  weight spanning trees.
\newblock {\em J. Parallel Distrib. Comput.}, 68(2):200--208, 2008.

\bibitem{LotkerPP01}
Zvi Lotker, Boaz Patt-Shamir, and David Peleg.
\newblock Distributed mst for constant diameter graphs.
\newblock In {\em 20th Annual ACM Symposium on Principles of Distributed
  Computing, (PODC 2001)}, pages 63--71. ACM, 2001.

\bibitem{ParkMHT90}
Jungho Park, Toshimitsu Masuzawa, Kenichi Hagihara, and Nobuki Tokura.
\newblock Distributed algorithms for reconstructing mst after topology change.
\newblock In {\em 4th International Workshop on Distributed Algorithms (WDAG
  1990)}, pages 122--132. LNCS, 1990.

\bibitem{ParkMHT92}
Jungho Park, Toshimitsu Masuzawa, Ken'ichi Hagihara, and Nobuki Tokura.
\newblock Efficient distributed algorithm to solve updating minimum spanning
  tree problem.
\newblock {\em Systems and Computers in Japan}, 23(3):1--12, 1992.

\bibitem{Peleg00}
David Peleg.
\newblock {\em Distributed computing: a locality-sensitive approach}.
\newblock Society for Industrial and Applied Mathematics, Philadelphia, PA,
  USA, 2000.

\bibitem{PelegR00}
David Peleg and Vitaly Rubinovich.
\newblock A near-tight lower bound on the time complexity of distributed
  minimum-weight spanning tree construction.
\newblock {\em SIAM J. Comput.}, 30(5):1427--1442, 2000.

\bibitem{PetitV07}
Franck Petit and Vincent Villain.
\newblock Optimal snap-stabilizing depth-first token circulation in tree
  networks.
\newblock {\em J. Parallel Distrib. Comput.}, 67(1):1--12, 2007.

\bibitem{PettieR02}
Seth Pettie and Vijaya Ramachandran.
\newblock An optimal minimum spanning tree algorithm.
\newblock {\em J. ACM}, 49(1):16--34, 2002.

\bibitem{Prencipe01}
Giuseppe Prencipe.
\newblock Instantaneous actions vs. full asynchronicity : Controlling and
  coordinating a set of autonomous mobile robots.
\newblock In {\em 7th Italian Conference on Theoretical Computer Science,
  (ICTCS 2001)}, volume 2202 of {\em Lecture Notes in Computer Science}, pages
  154--171. Springer, 2001.

\bibitem{Prim57}
R.~C. Prim.
\newblock Shortest connection networks and some generalizations.
\newblock {\em Bell System Technical Journal}, November:1389--1401, 1957.

\bibitem{PromelS97}
Hans~J{\"u}rgen Pr{\"o}mel and Angelika Steger.
\newblock Rnc-approximation algorithms for the steiner problem.
\newblock In {\em 14th Annual Symposium on Theoretical Aspects of Computer
  Science, (STACS 1997)}, volume 1200 of {\em Lecture Notes in Computer
  Science}, pages 559--570. Springer, 1997.

\bibitem{RajagopalanV99}
Sridhar Rajagopalan and V.~Vazirani.
\newblock On the bidirected cut relaxation for the metric steiner tree problem.
\newblock In {\em Tenth Annual ACM-SIAM Symposium on Discrete Algorithms,(SODA
  1999)}. ACM.

\bibitem{RaviMRRH93}
R.~Ravi, Madhav~V. Marathe, S.~S. Ravi, Daniel~J. Rosenkrantz, and Harry
  B.~Hunt III.
\newblock Many birds with one stone: multi-objective approximation algorithms.
\newblock In {\em 25th Annual ACM Symposium on Theory of Computing, (STOC
  1993)}, pages 438--447, 1993.

\bibitem{RobinsZ05}
Gabriel Robins and Alexander Zelikovsky.
\newblock Tighter bounds for graph steiner tree approximation.
\newblock {\em SIAM J. Discrete Math.}, 19(1):122--134, 2005.

\bibitem{Rovedakis09}
Stephane Rovedakis.
\newblock {\em Algorithmes auto-stabilisants de constructions d'arbres
  couvrants.}
\newblock PhD thesis, Université d'Evry Val d'Essonne, 2009.

\bibitem{Santoro06}
Nicola Santoro.
\newblock {\em Design and Analysis of Distributed Algorithms (Wiley Series on
  Parallel and Distributed Computing)}.
\newblock Wiley-Interscience, 2006.

\bibitem{SinghL07}
Mohit Singh and Lap~Chi Lau.
\newblock Approximating minimum bounded degree spanning trees to within one of
  optimal.
\newblock In {\em 39th Annual ACM Symposium on Theory of Computing, (STOC
  2007)}, pages 661--670. ACM, 2007.

\bibitem{Tarjan83}
Daniel~Dominic Sleator and Robert~Endre Tarjan.
\newblock A data structure for dynamic trees.
\newblock {\em J. Comput. Syst. Sci.}, 26(3):362--391, 1983.

\bibitem{Sollin61}
G.~Sollin.
\newblock Exposé au séminaire de c. berge, ihp. repris in extenso dans méthodes
  et modèles de la recherche opérationnelle.
\newblock volume~2, pages 33--45, 1961.

\bibitem{SuzukiY06}
Ichiro Suzuki and Masafumi Yamashita.
\newblock Erratum: Distributed anonymous mobile robots: Formation of geometric
  patterns.
\newblock {\em SIAM J. Comput.}, 36(1):279--280, 2006.

\bibitem{SuzukiY99}
Ichiro Suzuki and Musafumi Yamashita.
\newblock Distributed anonymous mobile robots: formation of geometric patterns.
\newblock {\em SIAM J. Comput.}, 28:1347--1363, 1999.

\bibitem{TakahashiM80}
H.~Takahashi and A.~Matsuyama.
\newblock An approximate solution for the steiner problem in graphs.
\newblock {\em Math. Jap.}, 24:573--577, 1980.

\bibitem{Tarjan79}
Robert~Endre Tarjan.
\newblock Applications of path compression on balanced trees.
\newblock {\em J. ACM}, 26:1--38, 1979.

\bibitem{Tel94}
Gerard Tel.
\newblock {\em Introduction to distributed algorithms}.
\newblock Cambridge University Press, New York, NY, USA, 1994.

\bibitem{TetaliW91}
Prasad Tetali and Peter Winkler.
\newblock On a random walk problem arising in self-stabilizing token
  management.
\newblock In {\em 10th Annual ACM Symposium on Principles of Distributed
  Computing, (PODC 1990)}, pages 273--280. ACM, 1991.

\bibitem{Tixeuil06}
Sébastien Tixeuil.
\newblock Vers l'auto-stabilisation des systèmes à grande echelle.
\newblock Habilitation à diriger les recherches, Université Paris-Sud XI, May
  2006.

\bibitem{WuWW86}
Ying-Fung Wu, Peter Widmayer, and C.~K. Wong.
\newblock A faster approximation algorithm for the steiner problem in graphs.
\newblock {\em Acta Inf.}, 23(2):223--229, 1986.

\bibitem{YamashitaK96ja}
Masafumi Yamashita and Tsunehiko Kameda.
\newblock Computing on anonymous networks: Part i-characterizing the solvable
  cases.
\newblock {\em IEEE Trans. Parallel Distrib. Syst.}, 7(1):69--89, 1996.

\bibitem{YamashitaK96jb}
Masafumi Yamashita and Tsunehiko Kameda.
\newblock Computing on anonymous networks: Part ii-decision and membership
  problems.
\newblock {\em IEEE Trans. Parallel Distrib. Syst.}, 7(1):90--96, 1996.

\bibitem{YamashitaS10}
Musafumi Yamashita and Ichiro Suzuki.
\newblock Characterizing geometric patterns formable by oblivious anonymous
  mobile robots.
\newblock {\em Theor. Comput. Sci.}, 411(26-28):2433--2453, 2008.

\bibitem{Zelikovsky93}
Alexander Zelikovsky.
\newblock An 11/6-approximation algorithm for the network steiner problem.
\newblock {\em Algorithmica}, 9(5):463--470, 1993.

\end{thebibliography}

\end{document}